\documentclass[conference]{IEEEtran}
\usepackage[utf8]{inputenc}
\usepackage[english]{babel}
\usepackage{graphicx}
\usepackage{stmaryrd}
\usepackage{amsmath}
\usepackage{amssymb}
\usepackage{amsfonts}
\usepackage{amsthm}
\usepackage{listings}
\usepackage{color}
\usepackage{proof}
\usepackage{mathpartir}
\usepackage{thm-restate}
\usepackage{comment}
\usepackage{mathtools}
\usepackage{framed}
\usepackage{multicol}
\usepackage[hidelinks]{hyperref}
\usepackage{url}
\usepackage[section]{placeins}
\usepackage{enumitem}
\usepackage[numbers,sort]{natbib}
\usepackage[nameinlink]{cleveref}
\usepackage{chngcntr}
\usepackage{microtype}
\usepackage{algorithm}
\usepackage{algpseudocode}

\usepackage{tikz}

\newcommand{\pie}[1]{%
\begin{tikzpicture}
    \draw (0,0) circle (1ex);\fill (0,-1ex) arc (-90:#1-90:1ex) -- (0,0) -- cycle;
\end{tikzpicture}%
}

\theoremstyle{plain}

\theoremstyle{definition}
\newtheorem{definition}{Definition}

\makeatletter
\newcommand{\oset}[3][0ex]{%
  \mathrel{\mathop{#3}\limits^{
    \vbox to#1{\kern-1\ex@
    \hbox{$\scriptstyle#2$}\vss}}}}
\makeatother

\newcommand{\xangle}[1]{\ensuremath{\langle #1 \rangle}}
\newcommand{\xbrace}[1]{\ensuremath{\{ #1 \}}}

\newcommand{\ltype}[2]{\ensuremath{#1 @ #2}}
\newcommand{\val}[2]{\ensuremath{\llparenthesis #1 \rrparenthesis_{#2}}}
\newcommand{\assign}{\ensuremath{=}}
\newcommand{\oblivassign}{\ensuremath{\;\texttt{?=}\;}}
\newcommand{\andexp}{\ensuremath{\texttt{\&\&}}}
\newcommand{\orexp}{\ensuremath{\texttt{||}}}
\newcommand{\caret}{\ensuremath{\texttt{\string^}}}
\newcommand{\sizeop}[1]{\ensuremath{{#1_{\textit{size}}}}}

\newcommand{\pc}{\ensuremath{\textnormal{\textit{pc}}}}
\newcommand{\pcstack}{\ensuremath{\overline{\pc}}}
\newcommand{\bit}{\ensuremath{b}}
\newcommand{\xbit}[1]{\ensuremath{\ifthenelse{\equal{#1}{0}}{0}{1}}}
\newcommand{\bitstack}{\ensuremath{\overline{\bit}}}
\newcommand{\ch}{\ensuremath{\textnormal{\textit{ch}}}}

\newcommand{\extends}{\ensuremath{\lessapprox}}
\newcommand{\phantomextends}{\ensuremath{\oset{\bullet}{\extends}}}
\newcommand{\lambdaextends}{\ensuremath{\extends_{\Lambda}}}
\newcommand{\lambdaphantomextends}{\ensuremath{\phantomextends_{\Lambda}}}

\newcommand{\maxpotential}{\ensuremath{q^\Lambda_{\textit{max}}}}

\newcommand{\Configuration}{\ensuremath{Q}}
\newcommand{\Configurationaux}{\ensuremath{\tilde{Q}}}
\newcommand{\Consumer}{\ensuremath{\textcolor[rgb]{0,0,1}{C}}}
\newcommand{\Consumeraux}{\ensuremath{\textcolor[rgb]{0,0,1}{\tilde{C}}}}
\newcommand{\consumer}[1]{
    \ensuremath{
        \textcolor[rgb]{0,0,1}{\boldsymbol (}
        #1
        \textcolor[rgb]{0,0,1}{\boldsymbol )}
    }
}
\newcommand{\Producer}{\ensuremath{\textcolor[rgb]{1,0,0}{P}}}
\newcommand{\Produceraux}{\ensuremath{\textcolor[rgb]{1,0,0}{\tilde{P}}}}
\newcommand{\producer}[1]{
    \ensuremath{
        \textcolor[rgb]{1,0,0}{\boldsymbol (}
        #1
        \textcolor[rgb]{1,0,0}{\boldsymbol )}^{\ch}
    }
}

\newcommand{\oblivio}{\textit{OblivIO}}

\newcommand{\eval}[2]{\ensuremath{\xangle{#1} \Downarrow #2}}

\newcommand{\inttype}{\ensuremath{\textnormal{\textit{int}}}}
\newcommand{\stringtype}{\ensuremath{\textnormal{\textit{string}}}}

\newcommand{\intenv}{\ensuremath{\pi}}
\newcommand{\networkstrategy}{\ensuremath{\omega}}
\newcommand{\hst}{\ensuremath{h}}
\newcommand{\timestamp}{\ensuremath{t}}

\newcommand{\ctx}[3]{\ensuremath{#1, #2, #3}}
\newcommand{\Ctx}[5]{\ensuremath{#1, #2, #3; #4; #5}}
\newcommand{\Ctxaux}[6]{\ensuremath{#1, #2, #3; #4; #5, #6}}

\newcommand{\ltriple}[3]{\ensuremath{#1; #2; #3}}

\newcommand{\xell}[1]{{\ensuremath{\ell_{#1}}}}
\newcommand{\ladv}{\xell{\textnormal{\textit{adv}}}}
\newcommand{\lval}{\xell{\textnormal{\textit{val}}}}
\newcommand{\lmode}{\xell{\textnormal{\textit{mode}}}}
\newcommand{\gammaequiv}[1]{\ensuremath{\approx^{\Gamma}_{#1}}}
\newcommand{\deltaequiv}[1]{\ensuremath{\approx^{\Delta}_{#1}}}
\newcommand{\piequiv}[1]{\ensuremath{\approx^{\Pi}_{#1}}}
\newcommand{\lambdaequiv}[1]{\ensuremath{\approx^{\Lambda}_{#1}}}

\newcommand{\longrightarrowdbl}{
    \longrightarrow
    \mkern-14mu
    \rightarrow
}

\newcommand{\xrightarrowdbl}[2][]{
    \xrightarrow[#1]{#2}
    \mkern-14mu
    \rightarrow
}

\definecolor{backgroundColor}{rgb}{1.0, 0.98, 0.95}
\definecolor{commentsColor}{rgb}{0.35, 0.35, 0.35}
\definecolor{keywordsColor}{rgb}{0.0, 0.0, 0.65}
\definecolor{stringColor}{rgb}{0.5, 0.0, 0.15}

\usepackage{stix} 

\def\istechnicalreport{} 

\title{OblivIO: Securing reactive programs by oblivious execution with bounded traffic overheads}

\author{
    \IEEEauthorblockN{Jeppe Fredsgaard Blaabjerg}
    \IEEEauthorblockA{Aarhus University\\jfblaa@cs.au.dk}
    \and
    \IEEEauthorblockN{Aslan Askarov}
    \IEEEauthorblockA{Aarhus University\\aslan@cs.au.dk}
}

\begin{document}

\thispagestyle{empty}
\setcounter{page}{1}
\pagestyle{plain}

\maketitle

\begin{abstract}
Traffic analysis attacks remain a significant problem for online security. Communication between nodes can be observed by network level attackers as it inherently takes place in the open. Despite online services increasingly using encrypted traffic, the shape of the traffic is not hidden.
To prevent traffic analysis, the shape of a system's traffic must be independent of secrets.

We investigate adapting the data-oblivious approach the reactive setting and present \oblivio{}, a secure language for writing reactive programs driven by network events. Our approach pads with dummy messages to hide which program sends are genuinely executed. We use an information-flow type system to provably enforce timing-sensitive noninterference. The type system is extended with potentials to bound the overhead in traffic introduced by our approach.
We address challenges that arise from joining data-oblivious and reactive programming and demonstrate the feasibility of our resulting language by developing an interpreter that implements security critical operations as constant-time algorithms.
\end{abstract}

\section{Introduction}
\label{sec:introduction}
%
%
Online communication between network nodes inherently takes place in the open. Even when using encrypted traffic, network level attackers can still observe the traffic shape including timing, bandwidth, and destination. Traffic analysis attacks are possible when secrets affect the traffic shape of a system and enable attackers to infer often highly detailed, sensitive information about user interactions across a wide range of online services \cite{chen2010side}.

Strategies for mitigating traffic analysis have predominantly been developed at the system-level \cite{cherubin2017website} where solutions can be applied to existing applications and services in a black-box fashion. However, this approach incurs significant -- if not practically infeasible -- overheads in latency and bandwidth in order to achieve a high degree of security \cite{dyer2012peek}. System-level approaches are program-agnostic by nature and hence performs some traffic padding even when no possible control flow would produce genuine traffic.

Traffic analysis is made possible by observable attributes of shape of traffic, such as time.
Timing leaks have been extensively studied in the programming languages community. Language-based approaches are particularly appealing as language-level information can make the enforcement precise.
Compared with general approaches for mitigating timing channels \cite{askarov2010predictive,bastys2020clockwork,russo2006closing,zhang2012language}, our focus on traffic analysis enables us to exploit dummy traffic as a means to reduce overheads introduced by the enforcement.

Timing channels are also a significant security concern for the cryptographic community. Almost all modern cryptography deals with timing side-channels by \textit{constant-time} programming \cite{cauligi2019towards}. Constant-time programs may neither branch nor access memory depending on secrets. This ensures that programs do not leak via their execution time. However, these restrictions make adhering to constant-time programming hard and requires developers to deviate from conventional programming practices \cite{almeida2016verifying}.
Data-oblivious languages \cite{zahur2015obliv,liu2015oblivm} ease the restrictions imposed by constant-time programming by introducing oblivious conditionals and oblivious memory accesses.
Oblivious conditionals execute both branches, but negate the unwanted side-effects of the non-chosen branch. The data-oblivious approach does not permit loops with secret guards as loop termination would cause a leak. Instead, the programmer can provide a public upper bound on the number of loop iterations and obliviously branch on secrets inside the loop.

The reactive programming model is a natural fit for online services and IoT devices. In this model, applications are written as a collection of event handlers that are triggered in response to observing associated events. This allows for scalable, reliable software, as it assumes no control over, or knowledge of, the number or timing of events \cite{nurkiewicz2016reactive}.
However, as shown by \citet{chen2010side}, many applications written in this model leak sensitive information. \Cref{lst:insufficient_funds} shows a simple handler for transferring money from one account to another, for example as part of an online transaction. If the amount to transfer exceeds the balance of the source account, no money is transferred and an error message is sent. This message will be observable if sent over standard internet protocols.

\begin{lstlisting}[float,floatplacement=!htbp,caption=Traffic leak,label=lst:insufficient_funds,mathescape]
TRANSFER(from: int,amount: int,to: int) {
    if amount <= balance[from]
    then {
        balance[from] -= amount;
        balance[to]     += amount;
    }
    else send(ERROR,(amount,balance[from]));
}
\end{lstlisting}

%
In this paper we consider the problem of mitigating traffic analysis in a reactive setting. We consider traffic leaks through the observable timing, size, and destination of messages. Our strategy for preventing traffic leaks is to adapt the data-oblivious approach to the reactive setting: execute both branches of sensitive conditionals, pad the traffic with dummy messages at send commands in non-chosen branches, and obliviously execute handlers upon receiving dummy messages. We implement this strategy in the design of \oblivio{}, a secure language for reactive programs, driven by network events. Programs written in the language are secure by construction, satisfying progress-sensitive, timing-sensitive noninterference, while incurring a bounded overhead in the number of dummy messages. The main contributions of this paper are:
\begin{itemize}
    \item We adapt the data-oblivious approach to the reactive setting and develop \oblivio{}, a statically typed language for writing reactive programs.
    \item We provide a security-labelled type system and prove that well-typed programs in \oblivio{} are secure against network level attackers, satisfying progress-sensitive, timing-sensitive noninterference.
    \item We bound the traffic overhead introduced by our approach by typing commands and network channels with potentials and show that the overhead is bounded by the potentials.
    \item We demonstrate the practicality of our approach by developing a real-time interpreter that implements oblivious branching and constant-time algorithms for security critical operations.
\end{itemize}

Our model builds upon the work of \citet{bohannon2009reactive} that gives the foundational model for noninterference for reactive programs. They develop a type system that enforces timing- and termination-insensitive noninterference. Their model does not consider leaks to network level attackers.

Like \citet{bohannon2009reactive}, we rely on static enforcement of noninterference. Static enforcement is well-suited for our problem as both the code and the types of reactions in the program are well-known and as we enforce progress-sensitive security. Here, we benefit from the precision of static analysis with regards to termination behaviour, which is difficult to achieve in purely dynamic setting. Many dynamic approaches use hybrid techniques where they inspect non-chosen branches of conditionals to recover this precision \cite{moore2011static,askarov2015hybrid}. Static enforcement would not be suitable if code or the types of reactions were not statically known.

The rest of this paper is structured as follows. In \Cref{sec:language} we discuss how we adapt the data-oblivious approach to the reactive setting and provide the semantics of \oblivio{}. The type system is presented in \Cref{sec:enforcement}. \Cref{sec:noninterference} presents the formal security condition and noninterference theorem, while \Cref{sec:enforcementoverhead} presents a bound on the traffic overhead introduced by our approach. In \Cref{sec:example} we demonstrate the language by example.
We discuss how we have implemented the language semantics in our interpreter in \Cref{sec:implementation}.
We discuss our approach and compare it with existing approaches in \Cref{sec:discussion} and provide related work in \Cref{sec:related_work} before we conclude in \Cref{sec:conclusion}.


\section{Language}
\label{sec:language}
In this section we provide brief background on reactive programs and data-obliviousness and discuss the terminology we will use in this paper. We then give our threat model and provide the semantics of \oblivio{}.

\subsection{Background and terminology}
\label{sec:background_and_terminology}
We briefly outline reactive programs, constant-time programming, and data-obliviousness, and the distinction between genuine and dummy traffic.

\subsubsection*{Reactive programs}
Reactive programs consist of handlers that are triggered by associated events, imposing no structure on the order or timing of events. This allows for reliable programs that scale easily by introducing new events and handlers.

Reactive programming is a popular model for online services, where client interaction is intermittent and irregular, due to its inherent flexibility, and languages such as JavaScript follow this model. Online services present radically different security challenges from applications running on a single machine or a closed network as communication channels use public, potentially compromised infrastructure.

\subsubsection*{Constant-time programming}
Constant-time programming is at the heart of almost all modern cryptography \cite{cauligi2019towards,libsodium,openssl}. Under this paradigm, programs may not branch on secrets, access secret memory locations, or use secrets in variable-time operations such as division. This ensures that programs are secure, but makes them difficult for developers to write \cite{almeida2016verifying}.

\subsubsection*{Data-oblivious programming}
Data-oblivious programming relaxes some of the restrictions imposed by constant-time programming. The principal idea of data-oblivious languages is oblivious conditionals \cite{zahur2015obliv,liu2015oblivm}. Instead of a branch in the control-flow, oblivious conditionals execute both branches and negate the unwanted side-effects in the non-chosen branch. This prevents branching-related timing differences and ensures that secrets do not affect whether a command is executed or whether an expression is evaluated.

\subsubsection*{Dummy messages}
A common strategy for mitigating traffic analysis is to introduce dummy messages. Dummy messages are additional, fake messages introduced by the enforcement mechanism \cite{diaz2010impact,apthorpe2019keeping,dyer2012peek,blaabjerg2021towards}.
Informally, dummy messages should only serve to hide which traffic is genuine and should not not alter the semantics of a system. This does not always hold in practice, e.g., as dummy messages may introduce overhead in latency or execution time that affects time reads. 

\subsection{A data-oblivious approach to reactive programming}
Extending the data-oblivious approach to programs with observable IO is not straightforward as observable side-effects cannot safely be suppressed.
If an assign statement occurs in a sensitive conditional, the assignment must be suitably padded in order to hide the genuine branch. Otherwise, the value would be unsafe to send at any later point, as a network level attacker could observe the size of the value and thus infer the genuine branch.
Suppressing messages in non-chosen branches is also not an option as the absence of traffic would be observed. Instead, any message sent in a non-chosen branch must be replaced a convincing dummy message. Additionally, such dummy message must be reacted to and handled in a convincing manner, including possibly generating new traffic.

Combining reactive and data-oblivious programming therefore introduces the need for bounding the number of dummy messages generated by a system. To demonstrate the problem, we consider the interaction between the two small programs below, where $\texttt{oblif}$ is a primitive for oblivious branching. For every message received, two new messages are sent. If this program were allowed, the amount of dummy traffic would grow exponentially over time, dominating real computation and crippling the utility of the system.

\noindent
\begin{minipage}[!htbp]{\columnwidth}
\begin{multicols}{2}
\begin{lstlisting}[captionpos=none,mathescape]
PING (x: int) {
    oblif x
    then send(PONG,1);
    else send(PONG,0);
}
\end{lstlisting}

\columnbreak

\begin{lstlisting}[captionpos=none,mathescape]
PONG (x: int) {
    oblif x
    then send(PING,1);
    else send(PING,0);
}
\end{lstlisting}
\end{multicols}
\end{minipage}

To address this issue, we take inspiration from static resource analysis, where types are annotated with potentials \cite{hofmann2003static,hoffmann2010amortized,hoffmann2012resource}. We discuss how we use potentials in \Cref{sec:enforcement} and show a bound on the overhead in traffic in \Cref{sec:enforcementoverhead}.

\subsection{Threat model}
We consider the security of a network node running a reactive program that processes incoming network messages sequentially using a single event loop. Every handler in the program defines a channel endpoint, and other nodes can send messages on the associated channel to trigger execution of the handler. We assume that all nodes run \oblivio{} programs. We consider a lattice $\mathcal{L}$ of security levels $\ell$ and assign levels to each channel. The lattice has distinguished bottom element $\bot$, corresponding to the network level.

We consider an active adversary who can be one of the other network nodes. The adversary knows the program being run and knows its initial secrets up to some level $\ladv$. We assume that the adversary can observe the presence, size, time, and channel of all network messages, can drop messages, and perform replay attacks.
We assume that the contents of network messages is hidden from unprivileged parties by encryption, and that the adversary can decrypt and read the contents of messages sent on channels up to level $\ladv$.

\subsection{Language}
\oblivio{} is a simple, statically typed language for data-oblivious, reactive programs. Programs written in \oblivio{} consist of a number of handlers that are triggered by associated events. We let events model network messages. Each handler in a program defines a channel endpoint, such that messages sent on the channel get processed by the handler. Events are processed sequentially and during execution, \oblivio{} programs change between consumer states, $\Consumer$, that wait for the next network message, and producer states, $\Producer$, that execute the appropriate handler for the current network message. We model incoming network messages using strategies, functions that map network traces to messages.

We let $n$ range over integer literals and $s$ range over string literals. We let $x$ range over variables. Values in \oblivio{} are written $\val{v}{z}$ and consist of a base value $v$ and a size $z$. Base values $v$ are either an integer or a string and sizes $z$ are non-negative integers. We assume an attacker that can observe the size of any network message sent. Therefore, we cannot easily protect secrets of arbitrary size. Instead, we settle for protecting secrets up to a publicly known upper bound. This upper bound can be any size greater than or equal to the actual size of the secret. That is, we assume that value $\val{v}{z}$ can be padded to $\val{v}{z'}$ for any $z' \geq z$.
We assume a function $\textit{size}$ for computing the size of base values $v$ and maintain well-formed values such that for any $\val{v}{z}$ we have $\textit{size}(v) \leq z$. We provide the formal definition in \Cref{sec:enforcement}.

At runtime, the system maintains a persistent, global store $\mu$ mapping variables to size-annotated values. Handlers define a variable $x$ for binding the value of a received message as a read-only value in local memory $m$. Local memory $m$ is cleared when the execution of the handler finishes.

We assume two distinct types of channels: local and network. Local channels model non-network channels, where the presence of messages is not observable to the attacker, e.g., keyboard input or sensor readings. 
We maintain a queue for each local channel to prevent the presence of messages on one channel from tainting reads on another. We let $\intenv$ denote a local environment: a mapping from local channels to a stream of value options. We let $\bullet$ denote that no value is available. We choose this representation as it makes it easier to bound the overhead in traffic (\Cref{sec:enforcementoverhead}).
Network channels model channels where communication can be observed. Thus, the presence of messages on all network channels is public. This allows network messages to be combined into a single queue. We model incoming network traffic using network strategies $\networkstrategy$, functions from network traces to messages of form $\ch(\timestamp,\bit,\val{v}{z})$. We choose this model over the equally expressive model of streams (\cite{clark2008non}) as it does not pre-compute future messages, thereby more intuitively capturing the interactive nature of network communication and allowing us to more faithfully state our overhead theorem (\Cref{theorem:enforcementoverhead}).

The distinction between local and network channels is motivated by the observation that messages with secret presence must be handled specially. Observable side-effects such as sending message or introducing latency would leak the presence of such message.
We therefore minimise the interface with messages with secret presence by not considering them events that handlers react to. Instead, we allow programs to read from local channels using input statements.

We now explain the formal semantics and non-standard features of the language.
We assume a security lattice $\mathcal{L}$ of security levels $\ell$ with bottom element $\bot$, lattice ordering $\sqsubseteq$, and least upper bound operation $\sqcup$. Level $\bot$ corresponds to the network level. We also refer to this level as \textit{public}.
We assume that all nodes run \oblivio{} programs. We assume that message contents can be sufficiently hidden, e.g., using encryption, but that their presence, size, and destination are publicly observable.

\Cref{def:syntax} presents the syntax of \oblivio{}. A program $p$ consists of a number of handlers $\ch(x) \xbrace{c}$, where $\ch$ is the network channel associated with the handler and variable $x$ binds the incoming message in local memory. Commands $c$ are largely standard, though with novel commands to facilitate oblivious execution. Commands $\texttt{stop}$ and $\texttt{pop}$ are only used internally and therefore not part of the syntax of the language.

\begin{figure}[!htbp]
\centering
\begin{align*}
    p \Coloneqq\;
    & \cdot \mid \ch(x) \xbrace{c}; p\\
    c \Coloneqq\;
    & \texttt{skip} \mid c;c \mid x \assign e \mid x \oblivassign e \mid x \oblivassign \texttt{input}(\ch, e)\\
    \mid\; &\texttt{send}(\ch,e) \mid \texttt{if $e$ then $c$ else $c$}\\
    \mid\; &\texttt{while $e$ do $c$} \mid \texttt{oblif $e$ then $c$ else $c$}\\
    e \Coloneqq\;
    &n \mid s \mid x \mid e \oplus e\\
    \oplus \Coloneqq\;
    &\texttt{+} \mid \texttt{-} \mid \texttt{*} \mid \texttt{=} \mid \texttt{!=} \mid \texttt{<} \mid \texttt{<=} \mid \texttt{>} \mid \texttt{>=} \mid \andexp \mid \orexp \mid \texttt{\caret}
\end{align*}
\caption{Syntax of the language}
\label{def:syntax}
\end{figure}

We use a big-step semantics for evaluating expressions (\Cref{def:semantics_expressions}). Expression $e$ are evaluated using local memory $m$ and global store $\mu$. For simplicity, memory $m$ is read-only in the presented model and binds just the current message value in the handler variable. We let local memory shadow the global store.

\begin{figure}[!htbp]
\centering
\begin{mathparpagebreakable}
\inferrule {
    \textit{size}(n) = z
} {
    \eval{n,m,\mu}{\val{n}{z}}
}
\and
\inferrule {
    \textit{size}(s) = z
} {
    \eval{s,m,\mu}{\val{s}{z}}
}
\and
\inferrule {
    x \in \textit{dom}(m)
} {
    \eval{x,m,\mu}{m(x)}
}
\and
\inferrule {
    x \notin \textit{dom}(m)
} {
    \eval{x,m,\mu}{\mu(x)}
}
\and
\inferrule {
    \eval{e_1,m,\mu}{\val{v_1}{z_1}} \\
    v_1 \oplus v_2 = v_3 \\\\
    \eval{e_2,m,\mu}{\val{v_2}{z_2}} \\
    z_1 \sizeop{\oplus} z_2 = z_3
} {
    \eval{e_1 \oplus e_2,m,\mu}{\val{v_3}{z_3}}
}
\end{mathparpagebreakable}
\caption{Semantics of evaluating expressions}
\label{def:semantics_expressions}
\end{figure}

We assume that binary operations $\oplus$ are total and have an associate operation $\sizeop{\oplus}$ for computing the size of the result from the sizes of the operands. We further assume that binary operations preserve well-formed values, that is, if $\textit{size}(v_1) \leq z_1$, $\textit{size}(v_2) \leq z_2$, and $v_1 \oplus v_2 = v_3$, then $\textit{size}(v_3) \leq z_1 \sizeop{\oplus} z_2$.
We discuss how $\sizeop{\oplus}$ can be implemented in \Cref{sec:implementation}.

\subsection{Command semantics}
Commands are evaluated using a small-step operational presented in \Cref{def:semantics_commands}.
\begin{figure*}[!htbp]
\centering
\begin{mathparpagebreakable}
\inferrule[Skip] { } {
    \xangle{\bitstack,\texttt{skip},m,\mu,\intenv,\hst}
    \longrightarrow
    \xangle{\bitstack,\texttt{stop},m,\mu,\intenv,\hst \Colon \textsf{skp}}
}
\and
\inferrule[Seq1] { 
    \xangle{\bitstack,c_1,m,\mu,\intenv,\hst}
    \xrightarrow{\alpha}
    \xangle{\bitstack',c'_1,m',\mu',\intenv',\hst'} \\
    c'_1 \neq \texttt{stop}
} {
    \xangle{\bitstack,c_1;c_2,m,\mu,\intenv,\hst}
    \xrightarrow{\alpha}
    \xangle{\bitstack',c'_1;c_2,m',\mu',\intenv',\hst'}
}
\and
\inferrule[Seq2] { 
    \xangle{\bitstack,c_1,m,\mu,\intenv,\hst}
    \xrightarrow{\alpha}
    \xangle{\bitstack',\texttt{stop},m',\mu',\intenv',\hst'}
} {
    \xangle{\bitstack,c_1;c_2,m,\mu,\intenv,\hst}
    \xrightarrow{\alpha}
    \xangle{\bitstack',c_2,m',\mu',\intenv',\hst'}
}
\and
\inferrule[Assign] { 
    \eval{e,m,\mu}{\val{v}{z}} \\
    \mu' = \mu[x \mapsto \val{v}{z}]
} {
    \xangle{\xbit{1} \Colon \bitstack,x \assign e,m,\mu,\intenv,\hst}
    \longrightarrow
    \xangle{\xbit{1} \Colon \bitstack,\texttt{stop},m,\mu',\intenv,\hst \Colon \textsf{asn}(x,e,z)}
}
\and
\inferrule[OblivAssign] {
    \mu(x) = \val{v_0}{z_0} \\
    \eval{e,m,\mu}{\val{v_1}{z_1}} \\
    z = \textit{max}(z_0,z_1) \\
    i =
    {\begin{cases}
        1
            &\textit{if } \bit = \xbit{1} \\
        0
            &\textit{if } \bit = \xbit{0}
    \end{cases}}
} {
    \xangle{\bit \Colon \bitstack,x \oblivassign e,m,\mu,\intenv,\hst}
    \longrightarrow
    \xangle{\bit \Colon \bitstack,\texttt{stop},m,\mu[x \mapsto \val{v_i}{z}],\intenv,\hst \Colon \textsf{casn}(x,e,z)}
}
\and
\inferrule[LocalInput] { 
    \mu(x) = \val{v_x}{z_x} \\
    \eval{e,m,\mu}{\val{n_e}{z_e}} \\
    z' = \textit{max}(z_x,n_e) \\
    v',\intenv' =
    {\begin{cases}
        v, \intenv[\ch \mapsto \textit{tl}]
            &\text{if }
            \bit = \xbit{1}
            \text{ and }
            \intenv(\ch) = \val{v}{z} \Colon \textit{tl}
            \text{ and } z \leq n_e \\
        v_x, \intenv[\ch \mapsto \textit{tl}]
            &\text{if }
            \bit = \xbit{1}
            \text{ and }
            \intenv(\ch) = \bullet \Colon \textit{tl} \\
        v_x, \intenv
            &\text{otherwise}
    \end{cases}}
} {
    \xangle{\bit \Colon \bitstack,x \oblivassign \texttt{input}(\ch,e),m,\mu,\intenv,\hst}
    \longrightarrow
    \xangle{\bit \Colon \bitstack,\texttt{stop},m,\mu[x \mapsto \val{v'}{z'}],\intenv',\hst \Colon \textsf{in}(x,\ch,e,z_x)}
}
\and
\inferrule[Send] { 
    \eval{e,m,\mu}{\val{v}{z}} \\
    \hst' = \hst \Colon \textsf{out}(\ch,e,z) \\
    \timestamp = \textit{time}(\hst')
} {
    \xangle{\bit \Colon \bitstack,\texttt{send}(\ch,e),m,\mu,\intenv,\hst}
    \xrightarrow{\overrightarrow{\ch}(\timestamp,\bit,\val{v}{z})}
    \xangle{\bit \Colon \bitstack,\texttt{stop},m,\mu,\intenv,\hst'}
}
\and
\inferrule[If] { 
    \eval{e,m,\mu}{\val{v}{z}} \\
    v \neq 0 \implies i = 1 \\
    v = 0 \implies i = 2
} {
    \xangle{\bitstack,\texttt{if $e$ then $c_1$ else $c_2$},m,\mu,\intenv,\hst}
    \longrightarrow \\\\
    \xangle{\bitstack,c_i,m,\mu,\intenv,\hst \Colon \textsf{br}(e,z,i)}
}
\and
\inferrule[While] {
    c' = \texttt{if $e$ then $c$; while $e$ do $c$ else skip}
} {
    \xangle{\xbit{1} \Colon \bitstack,\texttt{while $e$ do $c$},m,\mu,\intenv,\hst}
    \longrightarrow \\\\
    \xangle{\xbit{1} \Colon \bitstack,c',m,\mu,\intenv,\hst \Colon \textsf{whl}}
}
\and
\inferrule[Pop] { 
} {
    \xangle{\bit \Colon \bitstack,\texttt{pop},m,\mu,\intenv,\hst}
    \longrightarrow
    \xangle{\bitstack,\texttt{stop},m,\mu,\intenv,\hst \Colon \textsf{pop}}
}
\and
\inferrule[OblivIf] { 
    \bitstack = \bit \Colon \_ \\
    \eval{e,m,\mu}{\val{v}{z}} \\
    v \neq 0 \implies \bit_1 = \bit \land \bit_2 = \xbit{0} \\
    v = 0 \implies \bit_1 = \xbit{0} \land \bit_2 = \bit \\
    \hst' = \hst \Colon \textsf{obr}(e,z)
} {
    \xangle{\bitstack,\texttt{oblif $e$ then $c_1$ else $c_2$},m,\mu,\intenv,\hst}
    \longrightarrow
    \xangle{\bit_1 \Colon \bit_2 \Colon \bitstack,c_1;\texttt{pop};c_2;\texttt{pop},m,\mu,\intenv,\hst'}
}
\end{mathparpagebreakable}
\caption{Operational semantics of commands}
\label{def:semantics_commands}
\end{figure*}

Configuration $\xangle{\bitstack,c,m,\mu,\intenv,\hst}$ consists of a command $c$, memory $m$, store $\mu$, local environment $\intenv$, and history $\hst$.
We let history $\hst$ record the commands executed in a run and the variables used. This allows us to model that different instructions take different amounts of time as well as cache- and branch-related timing differences \cite{bastys2020clockwork,hedin2005timing}. We assume that execution time of instructions is affected by the size of values used, but not by their specific value. This assumption does not come for free, but requires careful handling of sensitive language primitives. Extending our language with pointers, arrays, or variable-time operations would require careful consideration. One straightforward strategy for sensitively indexing arrays is by linear scan over all array elements and using bit-masks to select only the element at the desired index.

To obtain high-resolution timestamps $\timestamp$, we assume a strictly increasing function $\textit{time}$ from histories to numeric values representing real time. That is, for all histories $\hst$ and history events $\textit{ev}$ we have $\textit{time}(\hst) < \textit{time}(\hst \Colon \textit{ev})$. Histories and high-resolution timestamps may appear too strong a formalism considering that we execute sensitive conditionals obliviously and therefore have that runs that agree on initial public state will agree on history and time. This level of accuracy is motivated by our attacker model and allows us to show the strong security guarantee enforced by our approach.

We let $\bitstack$ denote a stack of execution mode bits $\bit \in \xbrace{\xbit{1},\xbit{0}}$. For consistency with existing terminology \cite{liu2015oblivm} we say that execution takes place in either \textit{real} or \textit{phantom} mode. We let $\xbit{1}$ denote real mode and $\xbit{0}$ denote phantom mode. The execution mode is first set when triggering a handler. Genuine messages are handled in real mode while dummy messages are handled in phantom mode. The execution mode changes from real to phantom mode for the non-chosen branch when obliviously branching by $\texttt{oblif}$.

Transitions
\(
    \xangle{\bitstack,c,m,\mu,\intenv,\hst}
    \xrightarrow{\alpha}
    \xangle{\bitstack',c',m',\mu',\intenv',\hst'}
\)
denote a step from one configuration to another emitting output event $\alpha$. Output events $\alpha$ are possibly empty, denoted $\epsilon$, and are given by the following grammar:
\[
    \alpha \Coloneqq \epsilon
    \mid \overrightarrow{\ch}(\timestamp,\bit,\val{v}{z})
\]
Non-empty events correspond to network traffic and contain channel $\ch$, denoting which channel the message is sent on; high-resolution timestamp $\timestamp$, denoting when the message is sent; bit $\bit$, denoting the execution mode the message was sent under; and value $\val{v}{z}$. Timestamps $\timestamp$ are attacker observable giving us a strong attacker model (\Cref{sec:noninterference}). Intuitively, programs are noninterferent if related runs agree on the timestamps of all messages, which we can only ensure if their computational histories are the same.
Bit $\bit$ indicate whether a message is genuine or dummy. Messages produced in real mode $\bit = \xbit{1}$ are genuine, while messages produced in phantom mode $\bit = \xbit{0}$ are dummy. This enables the recipient of the message to execute the handler in the appropriate mode. We now explain the formal semantics and non-standard features.

\subsubsection*{If}
Standard conditionals are done by \texttt{if $e$ then $c_1$ else $c_2$}. The rule is largely standard, but to accurately model branch-related timing effects we append history $\hst$ with event $\textsf{br}(e,z,i)$, where $i$ is the chosen branch.

\subsubsection*{Oblif and Pop}
We introduce command \texttt{oblif $e$ then $c_1$ else $c_2$} for oblivious branching. When obliviously branching, two bits are pushed onto bit-stack $\bitstack$, one for each branch. The chosen branch continues with the current execution mode $\bit$, while the non-chosen branch is executed in phantom mode $\xbit{0}$. After executing each branch, command $\texttt{pop}$ removes the top element of the bit-stack.

\subsubsection*{Assign and oblivious assign}
Standard assignment is done by command $x \assign e$.
Standard assignment is intended for use only in real mode, and we do not give semantics to the command for phantom mode. Our type system enables us to ensure that the command is never reached in phantom mode (\Cref{sec:enforcement}).
We introduce a command $x \oblivassign e$ for oblivious assignment. The command conditionally updates the base value of $x$ depending on the execution mode and unconditionally pads the size of the value in the store to mask whether the base value was changed.

\subsubsection*{Input}
Programs sample local channels using non-blocking input command $x \oblivassign \texttt{input}(\ch, e)$.
Receives on channels with secret presence are restricted to non-blocking semantics as blocking would cause a leak \cite{sabelfeld2002static}.
The value of variable $x$ is only changed if executed in real mode and a message is available with size less than the evaluation of expression $e$. The size of $\mu(x)$ is unconditionally padded, similar to oblivious assignment.
The head of the message stream associated with the local channel is either $\val{v}{z}$, if a value is available, or $\bullet$, if no value is available.

\subsubsection*{Send}
Command $\texttt{send}(\ch,e)$ evaluates expression $e$ to obtain value $\val{v}{z}$.
The step emits event $\ch(\timestamp,\bit,\val{v}{z})$ capturing the network message; the value, when and to whom it was sent, and under what mode.

\subsubsection*{While}
Since the base values in store $\mu$ and memory $m$ are not modified while under phantom mode, executing loops in phantom mode would lead to non-termination as the value of guard expression $e$ could not be modified. For this reason, we restrict while-loops to real-mode execution.

\subsection{Program semantics}
A reactive program is at any point in one of two states. Consumer states $\consumer{p,\mu,\intenv,\networkstrategy,\hst,\tau}$ consist of program $p$, global store $\mu$, local environment $\intenv$, network strategy $\networkstrategy$, history $\hst$, and trace $\tau$. Consumer states, as the name implies, consumes network events and triggers the appropriate handlers. Producer states $\producer{p,\bitstack,c,m,\mu,\intenv,\networkstrategy,\hst,\tau}$ consist of program $p$, bit-stack $\bitstack$, command $c$, local memory $m$, global store $\mu$, local environment $\intenv$, network strategy $\networkstrategy$, history $\hst$, trace $\tau$, and are annotated with channel $\ch$ associated with the currently executing handler. We let $\Consumer$ range over consumer states, $\Producer$ range over producer states, and $\Configuration$ range over both consumer and producer states.

We extend events $\alpha$ with incoming messages $\overleftarrow{\ch}(\timestamp,\bit,\val{v}{z})$ and observed messages $\widetilde{\ch}(\timestamp,\bit,\val{v}{z})$:
\[
    \alpha \Coloneqq \ldots
    \mid \overleftarrow{\ch}(\timestamp,\bit,\val{v}{z})
    \mid \widetilde{\ch}(\timestamp,\bit,\val{v}{z})
\]
Observed messages $\widetilde{\ch}(\timestamp,\bit,\val{v}{z})$ correspond to the local node observing the network traffic between two remote nodes. Their inclusion enables us to bound the traffic overhead for the entire network (\Cref{sec:enforcementoverhead}).
We let $\leftrightsquigarrow$ range over $\xbrace{\leftarrow,\rightarrow,\sim}$.

Network traces $\tau$ are used as input for strategies $\networkstrategy$ to obtain the next network message and are given by the following grammar:
\[
    \tau \Coloneqq \epsilon \mid \tau \cdot \alpha
\]
By convention, we only modify the trace when appending non-empty events, that is, for any $\tau$ we have $\tau \cdot \epsilon = \tau$.

We now give semantics to the states and state transitions of a reactive program (\Cref{def:semantics_system}).

\begin{figure}[!htbp]
\centering
\begin{mathparpagebreakable}
\inferrule[CC] {
    \networkstrategy(\tau) = \ch(\timestamp,\bit,\val{v}{z}) \\
    (p)(\ch) \not\Downarrow 
} {
    \consumer{p,\mu,\intenv,\networkstrategy,\hst,\tau}
    \longrightarrow
    \consumer{p,\mu,\intenv,\networkstrategy,\hst,\tau \cdot \widetilde{\ch}(\timestamp,\bit,\val{v}{z})}
}
\and
\inferrule[CP] {
    \networkstrategy(\tau) = \ch(\timestamp,\bit,\val{v}{z}) \\
    (p)(\ch) \Downarrow c,x \\
    \hst' = \hst \Colon \textsf{hl}(\ch,\timestamp,z) \\
    \tau' = \tau \cdot \overleftarrow{\ch}(\timestamp,\bit,\val{v}{z})
} {
    \consumer{p,\mu,\intenv,\networkstrategy,\hst,\tau}
    \longrightarrow
    \producer{p,[\bit],c,[x \mapsto \val{v}{z}],\mu,\intenv,\networkstrategy,\hst', \tau'}
}
\and
\inferrule[PP] { 
    \xangle{\bitstack,c,m,\mu,\intenv,\hst}
    \xrightarrow{\alpha}
    \xangle{\bitstack',c',m',\mu',\intenv',\hst'}
} {
    \producer{p,\bitstack,c,m,\mu,\intenv,\networkstrategy,\hst,\tau}
    \longrightarrow
    \producer{p,\bitstack',c',m',\mu',\intenv',\networkstrategy,\hst',\tau \cdot \alpha}
}
\and
\inferrule[PC] { } {
    \producer{p,\bitstack,\texttt{stop},m,\mu,\intenv,\networkstrategy,\hst,\tau}
    \longrightarrow
    \consumer{p,\mu,\intenv,\networkstrategy,\hst \Colon \textsf{ret},\tau}
}
\end{mathparpagebreakable}
\caption{Operational semantics of system}
\label{def:semantics_system}
\end{figure}

Consumer state $\Consumer$ transitions depending on whether program $p$ defines a handler associated with that channel (\Cref{def:handler_selection}).

\begin{figure}[!htbp]
\centering
\begin{mathparpagebreakable}
\inferrule { } {
    (\ch (x) \xbrace{c}; p)(\ch) \Downarrow c,x
}
\and
\inferrule { 
    \ch \neq \ch' \\
    (p')(\ch') \Downarrow c',x'
} {
    (\ch(x) \xbrace{c}; p')(\ch') \Downarrow c',x'
}
%
%
%
%
\end{mathparpagebreakable}
\caption{Handler selection}
\label{def:handler_selection}
\end{figure}

If $p$ defines no associated handler, written $(p)(\ch)\not\Downarrow$, execution continues in consumer state $\Consumer'$, where the trace has been annotated with an observed message $\widetilde{\ch}(\timestamp,\bit,\val{v}{z})$. If $p$ defines an associated handler, written $(p)(\ch)\Downarrow c,x$, handler selection evaluates to command $c$ and variable $x$. Local memory $m$ is constructed by assigning message value $\val{v}{z}$ to variable $x$ and execution proceeds in producer state $\Producer$, executing command $c$. The producer state executes in the mode $\bit$ of the received message by using singleton bit-stack $[\bit]$.
A producer state $\Producer$ steps by the operational semantics of commands (\Cref{def:semantics_commands}), appending emitted events to the trace. Execution continues in producer state until reaching command $\texttt{stop}$, when it transitions back into consumer state.

\section{Enforcement}
\label{sec:enforcement}
In this section we provide and discuss the type system for \oblivio{}.
We type global stores $\mu$ using typing environment $\Gamma$. We type local channels using typing environment $\Pi$ and type network channels using typing environment $\Lambda$. Typing environments $\Gamma$, $\Pi$, and $\Lambda$ are static and do not change during execution. Local memory $m$ is typed using typing environment $\Delta$ which is computed per handler.
We give variables in store and memory, and local channels, a type of the form $\ltype{\sigma}{\ell}$, where $\sigma \in \xbrace{\inttype,\stringtype}$ and $\ell \in \mathcal{L}$. 
We let $q,r$ range over non-negative integer potentials and give network channels type of the form $\ltype{\sigma}{\ltriple{\lmode}{\lval}{q}}$, where $\lmode$ is the security level of the message mode, and $\lval$ is the security level of the message value. Potentials $q,r$ are inspired by static resource analysis \cite{hofmann2003static,hoffmann2010amortized,hoffmann2012resource}. Resource analysis commonly uses potentials to infer the resource bounds of a program. We use potentials $q$ to bound the number of dummy messages that may be generated by the handler of the channel.

\subsection{Typing of expressions}
We now present the type system for \oblivio{}. Expressions are typed $\Gamma;\Delta \vdash e : \ltype{\sigma}{\ell}$ (\Cref{def:type_system_expressions}). We type local memory using environment $\Delta$ and type global store using environment $\Gamma$. We let bindings in $\Delta$ shadow bindings in $\Gamma$. The rules are otherwise standard.

\begin{figure}[!htbp]
\centering
\begin{mathparpagebreakable}
\inferrule { } {
    \Gamma;\Delta \vdash n : \ltype{\inttype}{\ell}
}
\and
\inferrule { } {
    \Gamma;\Delta \vdash s : \ltype{\stringtype}{\ell}
}
\and
\inferrule {
    x \in \textit{dom}(\Delta)
} {
    \Gamma;\Delta \vdash x : \Delta(x)
}
\and
\inferrule { 
    x \notin \textit{dom}(\Delta)
} {
    \Gamma;\Delta \vdash x : \Gamma(x)
}
\and
\inferrule {
    \oplus : \sigma_1 \times \sigma_2 \rightarrow \sigma_3 \\\\
    \Gamma;\Delta \vdash e_1 : \ltype{\sigma_1}{\ell_1} \\
    \Gamma;\Delta \vdash e_2 : \ltype{\sigma_2}{\ell_2}
} {
    \Gamma;\Delta \vdash e_1 \oplus e_2 : \ltype{\sigma_3}{\ell_1 \sqcup \ell_2}
}
\end{mathparpagebreakable}
\caption{Typing rules for expressions}
\label{def:type_system_expressions}
\end{figure}

\subsection{Typing of commands}
Commands are typed $\Ctx{\Gamma}{\Pi}{\Lambda}{\Delta}{\pc} \vdash^q c$. Potential $q$ bounds the number of dummy messages that can be produced while executing command $c$, as well as any additional dummy messages that are transitively produced across the network when handling dummy messages produced by executing $c$. ASThis allows us to reason locally about the network-wide overhead in traffic.
The typing rules are provided in \Cref{def:type_system_commands}.

\begin{figure}[!htbp]
\centering
\begin{mathparpagebreakable}
\inferrule[T-Skip] { } {
    \Ctx{\Gamma}{\Pi}{\Lambda}{\Delta}{\pc} \vdash^q \texttt{skip}
}
\and
\inferrule[T-Assign] {
    x \notin \textit{dom}(\Delta) \\
    \Gamma(x) = \ltype{\sigma}{\ell_x} \\
    \Gamma;\Delta \vdash e : \ltype{\sigma}{\ell_e} \\
    \ell_e \sqsubseteq \ell_x
} {
    \Ctx{\Gamma}{\Pi}{\Lambda}{\Delta}{\bot} \vdash^q x \assign e
}
\and
\inferrule[T-Seq] {
    \Ctx{\Gamma}{\Pi}{\Lambda}{\Delta}{\pc} \vdash^{q_1} c_1 \\
    \Ctx{\Gamma}{\Pi}{\Lambda}{\Delta}{\pc} \vdash^{q_2} c_2
} {
    \Ctx{\Gamma}{\Pi}{\Lambda}{\Delta}{\pc} \vdash^{q_1 + q_2} c_1;c_2
}
\and
\inferrule[T-OblivAssign] {
    x \notin \textit{dom}(\Delta) \\
    \Gamma(x) : \ltype{\sigma}{\ell_x} \\
    \Gamma;\Delta \vdash e : \ltype{\sigma}{\ell_e} \\
    \ell_e \sqcup \pc \sqsubseteq \ell_x
} {
    \Ctx{\Gamma}{\Pi}{\Lambda}{\Delta}{\pc} \vdash^q x \oblivassign e
}
\and
\inferrule[T-LocalInput] {
    x \notin \textit{dom}(\Delta) \\
    \Gamma(x) : \ltype{\sigma}{\ell_x} \\
    \Pi(\ch) = \ltype{\sigma}{\ell_{\ch}} \\
    \Gamma;\Delta \vdash e : \ltype{\inttype}{\ell_e} \\
    \ell_e \sqcup \pc \sqsubseteq \ell_{\ch} \sqsubseteq \ell_x
} {
    \Ctx{\Gamma}{\Pi}{\Lambda}{\Delta}{\pc} \vdash^q x \oblivassign \texttt{input}(\ch,e)
}
\and
\inferrule[T-Send] {
    \Gamma;\Delta \vdash e : \ltype{\sigma}{\ell_e} \\
    \Lambda(\ch) = \ltype{\sigma}{\ltriple{\lmode}{\lval}{r}} \\
    \pc \sqsubseteq \lmode \\
    \ell_e \sqsubseteq \lval \\
    q' = 
    {\begin{cases}
        0
            &\textit{if } \pc = \bot \\
        1+r
            &\textit{otherwise}
    \end{cases}}
} {
    \Ctx{\Gamma}{\Pi}{\Lambda}{\Delta}{\pc} \vdash^{q+q'} \texttt{send}(\ch,e)
}
\and
\inferrule[T-If] {
    \Gamma;\Delta \vdash e : \ltype{\inttype}{\bot} \\
    \Ctx{\Gamma}{\Pi}{\Lambda}{\Delta}{\pc} \vdash^q c_1 \\
    \Ctx{\Gamma}{\Pi}{\Lambda}{\Delta}{\pc} \vdash^q c_2
} {
    \Ctx{\Gamma}{\Pi}{\Lambda}{\Delta}{\pc} \vdash^q \texttt{if $e$ then $c_1$ else $c_2$}
}
\and
\inferrule[T-While] {
    \Gamma;\Delta \vdash e : \ltype{\inttype}{\bot} \\
    \Ctx{\Gamma}{\Pi}{\Lambda}{\Delta}{\bot} \vdash^0 c
} {
    \Ctx{\Gamma}{\Pi}{\Lambda}{\Delta}{\bot} \vdash^q \texttt{while $e$ do $c$}
}
\and
\inferrule[T-OblivIf] {
    \Gamma;\Delta \vdash e : \ltype{\inttype}{\ell} \\
    \ell \neq \bot \\
    \Ctx{\Gamma}{\Pi}{\Lambda}{\Delta}{\pc \sqcup \ell} \vdash^{q_1} c_1 \\
    \Ctx{\Gamma}{\Pi}{\Lambda}{\Delta}{\pc \sqcup \ell} \vdash^{q_2} c_2
} {
    \Ctx{\Gamma}{\Pi}{\Lambda}{\Delta}{\pc} \vdash^{q_1 + q_2} \texttt{oblif $e$ then $c_1$ else $c_2$}
}
\end{mathparpagebreakable}
\caption{Typing of commands}
\label{def:type_system_commands}
\end{figure}

Our key insight is to restrict program commands with observable side-effects and provide safe, oblivious counterparts where possible.
In standard models, label $\pc$ tracks the security level of the control flow. That is not quite so in our model. The semantics of obliviously branching means that the control flow is the same for all executions that agree on the initial public state. However, the execution mode may differ. As such, $\pc$ tracks the security level of the execution mode in our model. We enforce that phantom mode only occurs for commands that type with non-public $\pc$.

\subsubsection*{Sequential composition}
Sequential composition $c_1;c_2$ types with potential $q_1+q_2$ if $c_i$ types with potential $q_i$, for $i=1,2$. As $q_1,q_2 \geq 0$ this prevents double spending of potential.

\subsubsection*{Assign and oblivious assign}
We allow standard, unconditional assignment only under public $\pc$ as we only give semantics to the command when executing in real mode.
We allow oblivious assignment under any $\pc$. Rule \textsc{T-OblivAssign} is similar to standard assignment rules in other models and is made safe by the padding semantics of \textsc{OblivAssign}.

\subsubsection*{If and oblif}
We restrict non-oblivious conditionals to public guards only and type \texttt{if $e$ then $c_1$ else $c_2$} with potential $q$ such that $q$ types both $c_1$ and $c_2$.
For branching on secrets, we use oblivious branching. The potential annotations in typing rule $\textsc{T-OblivIf}$ resemble those of \textsc{T-Seq} as both branches are executed and and both branches may produce traffic. We restrict oblivious branching to non-public guards to ensure that phantom computation only takes place under non-public $\pc$.

\subsubsection*{While}
Like other data-oblivious languages \cite{liu2015oblivm,zahur2015obliv}, we restrict while-loops to public guards. Unlike conditionals, there is no oblivious counterpart to while-loops. Intuitively, if the guard of a while-loop were secret, ending the loop at any point would cause a leak. A common technique in oblivious programming is to use a public upper-bound as the guard for the loop and obliviously branch on the secret inside the loop \cite{liu2015oblivm,zahur2015obliv}. \oblivio{} allows programmers to adopt this approach.
Rule \textsc{T-While} further restricts that $\texttt{while $e$ do $c$}$ must be typed with $\pc = \bot$ and that $c$ must be typed with potential $q=0$. The restriction on $\pc$ helps us ensure that computation in phantom mode terminates. We do not update base values in memory or store in phantom mode, thus guard $e$ would evaluate to the same value in every iteration of the loop if executed under phantom mode. The restriction on potential is motivated by bounding the number of dummy messages through the type system and prevents double spending of potential.

\subsubsection*{Send}
For sends, typing rule $\textsc{T-Send}$ enforces that the program counter label $\pc$ flows to context label $\lmode$ of the receiving channel, and that expression label $\ell_e$ flows to the value label $\lval$.
Our type system and semantic rules ensure that only commands that can be typed with non-public $\pc \neq \bot$ may be executed in phantom mode. Labelling channels with $\lmode = \bot$ therefore ensures that only genuine messages may be sent on the channel and allows the handler to perform non-oblivious assignments and while-loops even when $\lval \neq \bot$.

\textsc{T-Send} requires that $1+r$ potential is available if $\pc \neq \bot$, where $r$ is the potential available in the recipient handler of the message. This accounts for the message possibly being dummy as well as dummy messages produced transitively across the network in response. Because dummy messages require strictly greater potential than what is available in the recipient handler (as they can only be sent under non-public $\pc$), trace potential is strictly decreasing when sending dummy messages. This prevents handlers from sending dummy messages in an infinite loop. Note that loops are possible under $\pc = \bot$ where we are assured that messages are genuine.
We show that potentials bound the number of dummy messages in \Cref{sec:enforcementoverhead}.

\subsection{Typing of programs and systems}
Program $p$ is well-typed if the body of each handler is well-typed with respect to their respective channel (\Cref{def:type_system_program}).
\begin{figure}[!htbp]
\centering
\begin{mathparpagebreakable}
\inferrule { } {
    \ctx{\Gamma}{\Pi}{\Lambda} \vdash \cdot
}
\and
\inferrule {
    \Lambda(\ch) = \ltype{\sigma}{\ltriple{\lmode}{\lval}{q}} \\
    \Ctx{\Gamma}{\Pi}{\Lambda}{[x \mapsto \ltype{\sigma}{\lval}]}{\lmode} \vdash^q c \\
    \ctx{\Gamma}{\Pi}{\Lambda} \vdash p
} {
    \ctx{\Gamma}{\Pi}{\Lambda} \vdash \ch(x) \xbrace{c}; p
}
\end{mathparpagebreakable}
\caption{Typing of programs}
\label{def:type_system_program}
\end{figure}

We formally define well-formed values (\Cref{def:wellformed_value}). A value is well-formed if the size of its base value is less than or equal to the annotated public size bound.
\begin{definition}[Well-formed value]
\label{def:wellformed_value}
Value $\val{v}{z}$ is well-formed, written $\vdash \val{v}{z}$, if
\(
    \textit{size}(v) \leq z
\).
\end{definition}

We say that store $\mu$ is well-formed with respect to typing environment $\Gamma$ if the values bound in variables are well-formed and are of the correct type (\Cref{def:wellformed_store}).
\begin{definition}[Well-formed store w.r.t. typing environment]
\label{def:wellformed_store}
Store $\mu$ is well-formed w.r.t. typing environment $\Gamma$, written $\Gamma \vdash \mu$, if $\textit{dom}(\Gamma) = \textit{dom}(\mu)$ and for all $x \in \textit{dom}(\Gamma)$ we have
\begin{enumerate}
    \item $\Gamma(x) = \ltype{\inttype}{\ell} \implies \exists n,z : \mu(x) = \val{n}{z} \land \textit{size}(n) \leq z$
    \item $\Gamma(x) = \ltype{\stringtype}{\ell} \implies \exists s,z : \mu(x) = \val{s}{z} \land \textit{size}(s) \leq z$
\end{enumerate}
\end{definition}

Well-formedness of memory $m$ with respect to typing environment $\Delta$ is defined equivalently (\Cref{def:wellformed_memory}).
\begin{definition}[Well-formed memory w.r.t. typing environment]
\label{def:wellformed_memory}
Store $m$ is well-formed w.r.t. typing environment $\Delta$, written $\Delta \vdash m$, if $\textit{dom}(\Delta) = \textit{dom}(m)$ and for all $x \in \textit{dom}(\Delta)$ we have
\begin{enumerate}
    \item $\Delta(x) = \ltype{\inttype}{\ell} \implies \exists n,z : m(x) = \val{n}{z} \land \textit{size}(n) \leq z$
    \item $\Delta(x) = \ltype{\stringtype}{\ell} \implies \exists s,z : m(x) = \val{s}{z} \land \textit{size}(s) \leq z$
\end{enumerate}
\end{definition}

We say that local environment $\intenv$ is well-formed with respect to typing environment $\Pi$ if, for each channel, all values are well-formed and are of the correct type (\Cref{def:wellformed_internal_env}).
\begin{definition}[Well-formed local message environment w.r.t typing environment]
\label{def:wellformed_internal_env}
Define local environment $\intenv$ to be well-formed with respect to typing environment $\Pi$, written $\Pi \vdash \intenv$, if $\textit{dom}(\Pi)=\textit{dom}(\intenv)$ and for all $\ch \in \textit{dom}(\Pi)$ such that $\Pi(\ch) = \ltype{\sigma}{\ell}$ we have $\vdash_{\sigma} \intenv(\ch)$ defined by the following rules:
\begin{mathparpagebreakable}
\inferrule { } {
    \vdash_{\sigma} []
}
\and
\inferrule {
    \vdash_{\sigma} \textit{tl}
} {
    \vdash_{\sigma} \bullet \Colon \textit{tl}
}
\and
\inferrule {
    \textit{size}(n) \leq z \\
    \vdash_{\inttype} \textit{tl}
} {
    \vdash_{\inttype} \val{n}{z} \Colon \textit{tl}
}
\and
\inferrule {
    \textit{size}(s) \leq z \\
    \vdash_{\stringtype} \textit{tl}
} {
    \vdash_{\stringtype} \val{s}{z} \Colon \textit{tl}
}
\end{mathparpagebreakable}
\end{definition}

Next, we define the potential $q$ of trace $\tau$ with respect to typing environment $\Lambda$. Trace potential loosely bounds the number of future dummy messages that may result after observing a given trace. Genuine messages $\oset{\leftrightsquigarrow}{\ch}(\timestamp,\xbit{1},\val{v}{z})$ increase the potential of a trace by the annotated potential $r$ of channel $\ch$ in $\Lambda$, while dummy messages $\oset{\leftrightsquigarrow}{\ch}(\timestamp,\xbit{0},\val{v}{z})$ decrease the potential by one. The definition does not capture that potential is strictly decreasing for all sends under non-public $\pc$, but is nevertheless sufficient for showing our overhead theorem (\Cref{sec:enforcementoverhead}).

\begin{definition}[Trace potential] Define potential $q$ of a trace $\tau$ with respect to typing environment $\Lambda$, written $\Lambda \vdash \tau : q$, by the following rules:
\label{def:tracepotential}
\begin{mathparpagebreakable}
\inferrule{
}{
    \Lambda \vdash \epsilon : 0
}
\and
\inferrule{
    \Lambda \vdash \tau : q \\
    \Lambda(\ch) = \ltype{\_}{\ltriple{\_}{\_}{r}}
}{
    \Lambda \vdash \tau\; \cdot \oset{\leftrightsquigarrow}{\ch}(\timestamp,\xbit{1},\val{v}{z}) : q + r
}
\and
\inferrule{
    \Lambda \vdash \tau : q+1
}{
    \Lambda \vdash \tau\; \cdot \oset{\leftrightsquigarrow}{\ch}(\timestamp,\xbit{0},\val{v}{z}) : q
}
\end{mathparpagebreakable}
\end{definition}

Using \Cref{def:tracepotential}, we define well-formedness of network strategies $\networkstrategy$ with respect to typing environment $\Lambda$ (\Cref{def:wellformed_networkstrategy}). A network strategy is well-formed if messages $\ch(\timestamp,\bit,\val{v}{z})$ it produces are well-formed and of the correct type, and if dummy messages are only produced on non-public channels and only when the trace has sufficient potential for the dummy message and the annotated potential $r$ of channel $\ch$ in $\Lambda$.

We enforce the bound by a well-formedness condition for network strategies $\networkstrategy$ (\Cref{def:wellformed_networkstrategy}) and prove that it is enforced by handlers in well-typed programs. We provide this proof in
\ifdefined\istechnicalreport
\Cref{appendix:overhead}.
\else
the accompanying technical report.
\fi

\begin{definition}[Well-formed network strategy]
\label{def:wellformed_networkstrategy}
Network strategy $\networkstrategy$ is well-formed w.r.t. typing environment $\Lambda$, written $\Lambda \vdash \networkstrategy$, if for any $\tau$ such that $\networkstrategy(\tau) = \ch(\timestamp,\bit,\val{v}{z})$ and $\Lambda(\ch) = \ltype{\sigma}{\ltriple{\lmode}{\_}{r}}$ we have $\textit{size}(v) \leq z$ and
\begin{itemize}
    \item $\sigma = \inttype \implies \exists n: v = n$
    \item $\sigma = \stringtype \implies \exists s: v = s$
    \item $\bit = \xbit{0} \implies \lmode \neq \bot \land \Lambda \vdash \tau : q + 1 + r$
\end{itemize}
\end{definition}

We lift typing of programs, stores, local environments, and network strategies to typing of consumer states $\Consumer$, written $\ctx{\Gamma}{\Pi}{\Lambda} \vdash \Consumer$, in the straightforward way in \Cref{def:type_system_system}.
\begin{figure}[!htbp]
\centering
\begin{mathparpagebreakable}
\inferrule {
    \Gamma,\Pi,\Lambda \vdash p \\
    \Gamma \vdash \mu \\
    \Pi \vdash \intenv \\
    \Lambda \vdash \networkstrategy \\
} {
    \Gamma,\Pi,\Lambda \vdash \consumer{p,\mu,\intenv,\networkstrategy,\hst,\tau}
}
\end{mathparpagebreakable}
\caption{Typing of system}
\label{def:type_system_system}
\end{figure}

\section{Noninterference}
\label{sec:noninterference}
In this section define attacker knowledge, our security condition, and give our noninterference theorem that well-typed programs in \oblivio{} satisfy the security condition. We first define the equivalence relations necessary for our attacker knowledge definition and security condition.

We define stores $\mu_1,\mu_2$ to be equivalent up to level $\ladv$ with respect to typing environment $\Gamma$ (\Cref{def:store_equivalence}) if, for every variable, they agree on the values up to that level. We further require that the public sizes of values are the same.
\begin{definition}[Equivalence of store up to level]
\label{def:store_equivalence}
Stores $\mu_1$ and $\mu_2$ are equivalent up to level $\ladv$ w.r.t. typing environment $\Gamma$, written $\mu_1 \gammaequiv{\ladv} \mu_2$, if for all $x \in \textit{dom}(\Gamma)$ with $\Gamma(x) = \ltype{\sigma}{\ell}$ we have that if $\mu_1(x) = \val{v_1}{z_1}$ and $\mu_2(x) = \val{v_2}{z_2}$ then
\begin{enumerate}
    \item $ z_1 = z_2$
    \item $\ell \sqsubseteq \ladv \Rightarrow v_1 = v_2$
\end{enumerate}
\end{definition}
Equivalence of memories $m_1,m_2$ level $\ladv$ with respect to typing environment $\Delta$ is defined equivalently (\Cref{def:memory_equivalence}).
\begin{definition}[Equivalence of memory up to level]
\label{def:memory_equivalence}
Memories $m_1$ and $m_2$ are equivalent up to level $\ladv$ w.r.t. typing environment $\Delta$, written $m_1 \deltaequiv{\ladv} m_2$, if for all $x \in \textit{dom}(\Delta)$ s.t. $\Delta(x) = \ltype{\sigma}{\ell}$, we have that if $m_1(x) = \val{v_1}{z_1}$ and $m_2(x) = \val{v_2}{z_2}$ then
\begin{enumerate}
    \item $ z_1 = z_2$
    \item $\ell \sqsubseteq \ladv \Rightarrow v_1 = v_2$
\end{enumerate}
\end{definition}

We define local environments $\intenv_1,\intenv_2$ to be equivalent up to level $\ladv$ with respect to typing environment $\Pi$, written $\intenv_1 \piequiv{\ladv} \intenv_2$, by equality on the streams at all channels up level $\ladv$ (\Cref{def:internal_env_equivalence}).
\begin{definition}[Equivalence of local environment up to level]
\label{def:internal_env_equivalence}
Equivalence of local environments $\intenv_1$ and $\intenv_2$ up to level $\ladv$ w.r.t. typing environment $\Pi$, written $\intenv_1 \piequiv{\ladv} \intenv_2$, is defined by the following rule:
\begin{mathpar}
\inferrule { 
    \Pi(\ch) = \ltype{\sigma}{\ell} \\
    \ell \sqsubseteq \ladv \implies \intenv_1(\ch) = \intenv_2(\ch)
} {
    \intenv_1 \piequiv{\ladv} \intenv_2
}
\end{mathpar}
\end{definition}

Two events $\alpha_1,\alpha_2$ are equivalent up to level $\ladv$ (\Cref{def:event_equivalence}) if they agree on the properties observable at that level.
\begin{definition}[Equivalence of output events up to level]
\label{def:event_equivalence}
Equivalence of events $\alpha_1$ and $\alpha_2$, up to level $\ladv$ w.r.t. typing environment $\Lambda$, written $\alpha_1 \lambdaequiv{\ladv} \alpha_2$, is defined by as follows:
\begin{mathparpagebreakable}
\inferrule {
} {
    \epsilon \lambdaequiv{\ladv} \epsilon
}
\and
\inferrule {
    \Lambda(\ch) = \ltype{\sigma}{\ltriple{\lmode}{\lval}{q}} \\
    \lmode \sqsubseteq \ladv \implies \bit_1 = \bit_2 \\
    \lval \sqsubseteq \ladv \implies v_1 = v_2
} {
    \oset{\leftrightsquigarrow}{\ch}(\timestamp,\bit_1,\val{v_1}{z})
    \lambdaequiv{\ladv}
    \oset{\leftrightsquigarrow}{\ch}(\timestamp,\bit_2,\val{v_2}{z})
}
\end{mathparpagebreakable}
\end{definition}

We lift this definition to equivalence on traces, written $\tau_1 \lambdaequiv{\ladv} \tau_2$, in the straightforward way by point-wise equivalence (\Cref{def:trace_equivalence}).
\begin{definition}[Trace equivalence up to level]
\label{def:trace_equivalence}
Equivalence of traces $\tau_1,\tau_2$ up to level $\ladv$ w.r.t. typing environment $\Lambda$, written $\tau_1 \lambdaequiv{\ladv} \tau_2$, is defined by the following rules
\begin{mathpar}
\inferrule { } {
    \epsilon \lambdaequiv{\ladv} \epsilon
}
\and
\inferrule {
    \tau_1 \lambdaequiv{\ladv} \tau_2 \\
    \alpha_1 \lambdaequiv{\ladv} \alpha_2
} {
    \tau_1 \cdot \alpha_1 \lambdaequiv{\ladv} \tau_2 \cdot \alpha_2
}
\end{mathpar}
\end{definition}

We define external event queues $\networkstrategy_1,\networkstrategy_2$ to be equivalent up to level $\ladv$ with respect to typing environment $\Lambda$ (\Cref{def:networkstrategyequivalence}) if they agree on the channel, timestamp and size of messages for equivalent traces, as well as message mode and value of messages on channels observable at that level.

\begin{definition}[Equivalence of network strategies up to level]
\label{def:networkstrategyequivalence}
Network strategies $\networkstrategy_1$ and $\networkstrategy_2$ are equivalent up to level $\ladv$ w.r.t. typing environment $\Lambda$, written $\networkstrategy_1 \lambdaequiv{\ladv} \networkstrategy_2$, if for any $\ch$ s.t. $\Lambda(\ch) = \ltype{\sigma}{\ltriple{\lmode}{\lval}{q}}$ and any $\tau_1,\tau_2$ such that $\tau_1 \lambdaequiv{\ladv} \tau_2$ we have that if $\networkstrategy(\tau_1) = \ch_1(\timestamp_1,\bit_1,\val{v_1}{z_1})$ and $\networkstrategy(\tau_2) = \ch_2(\timestamp_2,\bit_2,\val{v_2}{z_2})$ then $\ch_1 = \ch_2$, $\timestamp_1 = \timestamp_2$, $z_1 = z_2$, and
\begin{itemize}
    \item $\lmode \sqsubseteq \ladv \implies \bit_1 = \bit_2$
    \item $\lval \sqsubseteq \ladv \implies v_1 = v_2$
\end{itemize}
\end{definition}

We lift the above equivalence definitions to define equivalence of consumer states, written $\Consumer_1 \approx^{\Gamma,\Pi,\Lambda}_{\ladv} \Consumer_2$, by straightforward lifting of the equivalences of the components (\Cref{def:equivalence_systems}).
\begin{definition}[Equivalence of consumer states] Define consumer states $\Consumer_1,\Consumer_2$ to be equivalent up to level $\ladv$, with respect to typing environments $\Gamma,\Pi,\Lambda$, written $\Consumer_1 \approx^{\Gamma,\Pi,\Lambda}_{\ladv} \Consumer_2$, by the following rule:
\label{def:equivalence_systems}
\begin{mathparpagebreakable}
\inferrule {
    \mu_1 \gammaequiv{\ladv} \mu_2 \\
    \intenv_1 \piequiv{\ladv} \intenv_2 \\
    \networkstrategy_1 \lambdaequiv{\ladv} \networkstrategy_2 \\
    \tau_1 \lambdaequiv{\ladv} \tau_2
} {
    \consumer{p,\mu_1,\intenv_1,\networkstrategy_1,\hst,\tau_1}
    \approx^{\Gamma,\Pi,\Lambda}_{\ladv}
    \consumer{p,\mu_2,\intenv_2,\networkstrategy_2,\hst,\tau_2}
}
\end{mathparpagebreakable}
\end{definition}

To simplify our security condition, we define the projection of the trace component from configuration $\Configuration$, written $\textit{trace}(\Configuration)$, in the straightforward way (\Cref{def:traceofconfiguration}).
\begin{definition}[Trace of configuration] Define the projection of the trace of configuration $\Configuration$, written $\textit{trace}(\Configuration)$, as follows:
\label{def:traceofconfiguration}
\begin{align*}
    \textit{trace}(\Configuration) =
    \begin{cases}
        \tau
            &\textit{if } \Configuration = \consumer{p,\mu,\intenv,\networkstrategy,\hst,\tau} \\
        \tau
            &\textit{if } \Configuration = \producer{p,\bitstack,c,m,\mu,\intenv,\networkstrategy,\hst,\tau}
    \end{cases}
\end{align*}
\end{definition}

We are now ready to define attacker knowledge (\Cref{def:attacker_knowledge}). Given typing environments $\Gamma,\Pi,\Lambda$, a consumer state $\Consumer$, and a level $\ladv$, we define attacker knowledge as the set of equivalent consumer states $\Consumer'$ that produce an equivalent trace when run.

\begin{definition}[Attacker knowledge]
\label{def:attacker_knowledge}
Given consumer state $\Consumer$ and trace $\tau$ such that
\(
    \Consumer
    \longrightarrow {\phantom{I}\mkern-18mu}^*\,
    \Configuration,
\)
with $\textit{trace}(\Configuration) = \tau$, define attacker knowledge at level $\ladv$ as follows:
\begin{align*}
    &k_{\Gamma,\Pi,\Lambda}(\Consumer,\tau,\ladv) \triangleq \\
    &\qquad
        \xbrace{ 
            \Consumer' \mid
            \Consumer \approx^{\Gamma,\Pi,\Lambda}_{\ladv} \Consumer'
            \land
            \Consumer'
            \longrightarrow {\phantom{I}\mkern-18mu}^*
            \Configuration'
            \land
            \tau \lambdaequiv{\ladv} \textit{trace}(\Configuration'_2)
        }
\end{align*}
\end{definition}

We use the attacker knowledge definition to define our timing-sensitive, progress-sensitive security condition (\Cref{def:security_condition}).

\begin{definition}[Progress-sensitive noninterference] 
\label{def:security_condition}
Given consumer state $\Consumer$ and a run
\(
    \Consumer
    \longrightarrow {\phantom{I}\mkern-18mu}^* \,
    \Configuration
\)
with $\textit{trace}(\Configuration)=\tau \cdot \alpha$, the run satisfies progress-sensitive noninterference if for all levels $\ladv$ we have
\(
    k_{\Gamma,\Pi,\Lambda}(\Consumer,\tau \cdot \alpha,\ladv) \supseteq k_{\Gamma,\Pi,\Lambda}(\Consumer,\tau,\ladv)
\).
\end{definition}

With our security condition defined, we are ready to state the soundness theorem for our type system (\Cref{theorem:soundness}).

\begin{restatable}[Soundness]{theorem}{soundness}
\label{theorem:soundness}
Given $\Gamma,\Pi,\Lambda$ and consumer state $\Consumer$ such that $\ctx{\Gamma}{\Pi}{\Lambda} \vdash \Consumer$. If
\(
    \Consumer
    \longrightarrow {\phantom{I}\mkern-18mu}^*\;
    \Configuration
\)
with $\textit{trace}(\Configuration) = \tau \cdot \alpha$ then
\(
    k_{\Gamma,\Pi,\Lambda}(\Consumer,\tau \cdot \alpha,\ladv) \supseteq k_{\Gamma,\Pi,\Lambda}(\Consumer,\tau,\ladv)
\).
\end{restatable}
\ifdefined\istechnicalreport
The proof is via an auxiliary model for \oblivio{} (\Cref{appendix:auxiliarymodel}) and is provided in \Cref{appendix:noninterference}.
\else
We provide the proof in the accompanying technical report.
\fi

\section{Enforcement overhead}
\label{sec:enforcementoverhead}
In this section we provide a bound on the traffic overhead introduced by \oblivio{}'s enforcement strategy. The bound is by a multiplicative factor given by typing environment $\Lambda$. We show this bound by considering two runs, one in the standard semantics given in \Cref{sec:language} and one in an unsafe semantics that suppresses all dummy messages, and bounding the difference in the lengths of the traces (\Cref{theorem:enforcementoverhead}). Our model accounts for observing network traffic between other nodes by events $\widetilde{\ch}(\timestamp,\bit,\val{v}{z})$ enabling us to bound the amount of dummy traffic across all network nodes.

The overhead in traffic produced by the enforcement can be categorised as either \textit{direct} (dummy messages), or \textit{indirect} (additional, genuine messages). As presented, \oblivio{} has no indirect overhead and produces the same genuine traffic under the standard and suppressing semantics. The language as presented contains no primitives that are affected by phantom computation, e.g., taking the size of values or taking the time. Such primitives could be introduced while still bounding direct overhead, but bounding indirect overhead would be difficult.

We first define the operational semantics of a system that suppressed dummy messages (\Cref{def:semantics_suppressing_system}). Rules \textsc{CC-Unsafe} and \textsc{CP-Unsafe} require that network strategy $\networkstrategy$ produces messages with mode bit $\xbit{1}$ to take a step, and rule \textsc{PP-Unsafe} filters all events that are not sends with mode bit $\xbit{1}$.

\begin{figure}[!htbp]
\centering
\begin{mathparpagebreakable}
\inferrule[CC-Unsafe] {
    \networkstrategy(\tau) = \ch(\timestamp,\xbit{1},\val{v}{z}) \\
    (p)(\ch) \not\Downarrow
} {
    \consumer{p,\mu,\intenv,\networkstrategy,\hst,\tau}
    \xrightarrow[\textit{unsafe}]{}
    \consumer{p,\mu,\intenv,\networkstrategy,\hst,\tau \cdot \widetilde{\ch}(\timestamp,\xbit{1},\val{v}{z})}
}
\and
\inferrule[CP-Unsafe] {
    \networkstrategy(\tau) = \ch(\timestamp,\xbit{1},\val{v}{z}) \\
    (p)(\ch) \Downarrow c,x \\
    \hst' = \hst \Colon \textsf{in}(\ch,\timestamp,z) \\
    \tau' = \tau \cdot \overleftarrow{\ch}(\timestamp,\xbit{1},\val{v}{z})
} {
    \consumer{p,\mu,\intenv,\networkstrategy,\hst,\tau}
    \xrightarrow[\textit{unsafe}]{}
    \producer{p,[\xbit{1}],c,[x \mapsto \val{v}{z}],\mu,\intenv,\networkstrategy,\hst',\tau'}
}
\and
\inferrule[PP-Unsafe] { 
    \xangle{\bitstack,c,m,\mu,\intenv,\hst}
    \xrightarrow{\alpha}
    \xangle{\bitstack',c',m',\mu',\intenv',\hst'} \\
    \tau' =
    {
    \begin{cases}
        \tau \cdot \alpha
            &\textit{if } \alpha = \overrightarrow{\ch}(t,\xbit{1},\val{v}{z}) \\
        \tau
            &\textit{otherwise}
    \end{cases}
    }
} {
    \producer{p,\bitstack,c,m,\mu,\intenv,\networkstrategy,\hst,\tau}
    \xrightarrow[\textit{unsafe}]{}
    \producer{p,\bitstack',c',m',\mu',\intenv',\networkstrategy,\hst',\tau'}
}
\and
\inferrule[PC-Unsafe] { } {
    \producer{p,\bitstack,\texttt{stop},m,\mu,\intenv,\networkstrategy,\hst,\tau}
    \xrightarrow[\textit{unsafe}]{}
    \consumer{p,\mu,\intenv,\networkstrategy,\hst \Colon \textsf{ret},\tau}
}
\end{mathparpagebreakable}
\caption{Operational semantics of suppressing system}
\label{def:semantics_suppressing_system}
\end{figure}

Next, we define extension relations for relating values, traces, and network strategies up to the effects of phantom.

We define the extension on values as equality up to padding of the extended value (\Cref{def:valueextension}).
\begin{definition}[Extension of values] Define value $\val{v_2}{z_2}$ to be an extension of value $\val{v_1}{z_1}$, written $\val{v_1}{z_1} \extends \val{v_2}{z_2}$, if $v_1 = v_2$ and $z_1 \leq z_2$.
\label{def:valueextension}
\end{definition}

We define trace $\tau_2$ to be a phantom extension of trace $\tau_1$ (\Cref{def:tracephantomextension}) if $\tau_1$ contains no dummy messages and, for all genuine messages, the traces agree on channels, bits, and values up to value extension.
\begin{definition}[Trace equivalence up to phantom] Define trace $\tau_2$ to be a phantom extension of trace $\tau_1$, written $\tau_1 \phantomextends \tau_2$, by the following rules:
\label{def:tracephantomextension}
\begin{mathparpagebreakable}
\inferrule{
} {
    \epsilon \phantomextends \epsilon
}
\and
\inferrule{
    \tau_1 \phantomextends \tau_2
} {
    \tau_1 \phantomextends \tau_2 \cdot \oset{\leftrightsquigarrow}{\ch}(\timestamp,\xbit{0},\val{v}{z})
}
\and
\inferrule{
    \tau_1 \phantomextends \tau_2 \\
    \val{v_1}{z_1} \extends \val{v_2}{z_2}
} {
    \tau_1 \cdot \oset{\leftrightsquigarrow}{\ch}(\timestamp,\xbit{1},\val{v_1}{z_1})
    \phantomextends
    \tau_2 \cdot \oset{\leftrightsquigarrow}{\ch}(\timestamp',\xbit{1},\val{v_2}{z_2})
}
\end{mathparpagebreakable}
\end{definition}

We define network strategy $\networkstrategy_2$ to be a phantom extension of network strategy $\networkstrategy_1$ with respect to typing environmtn $\Lambda$ (\Cref{def:networkstrategylambdaphantomextension}) if for related trace $\tau_1,\tau_2$ we have that if $\networkstrategy_2(\tau_2)$ returns a genuine message, then so does $\networkstrategy_1(\tau_1)$ and the values of the messages are related by extension.
\begin{definition}[Network strategy] Define network strategy $\networkstrategy_2$ to be a phantom extension of network strategy $\networkstrategy_1$ w.r.t typing environment $\Lambda$, written $\networkstrategy_1 \lambdaphantomextends \networkstrategy_2$, by the following rule:
\label{def:networkstrategylambdaphantomextension}
\begin{mathparpagebreakable}
\inferrule {
    \tau_1 \phantomextends \tau_2
    \land
    \networkstrategy_2(\tau_2) = \ch(\timestamp_2,\xbit{1},\val{v_2}{z_2})
    \implies \\\\
    \networkstrategy_1(\tau_1) = \ch(\timestamp_1,\xbit{1},\val{v_1}{z_1})
    \land
    \val{v_1}{z_1} \extends \val{v_2}{z_2}
} {
    \networkstrategy_1 \lambdaphantomextends \networkstrategy_2
}
\end{mathparpagebreakable}
\end{definition}

Finally, we define $\maxpotential$ as the maximum potential annotation of channels in typing environment $\Lambda$ (\Cref{def:maximumpotential}).
\begin{definition}[Maximum potential] Define maximum potential w.r.t $\Lambda$, written $\maxpotential$, as follows
\label{def:maximumpotential}
\[
    \maxpotential = \text{max}\xbrace{q \mid \ltype{\_}{\ltriple{\_}{\_}{q}} \in \text{range}(\Lambda)}
\]
\end{definition}

With the above definitions in place, we are ready to state our overhead theorem (\Cref{theorem:enforcementoverhead}). It says that given a configurations starting from the empty trace and considering a run in the suppressing semantics producing trace $\tau_1$, then a run in the standard semantics with any extended network strategy, will produce an extended trace $\tau_2$ that longer by at most a factor of $\maxpotential$.
\begin{restatable}[Overhead]{theorem}{enforcementoverhead}
\label{theorem:enforcementoverhead}
Consider $\consumer{p,\mu,\intenv,\networkstrategy_1,\hst,\epsilon}$ such that $\ctx{\Gamma}{\Pi}{\Lambda} \vdash \consumer{p,\mu,\intenv,\networkstrategy_1,\hst,\epsilon}$ and $\networkstrategy_2$ such that $\networkstrategy_1 \lambdaphantomextends \networkstrategy_2$. If we have a run in the unsafe semantics
\(
    \consumer{p,\mu,\intenv,\networkstrategy_1,\hst,\epsilon}
    \xrightarrow[\textit{unsafe}]{}{\phantom{I}\mkern-18mu}^*
    \Configuration_1
\)
with $\textit{trace}(\Configuration_1)=\tau_1$ then we have a run in the safe semantics
\(
    \consumer{p,\mu,\intenv,\networkstrategy_2,\hst,\epsilon}
    \longrightarrow {\phantom{I}\mkern-18mu}^*
    \Configuration_2
\)
with $\textit{trace}(\Configuration_2)=\tau_2$ such that $\tau_1 \phantomextends \tau_2$
and $|\tau_2| \leq |\tau_1| * (1 + \maxpotential)$.
\end{restatable}
\ifdefined\istechnicalreport
The proof is provided in \Cref{appendix:overhead} and is via the auxiliary model found in \Cref{appendix:auxiliarymodel}.
\else
The proof is provided in the accompanying technical report.
\fi

\section{Example: Auction service}
\label{sec:example}
In this section we demonstrate \oblivio{} by example. The example we shall use is a simple, secure auction service. Further examples can be found in 
\ifdefined\istechnicalreport
\Cref{appendix:examples}.
\else
the accompanying technical report.
\fi

We provide type annotations for handlers and variables with the convention that handler $\texttt{CH}_{\lmode} \; \$q \; (x: \sigma_{\lval}) \; \xbrace{c}$ at node $\texttt{NODE}$ is a handler for channel $\texttt{NODE/CH}$ such that
\[
    \Ctx{\Gamma}{\Pi}{\Lambda}{[x \mapsto \ltype{\sigma}{\lval}]}{\lmode} \vdash^q c
\]
Potential $q$ is zero where omitted.

The auction consists of users Alice and Bob, the auction house, and a simple timing service used by the auction house for timing rounds. We omit the code for Bob as it is equivalent to the code for Alice.

We consider users Alice and Bob to be trusted and only consider leaks to network level attackers. We therefore consider a simple two-point lattice $\xbrace{\texttt{L},\texttt{H}}$ with ordering $\texttt{L} \sqsubseteq \texttt{L}$, $\texttt{L} \sqsubseteq \texttt{H}$, and $\texttt{H} \sqsubseteq \texttt{H}$. All other flows are illegal.

\begin{lstlisting}[float=!htbp,floatplacement=!htbp,caption=Alice,label=lst:auction_alice,mathescape,escapechar=|]
ALICE // Node ID

var max_bid : int$_H$;

TO_LEAD$_H$ $\$1$ (bid : int$_H$) {
    oblif bid <= max_bid
    then send(AUCTIONHOUSE/ALICE_BID, bid); |\label{line:alicebid}|
    else skip;
}

...
\end{lstlisting}
Alice's client code is provided in \Cref{lst:auction_alice}. 
Messages on channel $\texttt{ALICE/TO\_LEAD}$ inform her of what she must bid to take the lead. The channel is typed with non-public context label and can receive dummy messages in case Alice is already leading.
When receiving a genuine message, Alice obliviously branches on whether the required bid is less than or equal to the maximum bid she is willing to make. If so, she makes the bid.
When receiving a dummy message, the send on \Cref{line:alicebid} will execute under phantom mode regardless of the value she receives.
The channel is typed with potential $1$ as it may produce a dummy message on channel \texttt{AUCTIONHOUSE/ALICE\_BID}, which is typed with potential $0$. When the auction finishes, Alice is notified of the winner on channel \texttt{AUCTIONHOUSE/AUCTION\_OVER\_NAME} and the winning bid on channel \texttt{AUCTIONHOUSE/AUCTION\_OVER\_BID}.

\begin{lstlisting}[float=!htbp,floatplacement=!htbp,caption=Auction timer,label=lst:auction_timer,mathescape]
AUCTIONTIMER // Node ID

var c : int$_L$;

BEGIN$_L$ (i : int$_L$) {
    c = i * 2000;
    while (c > 0) do {
        c = c - 1;
    }
    send(AUCTIONHOUSE/TICK, 0);
}
\end{lstlisting}
\Cref{lst:auction_timer} presents the code for the auction timer. It is a simple busy waiting loop that counts down to zero before sending a message on channel \texttt{AUCTIONHOUSE/TICK}.

\begin{lstlisting}[float=!htbp,floatplacement=!htbp,caption=Auction server,label=lst:auction_server,mathescape,escapechar=|]
AUCTIONSERVER // Node ID

var winner : string$_H$;
var winning_bid : int$_H$;
var round_counter : int$_L$;

ALICE_BID$_H$ (bid: int$_H$) {
    oblif winning_bid < bid
    then {
        winner ?= "Alice";
        winning_bid ?= bid;
    }
    else skip;
}

TICK$_L$ $\$4$ (dmy: int$_L$) {
    if round_counter > 0 |\label{line:round_counter_branching}|
    then {
        oblif winner != "Alice" |\label{line:alice_winner_branching}|
        then send(ALICE/TO_LEAD, winning_bid+1);
        else skip;
        ...
        round_counter = round_counter - 1;
        send(AUCTIONTIMER/BEGIN, 1);
    } else {
        send(ALICE/AUCTION_OVER_NAME,winner);
        send(ALICE/AUCTION_OVER_BID,winning_bid);
        ...
    }
}
\end{lstlisting}
\Cref{lst:auction_server} presents the auction server code. The server stores the current winner and the winning bid in variables with secret label $H$ and a round counter with public label $L$.
The server starts by sending messages to Alice and Bob and starts the auction timer. When receiving bids, the auction server checks whether the bid is greater than the current winning bid and if so updates winning bid and winner.

When receiving a message on channel \texttt{TICK} from the auction timer, the auction server checks how many rounds are remaining. As the number of rounds is public, the branching on \Cref{line:round_counter_branching} need not be oblivious.
On \Cref{line:alice_winner_branching}, the server obliviously branches on whether Alice is the current leader and if not, it sends her a message informing her what she must bid to take the lead.
Once the round counter reaches zero, the server informs Alice and Bob of the outcome of the auction.

\section{Implementation}
\label{sec:implementation}
In this section we demonstrate the practicality of \oblivio{} by developing an interpreter that implements the language semantics. The interpreter implements security critical operations as constant-time algorithms.

Best practice for constant-time programming is to write the code using low-level languages, as optimising compilers
may change security critical code and void the constant-time property \cite{simon2018you}. However, we argue that development of the interpreter acts as a sanity check on our semantics and ideas and demonstrates that they are, in principle, possible to implement in a low-level language. Our interpreter is written in OCaml, but contains no language specific features.

The interpreter represents integers in the straightforward way, assigning them fixed size.
Strings are represented as as tuples $\texttt{int*char array}$, where the the character array contains the padded string value and the integer component denotes the length of the prefix of the character array corresponding to the actual string. This choice allows us to access indices in arrays based on their padded sizes.
The interpreter supports arithmetic, boolean, and all comparison operations on integers. For strings, it supports comparisons \texttt{=} and \texttt{!=}, and concatenation $\caret$. The interpreter gives integers fixed size hence $\sizeop{\oplus}$ is defined trivially for all operations that return integer result. For concatenation of string $\val{s_1}{z_1},\val{s_2}{z_2}$, we define $z_1 \sizeop{\caret} z_2 = z_1+z_2$.

In the algorithms presented below, we shall use notation $\val{s}{z}$ for strings for consistency with the rest of the paper. The translation to the internal representation of the interpreter is straightforward. For simplicity, the presented algorithms assume integer representations of booleans and chars, leaving conversions implicit.

\Cref{alg:safeeq} presents data-oblivious string comparison. The algorithm returns $1$ if the strings are equal up to padding, otherwise $0$. That is, $\textsc{SafeEq}(\val{s_1}{z_1}, \val{s_2}{z_2})=1$ iff. $s_1 = s_2$. The algorithm stores whether any mismatch has been found in variable $x$, starting with comparing the secret lengths of the strings. The algorithm then checks for equality by taking the $\textit{xor}$ of the character values at every index $i$, masking out indices beyond the semantic strings by using bit $b$. This code is similar to code widely used in state of the art cryptographic libraries such as libsodium \cite{libsodium} and OpenSSL \cite{openssl}. For simplicity, the presented algorithm assumes that the public lengths $z_1,z_2$ are equal. These lengths are public and we assume that padding by a known amount does not leak any secrets.

\begin{algorithm}[!htb]
    \caption{Safe string comparison}
    \label{alg:safeeq}
    \begin{algorithmic}[1]
    \Require $z_1 = z_2$
    \Function{SafeEq}{$\val{s_1}{z_1},\val{s_2}{z_2}$}
        \State $x \gets \textit{size}(s_1) \textit{ lxor } \textit{size}(s_2)$
        \State $l \gets z_1$
        \For{$i = 0, \ldots, l-1$}
            \State $b \gets i < \textit{min}(\textit{size}(s_1),\textit{size}(s_2))$
            \State $x \gets x \textit{ lor } (b \textit{ land } (s_1[i] \textit{ lxor } s_2[i]))$
        \EndFor
        \State \Return $x = 0$
    \EndFunction
    \end{algorithmic}
\end{algorithm}

Our interpreter implements oblivious assignments using a deep-copy semantics and utilising algorithms for constant-time selection. We use a common algorithm for constant-time selection of integers $i,j$:
\[
    x = ((1 \textit{ xor } b) * i) \textit{ lor } (b * j)
\]
From this, we derive an algorithm for constant-time selection of strings (\Cref{alg:safeselect}).

\begin{algorithm}[!htbp]
    \caption{Safe string select}
    \label{alg:safeselect}
    \begin{algorithmic}[1]
    \Require $z_1 = z_2$
    \Function{SafeSelect}{$b,\val{s_1}{z_1},\val{s_2}{z_2}$}
        \For{$i = 0, \ldots, z_1-1$}
            \State $s[i] \gets ((1 \textit{ xor } b) * s_1[i]) \textit{ lor } (b * s_2[i])$
        \EndFor
        \State \Return $\val{s}{z_1}$
    \EndFunction
    \end{algorithmic}
\end{algorithm}

Function \textsc{SafeSelect} takes three arguments: selection bit $b$, and strings $\val{s_1}{z_1}$ and $\val{s_2}{z_2}$. If $b=0$ then $\val{s_1}{z_1}$ is returned and if $b=1$ then $\val{s_2}{z_2}$ is returned. We again assume for simplicity that $z_1 = z_2$. The selection algorithm selects from either $s_1$ or $s_2$ by multiplying with a bit-mask computed from $b$. This kind of bit-masking is ubiquitous in constant-time programming.

Next, we present our algorithm for string concatenation (\Cref{alg:safeconcat}).
\begin{algorithm}[!htbp]
    \caption{Safe string concatenation}
    \label{alg:safeconcat}
    \begin{algorithmic}[1]
    \Function{SafeConcat}{$\val{s_1}{z_1},\val{s_2}{z_2}$}
        \State $z \gets z_1 + z_2$
        \For{$i = 0, \ldots, z-1$}
            \State $s[i] \gets 0$
            \For{$j = 0, \ldots, z_1-1$}
                \State $c \gets s_1[j]$
                \State $b \gets (i = j) \textit{ land } (j < \textit{size}(s_1))$
                \State $s[i] \gets s[i] \textit{ lor } (b*c)$
            \EndFor
            \For{$j = 0, \ldots, z_2-1$}
                \State $c \gets s_2[j]$
                \State $b \gets (i = j + \textit{size}(s_1))$
                \State $s[i] \gets s[i] \textit{ lor } (b*c)$
            \EndFor
        \EndFor
        \State \Return $\val{s}{z}$
    \EndFunction
    \end{algorithmic}
\end{algorithm}

Function \textsc{SafeConcat} takes strings $\val{s_1}{z_1},\val{s_2}{z_2}$ as arguments and outputs $\val{s_1 \caret s_2}{z_1 + z_2}$, the concatenation of $s_1,s_2$ padded to size $z_1 + z_2$. 
This algorithm is our own invention and based on similar bit-masking as the algorithms above. To protect secret length $\textit{size}(s_1)$, the algorithm iterates over both input strings for each index in the output, leading to $\mathcal{O}((z_1+z_2)^2)$ time complexity.

\section{Discussion}
\label{sec:discussion}
We have shown how \oblivio{} can be used for writing secure, reactive programs, and we have demonstrated the feasibility of our approach by developing an interpreter that implements security critical operations as constant-time algorithms.

The core language of \oblivio{} is simple yet expressive and the bit-stack approach for oblivious branching lends itself to data-oblivious methods such as bit-masking. However, every security critical operation needs a secure implementation. Each additional feature we introduce to the language needs a functionally correct, constant-time implementation. Implementing constant-time algorithms is arduous and non-trivial and is made even harder by optimising compilers, that force developers of constant-time code to fight the compiler or write complicated algorithms in low-level languages \cite{simon2018you}. This state of affairs is clearly undesirable from a security perspective and stresses the need for secure compiler support that enable security properties of high-level code to be preserved during compilation.

\subsection{Defences against traffic analysis}
\label{sec:defences_against_traffic_analysis}
Traffic analysis attacks can broadly be divided into two categories: 1) \textit{anonymity}, inferring \textit{who} is participating in actions or communication, and 2) \textit{confidentiality}, inferring \textit{what} actions are performed or the contents of communication.
Various approaches exist for defending against traffic analysis. However, approaches developed for protecting against one category of attacks are not necessarily suitable for protecting against the other. Concealing \textit{who} is performing an action does not necessarily conceal \textit{what} action is performed, and concealing \textit{what} action is performed does not necessarily conceal \textit{who} is participating.

\begin{figure*}[!htb]
    \centering
\resizebox{\textwidth}{!}{%
    \begin{tabular}{l | c | c | c | c | c | c | c}
        \hline
        \textbf{Approach} & \textbf{Traffic overhead} & \textbf{Latency overhead} & \textbf{Bandwidth overhead} & \textbf{CPU-time overhead} & \textbf{Permissiveness} & \textbf{Ease of use} \\
        \hline
        
        \hline
        & & & & & &\\[-1em]
        \text{System-level} & \pie{0} & \pie{0} & \pie{180} & \pie{360} & \pie{360} & \pie{360} \\
        \hline
        & & & & & &\\[-1em]
        \text{Constant-time} & \pie{360} & \pie{360} & \pie{360} & \pie{180} & \pie{0} & \pie{0} \\ 
        \hline
        & & & & & &\\[-1em]
        \text{Data-oblivious} & \pie{180} & \pie{180} & \pie{180} & \pie{180} & \pie{180} & \pie{180} \\ 
        \hline
    \end{tabular}
}
    \caption{Comparison of different approaches}
    \label{fig:comparison}
\end{figure*}

\oblivio{} is designed for protecting confidentiality, protecting the \textit{what} of interactions. We discuss existing approaches and their trade-offs with respect to protecting confidentiality in more detail. The trade-offs are summarised in \Cref{fig:comparison}. We use $\scalebox{.7}{\pie{360}}$ to signify that the approach excels, $\scalebox{.7}{\pie{0}}$ to signify that the approach performs poorly, and $\scalebox{.7}{\pie{180}}$ to signify that the approach falls somewhere in between.

\subsubsection{System-level approaches}
The majority of defences against traffic analysis have been developed at the system level \cite{cherubin2017website} and rely on traffic morphing. Traffic morphing conceals genuine traffic by stretching out bursts of high activity and padding periods of low activity with dummy traffic.
As the approach is program-agnostic, it is general and can in principle be applied to existing systems. This makes it highly permissive. The approach also incurs effectively no overhead in computation time as only genuine traffic needs to be processed.
However, balancing traffic and latency overheads is challenging. In its simplest form, traffic morphing utilises constant rate padding, morphing and padding traffic to other nodes to a constant rate of fixed-size packets. Such constant-rate connections must be maintained for all other nodes the system observably sends traffic to.
Setting a low rate introduces significant latency overheads as messages are buffered, while a high rate introduces significant, if not prohibitive, amounts of dummy traffic \cite{dyer2012peek}.
Due to this trade-off between bandwidth and latency overheads, traffic morphing is most suitable for systems that either produce relatively constant rates of traffic to a fixed set of other nodes, or systems that can tolerate long network delays \cite{apthorpe2019keeping}.

While traffic morphing is in principle general and permissive, \citet{cherubin2017website} note that the approach may require substantial changes to applications and network stack, making deployment unfeasible in practice.

\subsubsection{Program-level approaches}
Compared to system-level approaches, program-level approaches for mitigating traffic analysis are less explored. The program-level approach enforces security by imposing restrictions on the program. This makes the approach less general and less permissive, but at potentially reduced overheads compared to system-level approaches. Program-level approaches can ensure that programs are safe by construction, and as all padding (if any) is performed at the level of the program, no changes are required to the network stack, as at all traffic is genuine at the level of the network.

\paragraph{Constant-time programming}
A simple, but restrictive, paradigm for eliminating timing leaks is constant-time programming. The approach restricts the use of secret data input for a number of standard language features such as branching, indexing arrays, and variable-time operations like division. To get around these restrictions, constant-time programs inline computation from conditional branches and make clever use of bit-masking to ensure correctness of the program.
Inlining branches incurs some overhead in computation time, but the approach introduces no direct overhead in traffic as no commands are conditionally executed depending on secrets.
If a system can be feasibly rewritten as constant-time programs this approach is preferential. However, as \citet{almeida2016verifying} note, adhering to constant-time programming is hard and requires developers to deviate from conventional programming practices.
This makes the approach suitable only for simple programs.

\paragraph{Data-oblivious, reactive programming}
Data-oblivious programming eases the restrictions imposed by constant-time programming by providing high-level support for writing constant-time code, thereby also easing the burden on developers. The principal idea of data-oblivious languages is oblivious conditionals. Oblivious conditionals execute both branches while negating unwanted side-effects of the non-chosen branch. This ensures that control flow and memory accesses are independent of secrets.
\oblivio{} extends this principle to the reactive setting, where programs receive and react to network messages. Our approach incurs an overhead in traffic by introducing dummy messages for send commands that are conditionally executed depending on secrets. This overhead is bounded by a multiplicative factor that is statically known from typing.
Our approach incurs a bandwidth overhead as the size of message values is padded to a public upper bound. Bounding the bandwidth overhead would require further restrictions to the language.
Our approach introduces an overhead in computation time by executing both branches of oblivious conditionals, but we restrict the language to ensure termination of non-chosen branches.
We do not provide a bound on the overhead in latency, but note that the number of dummy messages is bounded and handling them is guaranteed to terminate, hence we are guaranteed to respond to genuine messages eventually.
Compared to system-level approaches, our approach is particularly suitable for programs that do not produce constant rates of traffic to predetermined parties.
We discuss the main limitations of our approach below.

\subsection{Local channels}
Our model assumes that messages on local channels may be sampled in constant-time. This assumption is standard in the literature (e.g. \cite{sabelfeld2002static}). Incoming messages must be processed and prepared for the running application in shared memory, yet at the same time must not have any effect on the execution time of the application. This problem arises from the need to protect message presence. To fully achieve this, fixed-rate scheduling could be used for a separate thread, dedicated to process and prepare local input.

\subsection{Limitations}
Programs in \oblivio{} are static and functions are not first class. This simplifies our model and helps more clearly deliver the core concepts of our approach. However, dynamic features in reactive programs are becoming more and more prevalent, albeit at the cost of introducing leaks \cite{chen2010side}, and cloud computing sees
code being sent to remote servers to be run. Conceptually, the principles we have developed for \oblivio{} could be extended to handle public, dynamic code. Secret code entails secret control flow, which goes against the core design principle of the language. While it is straightforward to protect which of finitely many pieces of code is genuinely executed (by simply executing all but one in phantom mode) it is not clear how we could protect executing arbitrary code without severe restrictions.
The data-oblivious comparison and selection algorithms we present in \Cref{sec:implementation} can be extended to new data structures built from the value types we consider, such as pairs or lists. Here, functions types are again different. It is not clear how constant-time selection and padding could be extended to functions.

Our model assumes that it is public which handler (if any) is triggered by a network message. It is conceptually straightforward to hide this up to some degree, though at the cost of overhead in execution time and traffic, e.g., by setting up anonymity groups such that receiving an event for one handler would trigger the execution of all handlers in the group, with only the genuine handler executing in real mode.

Dynamically registering new secret handlers appears difficult. If a handler for a channel is registered in phantom mode, how should an incoming network message on that channel be handled? We could of course run the handler in phantom mode, but as the handler could produce network traffic this does not appear desirable in general.
If multiple handlers are potentially registered depending on secrets, all handlers would need to be run in order to protect the secrets.

Channels are not first class in our language. There is again no great conceptual difficulty in supporting standard operations on channels if it is not secret which channels are operated on. String identifiers for nodes and handlers could provide an intuitive model for channels, enabling padding and conditional assignment, but sending on conditionally bound channels would reveal which channel is bound as shown by the following example, where an attacker can infer the secret by observing whether the message goes to \texttt{ALICE} or \texttt{BOB}:

\begin{lstlisting}[captionpos=none]
oblif secret
then ch = ALICE/GREET
else ch = BOB/GREET;
send(ch,"Hello");
\end{lstlisting}

It may appear that an anonymous communication channel, possibly connecting via a set of relays, could be used to hide whom a message is sent to and thereby protect the secret. Unfortunately, it is not enough to send a single message for this program if the attacker knows the programs running at \texttt{ALICE} and \texttt{BOB}. Suppose \texttt{ALICE} replies to messages while \texttt{BOB} does not. An attacker armed with this knowledge could infer which node was sent to based on whether a reply is sent.

Our model does not allow real mode computation when handling dummy messages and does not permit genuine messages to be sent in response to dummy messages. Redefining the semantic rules to lift these restrictions would be straightforward, but would complicate reasoning about program correctness for developers and would make it difficult to bound the overhead in traffic.

\section{Related work}
\label{sec:related_work}
Previous work has applied language-based methods to the reactive programming model. \citet{bohannon2009reactive} develop a security type system enforcing noninterference for reactive programs. However, covert channels, such as timing channels, are outside of the model they consider.

\subsection{Secure multi execution and faceted values}
Another approach, which has gained traction in practice, is Secure Multi Execution (SME) \cite{devriese2010noninterference}.
As the name suggests, SME executes a code snippet multiple times -- once per security level -- carefully restricting the in- and output of each execution to only the legal channels. The promise of SME is that secure programs are not adversely affected by the mechanism, a property called \textit{transparency}.
Coupled with a scheduler that prioritises low-execution this approach can be applied in a black-box fashion to programs and protects against timing leaks. However, the low-priority scheduling discipline faces issues in the reactive setting as it does not extend to executing handlers for multiple events \cite{rafnsson2016secure}. Instead, reactive systems incorporating the SME principles such as FlowFox, a Firefox extension for secure information flow in JavaScript, settles for low-priority scheduling on a per-event basis \cite{de2012flowfox}. This compromise unfortunately introduces a leak, as high-execution of a handler for one event taints the low-execution of the handler for the next.

\citet{rafnsson2016secure} develop a fine-grained version of SME that models a network level attacker that observes when messages are sent. They employ a scheduler that enforces that low-runs never "outrun" high-runs to synchronise I/O between the runs.
Their transparency property states that timing-sensitive noninterfering programs are not adversely modified in their I/O behaviour.
\citet{almeida2016verifying} point out that adhering to constant-time programming requires expertise and forces developers to deviate from conventional programming practices.

Execution using faceted values \cite{austin2012multiple} has many of the same advantages and disadvantages as SME. While conditional assignment may intuitively seem related to faceted values, the above considerations for timing-channels in reactive programs under SME also apply here and make faceted values unsuitable. Our approach uses single, real values, regardless of execution mode, simplifying constant-time algorithms used for operations. Furthermore, our approach has the advantage of ruling out insecure programs through typing, rather than being limited to transparency for secure programs.

\subsection{Timing channels and reactive programs}
\citet{bastys2020clockwork} point out that remote attackers do not in general know when a program is started. They develop Clockwork, a monitor that rules out timing leaks in batch programs where starting time is not observed. They demonstrate the permissiveness of their model with the program $\texttt{if $h$ then $h_1$ := $h_2$; send($L$,$1$)}$, which is considered unsafe in many models, and indeed would not be accepted by the type system of \oblivio{}. However, without knowledge of when the program starts the program is safe. This observation does not easily extend to the reactive programs and the network level attackers we consider, simply because it is not possible to react to network messages that have not yet been received. In our model, it is known when an event arrives at a node, and failure to produce any expected responses is observable hence an attacker can easily infer that a system is not running.

\citet{blaabjerg2021towards} consider program-level mitigation of traffic analysis by separating traffic shape from traffic content. Their approach does not mitigate timing differences from sensitive conditionals hence secure programs must set up schedules for all future traffic before executing such conditionals. The approach therefore does not easily extend to the reactive setting.

\citet{mccall2018knowledge} consider how dynamic features in reactive programs, such as registering new handlers, can be used to leak information by abusing declassification policies. They design new SME rules to enforce separation of the declassification module from dynamically generated components.

\citet{vassena2019foundations} propose a dynamic information flow control parallel runtime system that supports deterministic parallel thread execution. Such methods could be used in \oblivio{} to support the parallel processing needed for handling input on local channels.

\subsection{Constant-time execution}
As discussed in \Cref{sec:discussion},
writing constant-time algorithms in a high-level language is not in general sufficient. Optimising compilers may rewrite the code, introduce branches, and otherwise break the constant-time guarantees. \citet{barthe2018secure} consider the problem of preserving side-channel countermeasures, such as constant-time code, during compilation, and present a framework for proving that a compilation pass preserves such countermeasures.

\citet{cauligi2019fact} present a DSL for writing constant-time cryptographic code and compiling it to LLVM bitcode. Their DSL allows programmers to write familiar, high-level code with variables annotated by security labels. Their compiler uses the security labels to transform unsafe behaviour that depends on secrets into constant-time code. They use the dudect analysis tool \cite{reparaz2017dude} to check that machine code compiled from their generated LLVM bitcode is constant-time.

\citet{dantas2018comparative} investigate timing analysis countermeasures in the presence of just-in-time compilation, such as in the JavaVM. Their empirical results indicate that static countermeasures fare worse in this setting, while dynamic countermeasures retain much of their effectiveness.

\subsection{Resource-awareness}
\citet{hofmann2003static} introduce linear potentials for analysing the resource consumption of programs with linear bounds. \citet{hoffmann2010amortized} extend the notion of potentials to polynomial potentials, allowing analysis of programs with polynomial bounds on resource consumption. \citet{krishnaswami2012higher} explores the notion of potentials in functional, reactive programming and present a language that statically bounds the size of the data-flow graph of reactive programs. \citet{dehesa2017verifying} explore a notion of resource-aware noninterference in a setting without IO, where the sizes of program values are known to the attacker.

\section{Conclusion}
\label{sec:conclusion}
We consider the problem of mitigating traffic analysis attacks against online services and applications written in the reactive programming model. We show that data-oblivious computation is a natural fit for preventing timing-channels in this setting when coupled with information-rich dummy traffic. 
We develop the language \oblivio{}, a language for data-oblivious, reactive programs. We prove that well-typed programs in the language are secure against traffic analysis by convincingly padding traffic with dummy messages, and we show that the overhead introduced by our approach is bounded. We demonstrate the expressiveness of our language by example and the practicality of the language by developing an interpreter that implements security critical operations as constant-time algorithms.

\section{Acknowledgements}
\label{sec:acknowledgements}
We thank the anonymous reviewers for their valuable suggestions for improving the paper.
This work was funded by the Danish Council Independent Research 
for the Natural Sciences (DFF/FNU, project 6108-00363).

\bibliography{bibliography}

\ifdefined\istechnicalreport
\clearpage

\onecolumn

\appendices
\crefalias{section}{appendix}
\crefalias{subsection}{appendix}

\section{Auxiliary model}
\label{appendix:auxiliarymodel}
This section provides an auxiliary model for \oblivio{} and we show adequacy and preservation for the auxiliary model. The auxiliary model is used for proving our soundness theorem from \Cref{sec:enforcement} (\Cref{theorem:soundness}) and overhead theorem from \Cref{sec:enforcementoverhead} (\Cref{theorem:enforcementoverhead}).

\Cref{def:semantics_commands_aux} presents the auxiliary semantics for commands. Program configurations in the auxiliary semantics augment the standard program configurations from \Cref{sec:language} with typing environments $\Gamma,\Delta$ and stack of $\pc$-labels $\pcstack$. These are used for tracking the $\pc$-level at runtime. Only the rules for oblivious branching and pop have been changed beyond simply changing to augmented configurations. The rule for oblivious branching uses $\Gamma,\Delta$ to obtain the security level of the guard expression and takes its least upper bound with the current $\pc$-label, which it pushes two copies of onto the stack -- one for each branch. The rule for pop now pops from both $\pcstack$ and $\bitstack$. This way, we relate each index in the $\bitstack$ to its corresponding security level in $\pcstack$, that is, $\pc$-stack does not describe the security level of the control flow, but of the execution mode.

\begin{figure}[!htbp]
\centering
\begin{mathparpagebreakable}
\inferrule[Skip-Aux] { } {
    \xangle{\Ctxaux{\Gamma}{\Pi}{\Lambda}{\Delta}{\pcstack}{q} \mid \bitstack,\texttt{skip},m,\mu,\intenv,\hst}
    \longrightarrowdbl
    \xangle{\Ctxaux{\Gamma}{\Pi}{\Lambda}{\Delta}{\pcstack}{q} \mid \bitstack,\texttt{stop},m,\mu,\intenv,\hst \Colon \textsf{skp}}
}
\and
\inferrule[\textsc{Seq1-Aux}] { 
    \xangle{\Ctxaux{\Gamma}{\Pi}{\Lambda}{\Delta}{\pcstack}{q} \mid \bitstack,c_1,m,\mu,\intenv,\hst}
    \xrightarrowdbl{\alpha}
    \xangle{\Ctxaux{\Gamma}{\Pi}{\Lambda}{\Delta}{\pcstack'}{q'} \mid \bitstack',c'_1,m',\mu',\intenv',\hst'} \\
    c'_1 \neq \texttt{stop}
} {
    \xangle{\Ctxaux{\Gamma}{\Pi}{\Lambda}{\Delta}{\pcstack}{q} \mid \bitstack,c_1;c_2,m,\mu,\intenv,\hst}
    \xrightarrowdbl{\alpha}
    \xangle{\Ctxaux{\Gamma}{\Pi}{\Lambda}{\Delta}{\pcstack'}{q'} \mid \bitstack',c'_1;c_2,m',\mu',\intenv',\hst'}
}
\and
\inferrule[\textsc{Seq2-Aux}] { 
    \xangle{\Ctxaux{\Gamma}{\Pi}{\Lambda}{\Delta}{\pcstack}{q} \mid \bitstack,c_1,m,\mu,\intenv,\hst}
    \xrightarrowdbl{\alpha}
    \xangle{\Ctxaux{\Gamma}{\Pi}{\Lambda}{\Delta}{\pcstack'}{q'} \mid \bitstack',\texttt{stop},m',\mu',\intenv',\hst'}
} {
    \xangle{\Ctxaux{\Gamma}{\Pi}{\Lambda}{\Delta}{\pcstack}{q} \mid \bitstack,c_1;c_2,m,\mu,\intenv,\hst}
    \xrightarrowdbl{\alpha}
    \xangle{\Ctxaux{\Gamma}{\Pi}{\Lambda}{\Delta}{\pcstack'}{q'} \mid \bitstack',c_2,m',\mu',\intenv',\hst'}
}
\and
\inferrule[Assign-Aux] { 
    \eval{e,m,\mu}{\val{v}{z}}
} {
    \xangle{\Ctxaux{\Gamma}{\Pi}{\Lambda}{\Delta}{\pcstack}{q} \mid \bit \Colon \xbit{1} \Colon \bitstack,x \assign e,m,\mu,\intenv,\hst}
    \longrightarrowdbl
    \xangle{\Ctxaux{\Gamma}{\Pi}{\Lambda}{\Delta}{\pcstack}{q} \mid \bit \Colon \xbit{1} \Colon \bitstack,\texttt{stop},m,\mu[x \mapsto \val{v}{z},\intenv,\hst \Colon \textsf{asn}(x,e,z)}
}
\and
\inferrule[OblivAssign-Aux] {
    \mu(x) = \val{v_0}{z_0} \\
    \eval{e,m,\mu}{\val{v_1}{z_1}} \\
    z = \textit{max}(z_0,z_1) \\
    i =
    {\begin{cases}
        1
            &\textit{if } \bit = \xbit{1} \\
        0
            &\textit{if } \bit = \xbit{0}
    \end{cases}}
} {
    \xangle{\Ctxaux{\Gamma}{\Pi}{\Lambda}{\Delta}{\pcstack}{q} \mid \bit \Colon \bitstack,x \oblivassign e,m,\mu,\intenv,\hst}
    \longrightarrowdbl
    \xangle{\Ctxaux{\Gamma}{\Pi}{\Lambda}{\Delta}{\pcstack}{q} \mid \bit \Colon \bitstack,\texttt{stop},m,\mu[x \mapsto \val{v_i}{z}],\intenv,\hst \Colon \textsf{casn}(x,e,z)}
}
\and
\inferrule[LocalInput-Aux] {
    \mu(x) = \val{v_x}{z_x} \\
    \eval{e,m,\mu}{\val{n_e}{z_e}} \\
    z' = \textit{max}(z_x,n_e) \\
    \hst' = \hst \Colon \textsf{in}(x,\ch,e,z') \\
    v',\intenv' =
    {\begin{cases}
        v, \intenv[\ch \mapsto \textit{tl}]
            &\text{if }
            \bit = \xbit{1}
            \text{ and }
            \intenv(\ch) = \val{v}{z} \Colon \textit{tl}
            \text{ and } z \leq n_e \\
        v_x, \intenv[\ch \mapsto \textit{tl}]
            &\text{if }
            \bit = \xbit{1}
            \text{ and }
            \intenv(\ch) = \bullet \Colon \textit{tl} \\
        v_x, \intenv
            &\text{otherwise}
    \end{cases}}
} {
    \xangle{\Ctxaux{\Gamma}{\Pi}{\Lambda}{\Delta}{\pcstack}{q} \mid \bit \Colon \bitstack,x \oblivassign \texttt{input}(\ch,e),m,\mu,\intenv,\hst}
    \longrightarrow
    \xangle{\Ctxaux{\Gamma}{\Pi}{\Lambda}{\Delta}{\pcstack}{q} \mid \bit \Colon \bitstack,\texttt{stop},m,\mu[x \mapsto \val{v'}{z'}],\intenv',\hst'}
}
\and
\inferrule[Send-Aux] { 
    \Lambda(\ch) = \ltype{\_}{\ltriple{\lmode}{\_}{r}}
    \bitstack = \bit \Colon \_ \\
    \eval{e,m,\mu}{\val{v}{z}} \\
    \hst' = \hst \Colon \textsf{out}(\ch,e,z) \\
    \timestamp = \textit{time}(\hst') \\
    q' =
    {\begin{cases}
        0
            &\textit{if } \lmode = \bot \\
        1 + r
            &\textit{otherwise}
    \end{cases}}
} {
    \xangle{\Ctxaux{\Gamma}{\Pi}{\Lambda}{\Delta}{\pcstack}{q+q'} \mid \bitstack,\texttt{send}(\ch,e),m,\mu,\intenv,\hst}
    \xrightarrowdbl{\overrightarrow{\ch}(\timestamp,\bit,\val{v}{z})}
    \xangle{\Ctxaux{\Gamma}{\Pi}{\Lambda}{\Delta}{\pcstack}{q} \mid \bitstack,\texttt{stop},m,\mu,\intenv,\hst'}
}
\and
\inferrule[If-Aux] { 
    \eval{e,m,\mu}{\val{v}{z}} \\
    v \neq 0 \implies i = 1 \\
    v = 0 \implies i = 2
} {
    \xangle{\Ctxaux{\Gamma}{\Pi}{\Lambda}{\Delta}{\pcstack}{q} \mid \bitstack,\texttt{if $e$ then $c_1$ else $c_2$},m,\intenv,\hst}
    \longrightarrowdbl
    \xangle{\Ctxaux{\Gamma}{\Pi}{\Lambda}{\Delta}{\pcstack}{q} \mid \bitstack,c_i,m,\mu,\intenv,\hst \Colon \textsf{br}(e,z,i)}
}
\and
\inferrule[While-Aux] {
    c' = \texttt{if $e$ then $c$; while $e$ do $c$ else skip}
} {
    \xangle{\Ctxaux{\Gamma}{\Pi}{\Lambda}{\Delta}{\pcstack}{q} \mid \xbit{1} \Colon \bitstack,\texttt{while $e$ do $c$},m,\mu,\intenv,\hst}
    \longrightarrowdbl
    \xangle{\Ctxaux{\Gamma}{\Pi}{\Lambda}{\Delta}{\pcstack}{q} \mid \xbit{1} \Colon \bitstack,c',m,\mu,\intenv,\hst \Colon \textsf{whl}}
}
\and
\inferrule[OblivIf-Aux] {
    \Gamma,\Delta \vdash e : \ltype{\sigma}{\ell} \\
    \pcstack = \pc \Colon \_ \\
    \bitstack = \bit \Colon \_ \\
    \eval{e,m,\mu}{\val{v}{z}} \\
    \pc' = \ell \sqcup \pc \\
    v \neq 0 \implies \bit_1 = \bit \land \bit_2 = \xbit{0} \\
    v = 0 \implies \bit_1 = \xbit{0} \land \bit_2 = \bit \\
    \pcstack' = \pc' \Colon \pc' \Colon \pcstack \\
    \bitstack' = \bit_1 \Colon \bit_2 \Colon \bitstack \\
    \hst' = \hst \Colon \textsf{obr}(e,z)
} {
    \xangle{\Ctxaux{\Gamma}{\Pi}{\Lambda}{\Delta}{\pcstack}{q} \mid \bitstack,\texttt{oblif e then $c_1$ else $c_2$},m,\intenv,\hst}
    \longrightarrowdbl
    \xangle{\Ctxaux{\Gamma}{\Pi}{\Lambda}{\Delta}{\pcstack'}{q} \mid \bitstack',c_1;\texttt{pop};c_2;\texttt{pop},m,\intenv,\hst'}
}
\and
\inferrule[Pop-Aux] { 
} {
    \xangle{\Ctxaux{\Gamma}{\Pi}{\Lambda}{\Delta}{\pc \Colon \pcstack}{q} \mid \bit \Colon \bitstack,\texttt{pop},m,\mu,\intenv,\hst}
    \longrightarrowdbl
    \xangle{\Ctxaux{\Gamma}{\Pi}{\Lambda}{\Delta}{\pcstack}{q} \mid \bitstack,\texttt{stop},m,\mu,\intenv,\hst \Colon \textsf{pop}}
}
\end{mathparpagebreakable}
\caption{Auxiliary semantics of commands}
\label{def:semantics_commands_aux}
\end{figure}

\begin{restatable}[Adequacy of commands]{lemma}{commandadequacy}
\label{lemma:commandadequacy} Consider $\Gamma,\Pi,\Lambda,\Delta,\pc,q$ and $\xangle{\bitstack,c,m,\mu,\intenv,\hst}$ such that $\Ctx{\Gamma}{\Pi}{\Lambda}{\Delta}{\pc} \vdash^q c$. We have
\[
    \xangle{\bitstack,c,m,\mu,\intenv,\hst}
    \xrightarrow{\alpha}
    \xangle{\bitstack',c',m',\mu',\intenv',\hst'}
\]
if and only if for all $\pcstack$ such that $\pcstack \vdash \bitstack$, there exists $\pcstack',q'$ such that $\pcstack' \vdash \bitstack'$ and $q \geq q'$ and
\[
    \xangle{\Ctxaux{\Gamma}{\Pi}{\Lambda}{\Delta}{\pcstack}{q} \mid \bitstack,c,m,\mu,\intenv,\hst}
    \xrightarrowdbl{\alpha}
    \xangle{\Ctxaux{\Gamma}{\Pi}{\Lambda}{\Delta}{\pcstack'}{q'} \mid \bitstack',c',m',\mu',\intenv',\hst'}
\]
\end{restatable}
\begin{IEEEproof}
We inspect each direction separately.

\begin{description}
    \item[From standard to auxiliary:]
    We proceed by induction on $c$.
    \begin{description}
        \item[Case $c$ is $\texttt{skip}$:]
        Done by picking $\pcstack'=\pcstack$.
        
        \item[Case $c$ is $c_1;c_2$:]
        By the induction hypothesis.
    
        \item[Case $c$ is $x \assign e$:]
        Done by picking $\pcstack'=\pcstack$.
    
        \item[Case $c$ is $x \oblivassign e$:]
        Done by picking $\pcstack'=\pcstack$.
            
        \item[Case $c$ is $x \oblivassign \texttt{input}(\ch,e)$:]
        Done by picking $\pcstack'=\pcstack$.
            
        \item[Case $c$ is $\texttt{send}(\ch,e)$:]
        By $\Ctx{\Gamma}{\Pi}{\Lambda}{\Delta}{\pc} \vdash^q c$ we have $\Lambda(\ch) = \ltype{\sigma}{\ltriple{\lmode}{\lval}{r}}$ such that $\lmode = \bot \implies q = q'$ and $\lmode \neq \bot \implies q = q' + 1 + r$.
        Done by picking $\pcstack'=\pcstack$.
            
        \item[Case $c$ is $\texttt{if $e$ then $c$ else $c$}$:]
        Done by picking $\pcstack'=\pcstack$.
            
        \item[Case $c$ is $\texttt{while $e$ do $c$}$:]
        Done by picking $\pcstack'=\pcstack$.
            
        \item[Case $c$ is $\texttt{oblif $e$ then $c$ else $c$}$:]
        By $\Ctx{\Gamma}{\Pi}{\Lambda}{\Delta}{\pc} \vdash^q c$ we have $\Gamma;\Delta \vdash e : \ltype{\sigma}{\ell}$.
        By the step in the standard semantics we have $\bitstack^\dprime = \bit \Colon \_$ for some $\bit$, and thus by assumption have 
        $\pcstack = \pc \Colon \_$ for some $\pc$. We are done by choosing as witness to the existential 
        $\pcstack' = \ell \sqcup \pc \Colon \ell \sqcup \pc \Colon \pcstack$.
            
        \item[Case $c$ is $\texttt{pop}$:]
        By the step in the standard semantics we have $\bitstack^\dprime = \bit \Colon \bitstack'$ for some $\bit$, and thus by assumption have $\pcstack = \pc \Colon \pcstack'$ for some $\pc$. Hence, we have a step in the auxiliary semantics and we are done.
    \end{description}

    \item[From auxiliary to standard:]
    By induction on $c$. All cases are trivial.
\end{description}
\end{IEEEproof}

The auxiliary type system (\Cref{def:type_system_commands_aux}) is similar to the standard type system, except using $\pcstack$ for tracking the security level of the execution mode. Most of the rules are straight forward and we only explain the rules for typing oblivious branching and pop. Unlike in the standard typing rules, the auxiliary typing rules contains a rule for command \texttt{pop} as the command appears at runtime. To show adequacy of the auxiliary typing, we restrict branches of oblivious branching to using $\pc$-stacks of size 1, and require that the stack is not empty after a pop. Intuitively, we disallow the branch itself from containing \texttt{pop} commands.

\begin{figure}[!htbp]
\centering
\begin{mathparpagebreakable}
\inferrule[T-Skip-Aux] { } {
    \Ctx{\Gamma}{\Pi}{\Lambda}{\Delta}{\pcstack} \vdash^q_{\textit{aux}} \texttt{skip} : \pcstack
}
\and
\inferrule[T-Seq-Aux] {
    \Ctx{\Gamma}{\Pi}{\Lambda}{\Delta}{\pcstack_1} \vdash^{q_1}_{\textit{aux}} c_1 : \pcstack_2 \\
    \Ctx{\Gamma}{\Pi}{\Lambda}{\Delta}{\pcstack_2} \vdash^{q_2}_{\textit{aux}} c_2 : \pcstack_3
} {
    \Ctx{\Gamma}{\Pi}{\Lambda}{\Delta}{\pcstack_1} \vdash^{q_1+q_2}_{\textit{aux}} c_1;c_2 : \pcstack_3
}
\and
\inferrule[T-Assign-Aux] {
    x \notin \textit{dom}(\Delta) \\
    \Gamma(x) = \ltype{\sigma}{\ell_x} \\
    \Gamma;\Delta \vdash e : \ltype{\sigma}{\ell_e} \\
    \ell_e \sqsubseteq \ell_x
} {
    \Ctx{\Gamma}{\Pi}{\Lambda}{\Delta}{[\bot]} \vdash^q_{\textit{aux}} x \assign e : [\bot]
}
\and
\inferrule[T-OblivAssign-Aux] {
    x \notin \textit{dom}(\Delta) \\
    \Gamma(x) : \ltype{\sigma}{\ell_x} \\
    \Gamma;\Delta \vdash e : \ltype{\sigma}{\ell_e} \\
    \ell_e \sqcup \pc \sqsubseteq \ell_x
} {
    \Ctx{\Gamma}{\Pi}{\Lambda}{\Delta}{\pc \Colon \pcstack} \vdash^q_{\textit{aux}} x \oblivassign e : \pc \Colon \pcstack
}
\and
\inferrule[T-LocalInput-Aux] {
    x \notin \textit{dom}(\Delta) \\
    \Gamma(x) : \ltype{\sigma}{\ell_x} \\
    \Pi(\ch) = \ltype{\sigma}{\ell_\ch} \\
    \Gamma;\Delta \vdash e : \ltype{\inttype}{\ell_e} \\
    \ell_e \sqcup \pc \sqsubseteq \ell_\ch \sqsubseteq \ell_x
} {
    \Ctx{\Gamma}{\Pi}{\Lambda}{\Delta}{\pc \Colon \pcstack} \vdash^q_{\textit{aux}} x \oblivassign \texttt{input}(\ch,e) : \pc \Colon \pcstack
}
\and
\inferrule[T-Send-Aux] {
    \Gamma;\Delta \vdash e : \ltype{\sigma}{\ell_e} \\
    \Lambda(\ch) = \ltype{\sigma}{\ltriple{\lmode}{\lval}{r}} \\
    \pc \sqsubseteq \lmode \\ 
    \ell_e \sqsubseteq \lval \\
    q' = 
    {\begin{cases}
        0
            &\textit{if } \pc = \bot \\
        1+r
            &\textit{otherwise}
    \end{cases}}
} {
    \Ctx{\Gamma}{\Pi}{\Lambda}{\Delta}{\pc \Colon \pcstack} \vdash^{q+q'}_{\textit{aux}} \texttt{send}(\ch,e) : \pc \Colon \pcstack
}
\and
\inferrule[T-If-Aux] {
    \Gamma;\Delta \vdash e : \ltype{\inttype}{\bot} \\
    \Ctx{\Gamma}{\Pi}{\Lambda}{\Delta}{\pcstack} \vdash^q_{\textit{aux}} c_1 : \pcstack \\
    \Ctx{\Gamma}{\Pi}{\Lambda}{\Delta}{\pcstack} \vdash^q_{\textit{aux}} c_2 : \pcstack
} {
    \Ctx{\Gamma}{\Pi}{\Lambda}{\Delta}{\pcstack} \vdash^q_{\textit{aux}} \texttt{if $e$ then $c_1$ else $c_2$} : \pcstack
}
\and
\inferrule[T-While-Aux] {
    \Gamma;\Delta \vdash e : \ltype{\inttype}{\bot} \\
    \Ctx{\Gamma}{\Pi}{\Lambda}{\Delta}{[\bot]} \vdash^0_{\textit{aux}} c : [\bot]
} {
    \Ctx{\Gamma}{\Pi}{\Lambda}{\Delta}{[\bot]} \vdash^q_{\textit{aux}} \texttt{while $e$ do $c$} : [\bot]
}
\and
\inferrule[T-OblivIf-Aux] {
    \Gamma;\Delta \vdash e : \ltype{\inttype}{\ell} \\
    \pcstack = \pc \Colon \pcstack' \\
    \Ctx{\Gamma}{\Pi}{\Lambda}{\Delta}{\pc \sqcup \ell \Colon \pcstack} \vdash^{q_1}_{\textit{aux}} c_1 : \pc \sqcup \ell \Colon \pcstack \\
    \Ctx{\Gamma}{\Pi}{\Lambda}{\Delta}{\pc \sqcup \ell \Colon \pcstack} \vdash^{q_2}_{\textit{aux}} c_2 : \pc \sqcup \ell \Colon \pcstack
} {
    \Ctx{\Gamma}{\Pi}{\Lambda}{\Delta}{\pcstack} \vdash^{q_1+q_2}_{\textit{aux}} \texttt{oblif $e$ then $c_1$ else $c_2$} : \pcstack
}
\and
\inferrule[T-Pop-Aux] {
    \pcstack \neq []
} {
    \Ctx{\Gamma}{\Pi}{\Lambda}{\Delta}{\pc \Colon \pcstack} \vdash^q_{\textit{aux}} \texttt{pop} : \pcstack
}
\end{mathparpagebreakable}
\caption{Auxiliary typing of commands}
\label{def:type_system_commands_aux}
\end{figure}

We show adequacy of the auxiliary type system in \Cref{lemma:commandtypingadequacy} and adequacy of the semantics for commands in in \Cref{lemma:commandadequacy}. Adequacy for commands relates the sizes of bit-stacks $\bitstack$ and $\pc$-stacks $\pcstack$.

\begin{restatable}[Adequacy of command typing]{lemma}{commandtypingadequacy}
\label{lemma:commandtypingadequacy}
Consider $\Gamma,\Delta,\Pi,\Lambda$, $\pc$, $q$, and command $c$. We have
\[
    \Ctx{\Gamma}{\Pi}{\Lambda}{\Delta}{\pc} \vdash^q c
\]
if and only if for all $\pcstack$
we have
\[
    \Ctx{\Gamma}{\Pi}{\Lambda}{\Delta}{\pc \Colon \pcstack} \vdash^q_{\text{aux}} c : \pc \Colon \pcstack
\]
\end{restatable}
\begin{IEEEproof}
We inspect each direction separately.
\begin{description}
    \item[Standard to auxiliary:]
    We proceed by induction on the typing judgement.
    \begin{description}
        \item[Case \textsc{T-Skip-Aux}:]
        Trivial.
        
        \item[Case \textsc{T-Seq}:]
        By two applications of the induction hypothesis.
        
        \item[Case \textsc{T-Assign}:]
        We observe by \textsc{T-Assign} that $\pc = \bot$. Done as there exists no $\pcstack \neq []$ such that $\bot \vdash \bot \Colon \pcstack$.
        
        \item[Case \textsc{T-OblivAssign}:]
        By observing that $[\pc]$ is a non-empty stack with $\pc$ as top element.
        
        \item[Case \textsc{T-LocalInput}:]
        By observing that $[\pc]$ is a non-empty stack with $\pc$ as top element.
        
        \item[Case \textsc{T-Send}:]
        By observing that $[\pc]$ is a non-empty stack with $\pc$ as top element.
        
        \item[Case \textsc{T-If}:]
        By two applications of the induction hypothesis.
        
        \item[Case \textsc{T-While}:]
        We observe by \textsc{T-While} that $\pc = \bot$. Done as there exists no $\pcstack \neq []$ such that $\bot \vdash \bot \Colon \pcstack$.
        
        \item[Case \textsc{T-OblivIf}:]
        By two applications of the induction hypothesis.
    \end{description}

    \item[Auxiliary to standard:]
    We proceed by induction on the typing judgement.
    \begin{description}
        \item[Case \textsc{T-Skip-Aux}:]
        Trivial.
        
        \item[Case \textsc{T-Seq-Aux}:]
        By two applications of the induction hypothesis.
        
        \item[Case \textsc{T-Assign-Aux}:]
        Trivial.
        
        \item[Case \textsc{T-OblivAssign-Aux}:]
        Trivial.
        
        \item[Case \textsc{T-LocalInput-Aux}:]
        Trivial.
        
        \item[Case \textsc{T-Send-Aux}:]
        Trivial.
        
        \item[Case \textsc{T-If-Aux}:]
        By two applications of the induction hypothesis.
        
        \item[Case \textsc{T-While-Aux}:]
        By application of the induction hypothesis.
        
        \item[Case \textsc{T-OblivIf-Aux}:]
        By two applications of the induction hypothesis.
        
        \item[Case \textsc{T-Pop-Aux}:]
        Holds vacuously.
    \end{description}
\end{description}
\end{IEEEproof}

We can now show preservation of commands (\Cref{lemma:commandpreservation}).

\begin{restatable}[Preservation of commands]{lemma}{commandpreservation}
\label{lemma:commandpreservation}
Consider $\pcstack^\dprime$ and $\xangle{\Ctxaux{\Gamma}{\Pi}{\Lambda}{\Delta}{\pcstack}{q} \mid \bitstack,c,m,\mu,\intenv,\hst}$ such that $\Delta \vdash m$, $\Gamma \vdash \mu$, $\Pi \vdash \intenv, \pcstack \vdash \bitstack$, and $\Ctx{\Gamma}{\Pi}{\Lambda}{\Delta}{\pcstack} \vdash^q_{\textit{aux}} c : \pcstack^\dprime$. If
\[
    \xangle{\Ctxaux{\Gamma}{\Pi}{\Lambda}{\Delta}{\pcstack}{q} \mid \bitstack,c,m,\mu,\intenv,\hst}
    \xrightarrow{\alpha}
    \xangle{\Ctxaux{\Gamma}{\Pi}{\Lambda}{\Delta}{\pcstack'}{q'} \mid \bitstack',c',m',\mu',\intenv',\hst'}
\]
then $\Delta \vdash m'$, $\Gamma \vdash \mu'$, $\Pi \vdash I'$, $\pcstack' \vdash \bitstack'$, and one of the following holds
\begin{enumerate}
    \item $c'$ = \texttt{stop} and $\pcstack' = \pcstack^\dprime$
    \item or $\Ctx{\Gamma}{\Pi}{\Lambda}{\Delta}{\pcstack'} \vdash^{q'}_{\textit{aux}} c' : \pcstack^\dprime$
\end{enumerate}
\end{restatable}
\begin{IEEEproof} The proof is by induction on the structure of $c$.
\begin{description}
    \item[Case $c$ is $\texttt{skip}$:]
    By \textsc{Skip-Aux} we are done by the first item as $c'$ = $\texttt{stop}$ and $\pcstack = \pcstack' = \pcstack^\dprime$.
    
    \item[Case $c$ is $c_1;c_2$:]
    By \textsc{T-Seq-Aux} we have $q=q_1 + q_2$ such that (1) $\Ctx{\Gamma}{\Pi}{\Lambda}{\Delta}{\pcstack_1} \vdash^{q_1}_{\text{aux}} c_1 : \pcstack_2$ and (2) $\Ctx{\Gamma}{\Pi}{\Lambda}{\Delta}{\pcstack_2} \vdash^{q_2}_{\text{aux}} c_2 : \pcstack_3$. By $\textsc{Seq-Aux}$ we case on $c' = c_2$.
    \begin{description}
        \item[Case $c' = c_2$:]
        By $\textsc{Seq2-Aux}$ and the induction hypothesis that $\pcstack' = \pcstack_2$ hence by hence we are done by (2).
        
        \item[Case $c' \neq c_2$:]
        By the induction hypothesis we have $\Ctx{\Gamma}{\Pi}{\Lambda}{\Delta}{\pcstack'} \vdash^{q'_1}_{\text{aux}} c' : \pcstack_2$ such that $q' = q'_1 + q_2$ and we are done by (2) and \textsc{T-Seq-Aux}.
    \end{description}
    
    \item[Case $c$ is $x \assign e$:]
    By the step we have $\bitstack = \bit \Colon \bitstack'$.
    We case on $\bit = 0$:
    \begin{description}
        \item[Case $\bit = 0$:]
        We have $\mu' = \mu$ and are done by the first item.
        
        \item[Case $\bit \neq 0$:]
        We have $\eval{e,m,\mu}{\val{v}{z}}$ such that $\mu'=\mu[x \mapsto \val{v}{z}]$. By \textsc{T-Assign-Aux} we have $\Gamma(x) = \ltype{\sigma}{\ell_x}$ and $\Gamma;\Delta \vdash e : \ltype{\sigma}{\ell_e}$ and therefore $\Gamma \vdash \mu[x \mapsto \val{v}{z}]$, and we are done by the first item.
    \end{description}

    \item[Case $c$ is $x \oblivassign e$:]
    By \textsc{T-OblivAssign-Aux} we have $\Gamma(x) = \ltype{\sigma}{\ell_x}$ and $\Gamma;\Delta \vdash e : \ltype{\sigma}{\ell_e}$. Hence, given $\mu(x) = \val{v_0}{z_0}$ and $\eval{e,m,\mu}{\val{v_1}{z_1}}$ and letting $z=\textit{max}(z_0,z_1)$ we have $\Gamma \vdash \mu[x \mapsto \val{v_i}{z}]$ for $i=0,1$.
    By \textsc{OblivAssign-Aux} we are done by the first item as $c'$ = $\texttt{stop}$ and $\pcstack = \pcstack' = \pcstack^\dprime$.
        
    \item[Case $c$ is $x \oblivassign \texttt{input}(\ch,e)$:]
    By \textsc{T-Input-Aux} we have (1) $\Gamma(x) = \ltype{\sigma}{\ell_x}$ and (2) $\Pi(\ch) = \ltype{\sigma}{\ell_\ch}$.
    We consider the each of the three cases for $v^\dprime,\intenv^\dprime$, and are trivially done in the third case.
    In the first two cases we have by $\Pi \vdash \intenv$, (2) and $z' = \textit{max}(z_x,n)$ that $\Gamma \vdash \mu[x \mapsto \val{v^\dprime}{z'}]$ and $\Pi \vdash \intenv^\dprime$.
    By \textsc{Input-Aux} we are done by the first item as $c'$ = $\texttt{stop}$ and $\pcstack = \pcstack' = \pcstack^\dprime$.
        
    \item[Case $c$ is $\texttt{send}(\ch,e)$:]
    By \textsc{Send-Aux} we are done by the first item as $c'$ = $\texttt{stop}$ and $\pcstack = \pcstack' = \pcstack^\dprime$.
        
    \item[Case $c$ is $\texttt{if $e$ then $c_1$ else $c_2$}$:]
    By \textsc{T-If-Aux} we have (1) $\Ctx{\Gamma}{\Pi}{\Lambda}{\Delta}{\pcstack} \vdash^{q_1+q_2}_{\text{aux}} c : \pcstack$, (2) $\Ctx{\Gamma}{\Pi}{\Lambda}{\Delta}{\pcstack} \vdash^{q_1}_{\text{aux}} c_1 : \pcstack$, and (3) $\Ctx{\Gamma}{\Pi}{\Lambda}{\Delta}{\pcstack} \vdash^{q_2}_{\text{aux}} c_2 : \pcstack$.
    
    Hence $\pcstack' = \pcstack^\dprime = \pcstack$. We case on the branch taken and are done by (2) and (3) respectively.
        
    \item[Case $c$ is $\texttt{while $e$ do $c$}$:]
    By \textsc{T-While-Aux} we have (1) $\Gamma;\Delta \vdash_{\text{aux}} e : \ltype{\inttype}{\bot}$, and (2) $\Ctx{\Gamma}{\Pi}{\Lambda}{\Delta}{[\bot]} \vdash^0_{\text{aux}} c : [\bot]$. Hence, we are done as we can conclude
    \[
        \Ctx{\Gamma}{\Pi}{\Lambda}{\Delta}{[\bot]} \vdash^0_{\text{aux}} \texttt{if $e$ then $c$; while $e$ do $c$ else skip} : [\bot]
    \]
        
    \item[Case $c$ is $\texttt{oblif $e$ then $c_1$ else $c_2$}$:]
    By \textsc{T-Oblif-Aux} we have (1) $\Ctx{\Gamma}{\Pi}{\Lambda}{\Delta}{\pcstack} \vdash^{q_1+q_2}_{\text{aux}} \texttt{oblif $e$ then $c_1$ else $c_2$} : \pcstack$, (2) $\pcstack = \pc \Colon \_$, (3) $\Gamma;\Delta \vdash e : \ltype{\inttype}{\ell}$ s.t. $\ell \neq \bot$, (4) $\Ctx{\Gamma}{\Pi}{\Lambda}{\Delta}{\pc \sqcup \ell \Colon \pcstack} \vdash^{q_1}_{\text{aux}} c_1 : \pc \sqcup \ell \Colon \pcstack$, and (5) $\Ctx{\Gamma}{\Pi}{\Lambda}{\Delta}{\pc \sqcup \ell \Colon \pcstack} \vdash^{q_2}_{\text{aux}} c_2 : \pc \sqcup \ell \Colon \pcstack$.
    
    By \textsc{OblivIf-Aux} we must show
    \[
        \Ctx{\Gamma}{\Pi}{\Lambda}{\Delta}{\pc \sqcup \ell \Colon \pc \sqcup \ell \Colon \pcstack}
        \vdash^{q_1+q_2}_{\text{aux}} c_1;\texttt{pop};c_2;\texttt{pop} : \pcstack
    \]
    This holds by (4), \textsc{T-Seq-Aux}, \textsc{T-Pop-Aux}, \textsc{T-Seq-Aux}, (5), and \textsc{T-Seq-Aux} and \textsc{T-Pop-Aux} again.
        
    \item[Case $c$ is $\texttt{pop}$:]
    We are done by \textsc{Pop-Aux} observing that $c = \texttt{stop}$ and $\pcstack' = \pcstack^\dprime$.
\end{description}
\end{IEEEproof}

Before defining auxiliary typing of systems, we first define two well-formedness criteria. We say that $\pc$-stack $\pcstack$ is well-formed with respect to level $\pc$ if $\pc$ is the bottom element of the stack and higher elements in the stack are not $\bot$ and are at least as secret as the elements below (\Cref{def:wellformedpcstack}). This definition gives us that if $\pc \vdash \pcstack$ for some $\pc \neq \bot$, then all elements of $\pcstack$ are at least as secret as $\pc$ and hence non-public.
\begin{definition}[Well-formed $\pc$-stack] Define $\pc$-stack $\pcstack$ to be well-formed w.r.t. level $\pc$, written $\pc \vdash \pcstack$, by the following rules:
\label{def:wellformedpcstack}
\begin{mathparpagebreakable}
\inferrule{
}{
    \pc \vdash [\pc]
}
\and
\inferrule{
    \pc' \neq \bot \\
    \pc'' \sqsubseteq \pc' \\
    \pc \vdash \pc'' \Colon \pcstack
}{
    \pc \vdash \pc' \Colon \pc'' \Colon \pcstack
}
\end{mathparpagebreakable}
\end{definition}

We say that bit-stack $\bitstack$ is well-formed with respect to $\pc$-stack $\pcstack$ if they are the same size, the bottom element of the bit-stack is $\xbit{0}$ only if the bottom element of the $\pc$-stack is non-public, and for any additional layer of the stacks, the element in $\pcstack$ is non-public \Cref{def:wellformedbitstack}. This definition helps us maintain the invariant that phantom computation can only take place under non-public guards and that dummy messages can only be sent on channels with non-public mode label $\lmode$.
\begin{definition}[Well-formed bit-stack] Define bit-stack $\bitstack$ to be well-formed w.r.t. $\pc$-stack $\pcstack$, written $\pcstack \vdash \bitstack$, by the following rules:
\label{def:wellformedbitstack}
\begin{mathparpagebreakable}
\inferrule{
    \pc = \bot \implies \bit = \xbit{1}
}{
    [\pc] \vdash [\bit]
}
\and
\inferrule{
    \pc \neq \bot \\
    \pcstack \vdash \bitstack
}{
    \pc \Colon \pcstack \vdash \bit \Colon \bitstack
}
\end{mathparpagebreakable}
\end{definition}

Next, we define shorthand notation for states of the auxiliary semantics in \Cref{def:notation_aux}. The states augment the states in the standard semantics. Consumer states are augmented with typing environments and producer states are augmented with typing environments and $\pc$-stack.

\begin{definition}[Notation (Aux)]
\label{def:notation_aux}
\begin{align*}
    \Consumeraux \Coloneqq 
        &\;\consumer{\ctx{\Gamma}{\Pi}{\Lambda} \mid p,\mu,\intenv,\networkstrategy,\hst,\tau} \\
    \Produceraux \Coloneqq
        &\;\producer{\Ctxaux{\Gamma}{\Pi}{\Lambda}{\Delta}{\pcstack}{q} \mid p,\bitstack,c,m,\mu,\intenv,\networkstrategy,\hst,\tau} \\
    \Configurationaux \Coloneqq
        &\;\Consumeraux \mid \Produceraux
\end{align*}
\end{definition}

We define function $\textit{trace}(\Configurationaux)$ for projecting the trace of an auxiliary configuration in the straightforward way.
\begin{definition}[Trace of auxiliary configuration] Define the projection of the trace of auxiliary configuration $\Configurationaux$, written $\textit{trace}(\Configurationaux)$, as follows:
\label{def:traceofauxiliaryconfiguration}
\begin{align*}
    \textit{trace}(\Configurationaux) =
    \begin{cases}
        \tau
            &\textit{if } \Configurationaux = \consumer{\ctx{\Gamma}{\Pi}{\Lambda} \mid p,\mu,\intenv,\networkstrategy,\hst,\tau} \\
        \tau
            &\textit{if } \Configurationaux = \producer{\Ctxaux{\Gamma}{\Pi}{\Lambda}{\Delta}{\pcstack}{q} \mid p,\bitstack,c,m,\mu,\intenv,\networkstrategy,\hst,\tau}
    \end{cases}
\end{align*}
\end{definition}

\Cref{def:semantics_system_aux} presents the auxiliary semantics for systems. The rules are straight forward from the introduction of $\pcstack$ to the semantics of commands. Here we make use of the distinction between the security level of the value being sent and the security level of the mode indicator. The producer state does not need to protect whether it is running in normal or phantom mode against attackers at some level $\ladv$ if the indicator bit in the message is known at that level.

\begin{figure}[!htbp]
\centering
\begin{mathparpagebreakable}
\inferrule[CC-Aux] {
    \networkstrategy(\tau) = \ch(\timestamp,\bit,\val{v}{z}) \\
    (p)(\ch) \not\Downarrow
} {
    \consumer{\ctx{\Gamma}{\Pi}{\Lambda} \mid p,\mu,\intenv,\networkstrategy,\hst,\tau}
    \longrightarrowdbl
    \consumer{\ctx{\Gamma}{\Pi}{\Lambda} \mid p,\mu,\intenv,\networkstrategy,\hst,\tau \cdot \ch(\timestamp,\bit,\val{v}{z})}
}
\and
\inferrule[CP-Aux] {
    \networkstrategy(\tau) = \ch(\timestamp,\bit,\val{v}{z}) \\
    (p)(\ch) \Downarrow c,x \\
    \Lambda(\ch) = \ltype{\sigma}{\ltriple{\lmode}{\lval}{q}} \\
    \pcstack = [\lmode] \\
    \bitstack = [\bit] \\
    \Delta = [x \mapsto \ltype{\sigma}{\lval}] \\
    m = [x \mapsto \val{v}{z}]
} {
    \consumer{\ctx{\Gamma}{\Pi}{\Lambda} \mid p,\mu,\intenv,\networkstrategy,\hst,\tau}
    \longrightarrowdbl
    \producer{\Ctxaux{\Gamma}{\Pi}{\Lambda}{\Delta}{\pcstack}{q} \mid p,\bitstack,c,m,\mu,\intenv,\networkstrategy,\hst \Colon \textsf{ok}(\ch,\timestamp,z), \tau \cdot \ch(\timestamp,\bit,\val{v}{z}) }
}
\and
\inferrule[PP-Aux] { 
    \xangle{\Ctxaux{\Gamma}{\Pi}{\Lambda}{\Delta}{\pcstack}{q} \mid \bitstack,c,m,\mu,\intenv,\hst}
    \xrightarrowdbl{\alpha}
    \xangle{\Ctxaux{\Gamma}{\Pi}{\Lambda}{\Delta}{\pcstack'}{q'} \mid \bitstack',c',m',\mu',\intenv',\hst'}
} {
    \producer{\Ctxaux{\Gamma}{\Pi}{\Lambda}{\Delta}{\pcstack}{q} \mid p,\bitstack,c,m,\mu,\intenv,\networkstrategy,\hst,\tau}
    \longrightarrowdbl
    \producer{\Ctxaux{\Gamma}{\Pi}{\Lambda}{\Delta}{\pcstack'}{q'} \mid p,\bitstack',c',m',\mu',\intenv',\networkstrategy,\hst',\tau \cdot \alpha}
}
\and
\inferrule[PC-Aux] {
} {
    \producer{\Ctxaux{\Gamma}{\Pi}{\Lambda}{\Delta}{\pcstack}{q} \mid p,\bitstack,\texttt{stop},m,\mu,\intenv,\networkstrategy,\hst,\tau}
    \longrightarrowdbl
    \consumer{\ctx{\Gamma}{\Pi}{\Lambda} \mid p,\mu,\intenv,\networkstrategy,\hst \Colon \textsf{ret},\tau}
}
\end{mathparpagebreakable}
\caption{Auxiliary semantics of system}
\label{def:semantics_system_aux}
\end{figure}

\Cref{def:type_system_program_aux} and \Cref{def:type_system_system_aux} present the typing rules for respectively auxiliary programs and systems. As we use the auxiliary model for reasoning about the program at runtime, we introduce a typing rule for consumer states.

\begin{figure}[!htbp]
\centering
\begin{mathpar}
\inferrule { } {
    \Gamma,\Pi,\Lambda \vdash_\text{aux} \cdot
}
\and
\inferrule {
    \Lambda(\ch) = \ltype{\sigma}{\ltriple{\lmode}{\lval}{q}} \\
    \Gamma,\Pi,\Lambda \vdash_\text{aux} p \\
    \Gamma,[x \mapsto \ltype{\sigma}{\lval}],\Pi,\Lambda,[\lmode] \vdash^q c : [\lmode]
} {
    \Gamma,\Pi,\Lambda \vdash_\text{aux} \ch(x) \xbrace{c}; p
}
\end{mathpar}
\caption{Auxiliary typing of programs}
\label{def:type_system_program_aux}
\end{figure}

\begin{figure}[!htbp]
\centering
\begin{mathpar}
\inferrule {
    \Gamma,\Pi,\Lambda \vdash_\text{aux} p \\
    \Gamma \vdash \mu \\
    \Pi \vdash \intenv \\
    \Lambda \vdash \networkstrategy
} {
    \vdash_\text{aux} \consumer{\ctx{\Gamma}{\Pi}{\Lambda} \mid p,\mu,\intenv,\networkstrategy,\hst,\tau}
}
\and
\inferrule {
    \Gamma,\Pi,\Lambda \vdash_\text{aux} p \\
    \Lambda(\ch) = \ltype{\sigma}{\ltriple{\lmode}{\lval}{q+r}} \\
    \Gamma \vdash \mu \\
    \Delta \vdash m \\
    \Pi \vdash \intenv \\
    \Lambda \vdash \networkstrategy \\
    \lmode \vdash \pc \Colon \pcstack \\
    \pc \Colon \pcstack \vdash \bitstack \\
    \Ctx{\Gamma}{\Pi}{\Lambda}{\Delta}{\pc \Colon \pcstack} \vdash^{q}_\text{aux} c : [\lmode]
} {
     \vdash_\text{aux} \producer{\Ctxaux{\Gamma}{\Pi}{\Lambda}{\Delta}{\pc \Colon \pcstack}{q} \mid p,\bitstack,c,m,\mu,\intenv,\networkstrategy,\hst,\tau}
}
\end{mathpar}
\caption{Auxiliary typing of system}
\label{def:type_system_system_aux}
\end{figure}

\Cref{lemma:typingadequacy} shows adequacy of auxiliary typing of systems.

\begin{restatable}[Adequacy of typing]{lemma}{typingadequacy}
\label{lemma:typingadequacy}
Consider $\Gamma,\Pi,\Lambda$ and $\consumer{p,\mu,\intenv,\networkstrategy,\hst,\tau}$. We have
\[
    \ctx{\Gamma}{\Pi}{\Lambda} \vdash \consumer{p,\mu,\intenv,\networkstrategy,\hst,\tau}
\]
if and only if
\[
    \vdash_{\text{aux}} \consumer{\ctx{\Gamma}{\Pi}{\Lambda} \mid p,\mu,\intenv,\networkstrategy,\hst,\tau}
\]
\end{restatable}
\begin{IEEEproof}
We inspect each direction separately.
\begin{description}
    \item[Case standard to auxiliary:]
    By $\ctx{\Gamma}{\Pi}{\Lambda} \vdash \consumer{p,\mu,\intenv,\networkstrategy,\hst}$ we have (1) $\ctx{\Gamma}{\Pi}{\Lambda} \vdash p$, (2) $\Gamma \vdash \mu$, (3) $\Pi \vdash \intenv$, and (4) $\Lambda \vdash \networkstrategy$. By (2), (3), and (4) it suffices to show $\ctx{\Gamma}{\Pi}{\Lambda} \vdash_{\text{aux}} p$.
    
    We proceed by induction on $p$.
    \begin{description}
        \item[Case $p$ is $\cdot$:]
        Trivial.
        
        \item[Case $p$ is $\ch(x) \xbrace{c}; p'$:]
        We are done by \Cref{lemma:commandtypingadequacy} and the induction hypothesis.
    \end{description}
    
    \item[Case auxiliary to standard:]
    By $\vdash_{\text{aux}} \consumer{\ctx{\Gamma}{\Pi}{\Lambda} \mid p,\mu,\intenv,\networkstrategy,\hst}$ we have (1) $\ctx{\Gamma}{\Pi}{\Lambda} \vdash p$, (2) $\Gamma \vdash \mu$, (3) $\Pi \vdash \intenv$, and (4) $\Lambda \vdash \networkstrategy$. By (2), (3), and (4) it suffices to show $\ctx{\Gamma}{\Pi}{\Lambda} \vdash p$.
    
    We proceed by induction on $p$.
    \begin{description}
        \item[Case $p$ is $\cdot$:]
        Trivial.
        
        \item[Case $p$ is $\ch(x) \xbrace{c}; p'$:]
        We are done by \Cref{lemma:commandtypingadequacy} and the induction hypothesis.
    \end{description}
\end{description}
\end{IEEEproof}

\begin{restatable}[Preservation]{lemma}{preservation}
\label{lemma:preservation}
Given $\Configurationaux$ such that $\vdash_\text{aux} \Configurationaux$, if
\[
    \Configurationaux
    \xrightarrowdbl{\alpha}
    \Configurationaux'
\]
then $\vdash_\text{aux} \Configurationaux'$.
\end{restatable}
\begin{IEEEproof}
We proceed by casing on stepping relation $\xrightarrowdbl{\alpha}$.
\begin{description}
    \item[Case \textsc{CC-Aux}:]
    Trivial.
    
    \item[Case \textsc{CP-Aux}:]
    We have $\Configurationaux = \consumer{\ctx{\Gamma}{\Pi}{\Lambda} \mid p,\mu,\intenv,\networkstrategy,\hst,\tau}$. We have by assumption that $\ctx{\Gamma}{\Pi}{\Lambda} \vdash_\text{aux} \consumer{\ctx{\Gamma}{\Pi}{\Lambda} \mid p,\mu,\intenv,\networkstrategy,\hst,\tau}$ and hence (1) $\Gamma \vdash \mu$, (2) $\Pi \vdash \intenv$, (3) $\Lambda \vdash \networkstrategy$, and (4) $\ctx{\Gamma}{\Pi}{\Lambda} \vdash_\text{aux} p$.
    
    Let $\networkstrategy(\tau) = \ch(\timestamp,\bit,\val{v}{z})$. We proceed by induction on $p$.
    \begin{description}
        \item[Case $p$ is $\ch(x) \xbrace{c}; p'$:]
        By $\ctx{\Gamma}{\Pi}{\Lambda} \vdash_{\text{aux}} \ch(x) \xbrace{c}; p'$ we have (5) $\Lambda(\ch) = \ltype{\sigma}{\ltriple{\lval}{\lmode}{q}}$, (6) $\Delta=[x \mapsto \ltype{\sigma}{\lval}]$, and (7) $\Ctx{\Gamma}{\Pi}{\Lambda}{\Delta}{[\lmode]} \vdash^q c : [\lmode]$.
        We have $m = [x \mapsto \val{v}{z}]$, and therefore by (6) we have $\Delta \vdash m$.
        
        We have $\bitstack=[\bit]$ and $\pcstack=[\lmode]$.
        We must show $\vdash_{\text{aux}} \producer{\Ctxaux{\Gamma}{\Pi}{\Lambda}{\Delta}{\pcstack}{q} \mid c, m,\mu,\intenv,\networkstrategy,\hst \Colon \textsf{hl}(\ch,t,z), \tau \cdot \overleftarrow{\ch}(\timestamp,\bit,\val{v}{z})}$,
        which we have by (4), (5), (7) and the induction hypothesis.
        
        \item[Case $p$ is $\ch'(x) \xbrace{c}; p'$:]
        By the induction hypothesis.
    \end{description}
    
    \item[Case \textsc{PP-Aux}:]
    By \Cref{lemma:commandpreservation}.
    
    \item[Case \textsc{PC-Aux}:]
    By \Cref{lemma:commandpreservation}.
\end{description}
\end{IEEEproof}

Finally, we show adequacy of the auxiliary system semantics (\Cref{lemma:adequacy}).
\begin{restatable}[Adequacy]{lemma}{adequacy}
\label{lemma:adequacy}
Consider $\Gamma,\Pi,\Lambda$ and $\consumer{p,\mu,\intenv,\networkstrategy,\hst,\tau}$ such that $\ctx{\Gamma}{\Pi}{\Lambda} \vdash \consumer{p,\mu,\intenv,\networkstrategy,\hst,\tau}$. We have
\begin{enumerate}
    \item 
    \[
        \consumer{p,\mu,\intenv,\networkstrategy,\hst,\tau}
        \longrightarrow {\phantom{I}\mkern-14mu}^n \;
        \consumer{p,\mu',\intenv',\networkstrategy,\hst',\tau'}
    \]
    if and only if
    \[
        \consumer{\ctx{\Gamma}{\Pi}{\Lambda} \mid p,\mu,\intenv,\networkstrategy,\hst,\tau}
        \longrightarrowdbl {\phantom{I}\mkern-14mu}^n \;
        \consumer{\ctx{\Gamma}{\Pi}{\Lambda} \mid p,\mu',\intenv',\networkstrategy,\hst',\tau'}
    \]
    and $\vdash_{\text{aux}} \consumer{\ctx{\Gamma}{\Pi}{\Lambda} \mid p,m',\intenv',\networkstrategy,\hst',\tau'}$
    \item
    \[
        \consumer{p,\mu,\intenv,\networkstrategy,\hst,\tau}
        \longrightarrow {\phantom{I}\mkern-14mu}^n \;
        \producer{p,\bitstack,c,m,\mu',\intenv',\networkstrategy,\hst',\tau'}
    \]
    if and only if there exists $\Delta,\pcstack,q$ such that $\pcstack \vdash \bitstack$ and
    \[
        \consumer{\ctx{\Gamma}{\Pi}{\Lambda} \mid p,\mu,\intenv,\networkstrategy,\hst,\tau}
        \longrightarrowdbl {\phantom{I}\mkern-14mu}^n \;
        \producer{\Ctxaux{\Gamma}{\Pi}{\Lambda}{\Delta}{\pcstack}{q} \mid p,\bitstack,c,m,\mu',\intenv',\networkstrategy,\hst',\tau'}
    \]
    and $\vdash_{\text{aux}} \producer{\Ctxaux{\Gamma}{\Pi}{\Lambda}{\Delta}{\pcstack}{q} \mid p,\bitstack,c,m,\mu',\intenv',\networkstrategy,\hst',\tau'}$
\end{enumerate}
\end{restatable}
\begin{IEEEproof}
We investigate each direction separately.

\begin{description}
    \item[Case standard to auxiliary:]
    The proof is by induction in $n$.
    \begin{description}
        \item[Case $n=0$:]
        We investigate the two cases. The first one is trivial, the second one holds vacuously.
        
        \item[Case $n=k+1$:]
        By the induction hypothesis we have
        \begin{enumerate}
            \item 
            \[
                \consumer{p,\mu,\intenv,\networkstrategy,\hst,\tau}
                \longrightarrow {\phantom{I}\mkern-18mu}^k \;
                \consumer{p,\mu^\dprime,\intenv^\dprime,\networkstrategy,\hst^\dprime,\tau^\dprime}
            \]
            if and only if
            \[
                \consumer{\ctx{\Gamma}{\Pi}{\Lambda} \mid p,\mu,\intenv,\networkstrategy,\hst,\tau}
                \longrightarrowdbl {\phantom{I}\mkern-18mu}^k \;
                \consumer{\ctx{\Gamma}{\Pi}{\Lambda} \mid p,\mu^\dprime,\intenv^\dprime,\networkstrategy,\hst^\dprime,\tau^\dprime}
            \]
            and $\vdash_{\text{aux}} \consumer{\Gamma,\Pi,\Lambda \mid p,\mu^\dprime,\intenv^\dprime,\networkstrategy,\hst^\dprime,\tau^\dprime}$
            \item
            \[
                \consumer{p,\mu,\intenv,\networkstrategy,\hst,\tau}
                \longrightarrow {\phantom{I}\mkern-18mu}^k \;
                \producer{p,\bitstack',c',m',\mu^\dprime,\intenv^\dprime,\networkstrategy,\hst^\dprime,\tau^\dprime}
            \]
            if and only if there exists $\Delta',\pcstack',q'$ such that $\pcstack' \vdash \bitstack'$ and
            \[
                \consumer{\ctx{\Gamma}{\Pi}{\Lambda} \mid p,m,\intenv,\networkstrategy,\hst,\tau}
                \longrightarrowdbl {\phantom{I}\mkern-18mu}^k \;
                \producer{\Ctxaux{\Gamma}{\Pi}{\Lambda}{\Delta'}{\pcstack'}{q'} \mid p,\bitstack',c',m',\mu^\dprime,\intenv^\dprime,\networkstrategy,\hst^\dprime,\tau^\dprime}
            \]
            and $\vdash_{\text{aux}} \producer{\Ctxaux{\Gamma}{\Pi}{\Lambda}{\Delta'}{\pcstack'}{q'} \mid p,\bitstack',c',m',\mu^\dprime,\intenv^\dprime,\networkstrategy,\hst^\dprime,\tau^\dprime}$
        \end{enumerate}
        We proceed by casing on stepping relation $\longrightarrow$ for step $k+1$:
        \begin{description}
            \item[Case \textsc{CC}:]
                Trivial.
            
            \item[Case \textsc{CP}:]
                By \textsc{CP}, we have (1) $\networkstrategy(\tau) = \ch(\timestamp,\bit,\val{v}{z})$, (2) $(p)(\ch) \Downarrow c,x$, and (3) $m=[x \mapsto \val{v}{z}]$. By $\ctx{\Gamma}{\Pi}{\Lambda} \vdash \consumer{p,\mu,\intenv,\networkstrategy,\hst,\tau}$ we have (4) $\bitstack=[\bit]$ and (5) $\Lambda(\ch)=\ltype{\sigma}{\ltriple{\lval}{\lmode}{q}}$.
                
                By (1), (2), (3), (4), and (5) we therefore have that the run in the auxiliary semantics steps by \textsc{CP-Aux} to $\pcstack=[\lmode]$, $\bitstack=[\bit]$ and thus $\pcstack \vdash \bitstack$.
                We are done by \Cref{lemma:preservation} and IH.
            
            \item[Case \textsc{PP}:]
                By IH and \Cref{lemma:commandadequacy}.
            
            \item[Case \textsc{PC}:]
                Trivial.
        \end{description}
    \end{description}
    
    \item[Case auxiliary to standard:]
    The proof is by induction in $n$.
    \begin{description}
        \item[Case $n=0$:]
        We investigate the two cases. The first one is trivial, the second one holds vacuously.
        
        \item[Case $n=k+1$:]
        We proceed by casing on stepping relation $\longrightarrowdbl$ for the last step. All cases are trivial.
    \end{description}
\end{description}
\end{IEEEproof}
\clearpage
\section{Noninterference}
\label{appendix:noninterference}
This section shows noninterference for well-typed \oblivio{} programs.

We show noninterference for expressions (\Cref{lemma:noninterferenceexpressions}). The proof is straight forward by induction on expressions $e$.

\begin{restatable}[Noninterference expressions]{lemma}{noninterferenceexpressions}
\label{lemma:noninterferenceexpressions}
Given an attacker level $\ladv$, typing environments $\Gamma,\Delta$ and two pairs of memories and stores $m_1,\mu_1$ and $m_2,\mu_2$ such that $m_1 \deltaequiv{\ladv} m_2$ and $\mu_1 \gammaequiv{\ladv} \mu_2$, and an expression $e$ such that $\Gamma;\Delta \vdash e : \ltype{\sigma}{\ell}$ and $\eval{e,m_1,\mu_1}{\val{v_1}{z_1}}$ and $\eval{e,m_2,\mu_1}{\val{v_2}{z_2}}$, then we have that
\begin{itemize}
    \item $z_1 = z_2$
    \item $\ell \sqsubseteq \ladv \implies v_1 = v_2$
\end{itemize}
\end{restatable}
\begin{IEEEproof}
The proof is by induction in $e$.
\begin{description}
    \item[Case $e$ is $n$:]
    Immediate.
    
    \item[Case $e$ is $s$:]
    Immediate.

    \item[Case $e$ is $x$:]
    By definition of $m_1 \deltaequiv{\ladv} m_2$ and $\mu_1 \gammaequiv{\ladv} \mu_2$.

    \item[Case $e$ is $e_1 \oplus e_2$:]
    By two applications of the induction hypothesis.
\end{description}
\end{IEEEproof}

Before we can show noninterference of commands (\Cref{lemma:noninterferencecommands}), we first need to define equivalence of bit-stacks (\Cref{def:bit_stack_equivalence}).

\begin{definition}[Bit stack equivalence]
\label{def:bit_stack_equivalence} Bit-stacks $\bitstack_1$ and $\bitstack_2$ are related at level $\ladv$ w.r.t. $\pc$-stack $\pcstack$, written $\bitstack_1 \approx^{\pcstack}_{\ladv} \bitstack_2$, if they agree on the bits at all indices where the corresponding index in $\pcstack$ flows to to $\ladv$.
\[
\inferrule { } {
    [] \approx^{[]}_{\ladv} []
}
\hspace{3em}
\inferrule {
    \pc \sqsubseteq \ladv \implies \bit_1 = \bit_2 \\
    \bitstack_1 \approx^{\pcstack}_{\ladv} \bitstack_2
} {
    \bit_1 :: \bitstack_1 \approx^{\pc \Colon \pcstack}_{\ladv} \bit_2 \Colon \bitstack_2
}
\]
\end{definition}

Using bit-stack equivalence, we are ready to show noninterference for commands (\Cref{lemma:noninterferencecommands}). The oblivious model makes the noninterference statement simple, as two related runs always step to the same command.

\begin{restatable}[Noninterference commands]{lemma}{noninterferencecommands}
\label{lemma:noninterferencecommands}
Given $\Gamma,\Pi,\Lambda,\Delta,\pcstack,q$ and a command $c$, such that $\Ctx{\Gamma}{\Pi}{\Lambda}{\Delta}{\pcstack} \vdash^q c : \pcstack^\dprime$ and $\bitstack_1 \approx^{\pcstack}_{\ladv} \bitstack_2$, $m_1 \deltaequiv{\ladv} m_2$ $\mu_1 \gammaequiv{\ladv} \mu_2$, and $\intenv_1 \piequiv{\ladv} \intenv_2$. If
\[
    \xangle{\Ctxaux{\Gamma}{\Pi}{\Lambda}{\Delta}{\pcstack}{q} \mid \bitstack_1,c,m_1,\intenv_1,\hst}
    \xrightarrow{\alpha_1}
    \xangle{\Ctxaux{\Gamma}{\Pi}{\Lambda}{\Delta}{\pcstack'}{q'} \mid \bitstack'_1,c',m'_1,\intenv'_1,\hst'}
\]
then
\[
    \xangle{\Ctxaux{\Gamma}{\Pi}{\Lambda}{\Delta}{\pcstack}{q} \mid \bitstack_2,c,m_2,\intenv_2,\hst}
    \xrightarrow{\alpha_2}
    \xangle{\Ctxaux{\Gamma}{\Pi}{\Lambda}{\Delta}{\pcstack'}{q'} \mid \bitstack'_2,c',m'_2,\intenv'_2,\hst'}
\]
and 
\begin{enumerate}
    \item $\alpha_1 \lambdaequiv{\ladv} \alpha_2$
    \item $\bitstack'_1 \approx^{\pcstack'}_{\ladv} \bitstack'_2$
    \item $m'_1 \deltaequiv{\ladv} m'_2$
    \item $\mu'_1 \gammaequiv{\ladv} \mu'_2$
    \item $\intenv'_1 \piequiv{\ladv} \intenv'_2$
\end{enumerate}
\end{restatable}
\begin{IEEEproof} The proof is by induction on $c$.
\begin{description}
    \item[Case $c$ is $\texttt{skip}$:]
    By \textsc{Skip-Aux} we have $\alpha_1 = \epsilon = \alpha_2$, hence $\alpha_1 \lambdaequiv{\ladv} \alpha_2$.
    
    \item[Case $c$ is $c_1;c_2$:]
    By the induction hypothesis.

    \item[Case $c$ is $x \assign e$:]
    By \textsc{T-Assign-Aux} we have $\pcstack = \pcstack^\dprime$ and their top element is $\bot$, hence by $\bitstack_1 \approx^{\pcstack}_{\ladv} \bitstack_2$ we have $\bitstack_1 = \bit \Colon \_$ and $\bitstack_2 = \bit \Colon \_$.
    We have $\Gamma(x) = \ltype{\sigma}{\ell_x}$ and $\Gamma;\Delta \vdash e : \ltype{\sigma}{\ell_e}$ such that $\ell_e \sqsubseteq \ell_x$.
    
    We case on $\bit$:
    \begin{description}
        \item[Case $\bit = 0$:]
        Trivial.
        
        \item[Case $\bit = 1$:]
        We case on $\ell_e \sqsubseteq \ladv$, in both cases we are done by \Cref{lemma:noninterferenceexpressions} and \Cref{def:store_equivalence}.
    \end{description}

    \item[Case $c$ is $x \oblivassign e$:]
    By \textsc{T-LocalAssign-Aux} we have (1) $\pcstack = \pc \Colon \pcstack^\dprime$, (2) $\Gamma(x) = \ltype{\sigma}{\ell_x}$, (3) $\Gamma;\Delta \vdash e : \ltype{\sigma}{\ell_e}$, and (4) $\ell_e \sqcup \pc \sqsubseteq \ell_x$.
    
    We case on $\ell_x \sqsubseteq \ladv$.
    
    \begin{description}
        \item[Case $\ell_x \sqsubseteq \ladv$:]
        If $\ell_x \sqsubseteq \ladv$ then by (2) and \Cref{def:store_equivalence}, we have $\eval{x,m_1,\mu_1}{\val{v_0}{z_0}}$ and $\eval{x,m_2,\mu_2}{\val{v_0}{z_0}}$ for some $\val{v_0}{z_0}$. By (4) transitivity of $\sqsubseteq$ we have $\ell_e \sqsubseteq \ladv$ and hence by (3) and \Cref{lemma:noninterferenceexpressions} we have $\eval{e,m_1,\mu_1}{\val{v_1}{z_1}}$ and $\eval{e,m_2,\mu_2}{\val{v_1}{z_1}}$ for some $\val{v_1}{z_1}$. By (1), (4) and transitivity we also have $\pc \sqsubseteq \ladv$, and hence by \Cref{def:bit_stack_equivalence} have $\bitstack_1 = \bit \Colon \bitstack^\dprime_1$ and $\bitstack_2 = \bit \Colon \bitstack^\dprime_2$. In particular, both runs agree on $z=\textit{max}(z_0,z_1)$ and $i=\bit$ hence we have $\mu'_1 = \mu_1[x \mapsto \val{v_i}{z}]$ and $\mu'_2 = \mu_2[x \mapsto \val{v_i}{z}]$, and by \Cref{def:store_equivalence} $\mu'_1 \gammaequiv{\ladv} \mu'_2$.
    
        \item[Case $\ell_x \not\sqsubseteq \ladv$:]
        If $\ell_x \not\sqsubseteq \ladv$, we have by (2) and \Cref{def:store_equivalence} for the two runs $x$ as $\mu_1(x) = \val{v_0}{z_0}$ and $\mu_2(x) = \val{v'_0}{z_0}$. By (3) and \Cref{lemma:noninterferenceexpressions} we have $\eval{e,m_1,\mu_1}{\val{v_1}{z_1}}$ and $\eval{e,m_2,\mu_2}{\val{v'_1}{z_1}}$. In particular, the runs agree on sizes $z_0,z_1$, hence they agree on $z = \textit{max}(z_0,z_1)$. From this, and as $\ell_x \not\sqsubseteq \ladv$, we have $\mu'_1 \gammaequiv{\ladv} \mu'_2$ by \Cref{def:store_equivalence}.
    \end{description}
        
    \item[Case $c$ is $x \oblivassign \texttt{input}(\ch,e)$:]
    By \textsc{T-LocalInput-Aux} we have (1) $\Gamma(x) = \ltype{\sigma}{\ell_x}$, (2) $\Pi(\ch) = \ltype{\sigma}{\ell_\ch}$, (3) $\Gamma;\Delta \vdash e : \ltype{\inttype}{\ell_e}$, (4) $\ell_e \sqcup \pc \sqsubseteq \ell_\ch$, and (5) $\ell_\ch \sqsubseteq \ell_x$.
    
    We case on $\ell_x \sqsubseteq \ladv$.
    
    \begin{description}
        \item[Case $\ell_x \sqsubseteq \ladv$:]
        By \Cref{lemma:noninterferenceexpressions} we have
        $\mu_1(x) = \val{v_x}{z_x}$ and
        $\mu_2(x) = \val{v_x}{z_x}$
        and by transitivity of $\sqsubseteq$ and \Cref{lemma:noninterferenceexpressions} have
        $\eval{e,m_1,\mu_1}{\val{n_e}{z_e}}$ and
        $\eval{e,m_2,\mu_2}{\val{n_e}{z_e}}$.
        By the step if the first run, we have $\bitstack_1 = \bit \Colon \_$, and as $\pc \sqsubseteq \ladv$ we have by \Cref{def:bit_stack_equivalence} that $\bitstack_2 = \bit \Colon \_$. We have in both runs $\hst' = \hst \Colon \textsf{in}(x,\ch,e,\textit{max}(z_x,n_e))$
        
        By transitivity of $\sqsubseteq$ we have $\ell_\ch \sqsubseteq \ladv$, we therefore by $\intenv_1 \piequiv{\ladv} \intenv_2$ we have $v^\dprime_1 = v^\dprime_2$ and $\intenv^\dprime_1 \piequiv{\ladv} \intenv^\dprime_2$.
        
        By $\mu_1 \gammaequiv{\ladv} \mu_2$ and \Cref{def:store_equivalence} we have $\mu_1[x \mapsto \val{v^\dprime_1}{z}] \gammaequiv{\ladv} \mu_2[x \mapsto \val{v^\dprime_2}{z}]$.
    
        \item[Case $\ell_e \not\sqsubseteq \ladv$:]
        By $\mu_1 \gammaequiv{\ladv} \mu_2$ we have 
        $\mu_1(x) = \val{v_x}{z_x}$ and
        $\mu_2(x) = \val{v'_x}{z_x}$.
        
        By \Cref{lemma:noninterferenceexpressions} we have
        $\eval{e,m_1,\mu_1}{\val{n_e}{z_e}}$ and
        $\eval{e,m_2,\mu_2}{\val{n_e}{z_e}}$.
        
        We have in both runs $\hst' = \hst \Colon \textsf{in}(x,\ch,e,\textit{max}(z_x,n_e))$.
    
        By the step if the first run we have $\bitstack_1 = \bit_1 \Colon \_$, thus by \Cref{def:bit_stack_equivalence} we have that $\bitstack_2 = \bit_2 \Colon \_$.
        
        By $\intenv_1 \piequiv{\ladv} \intenv_2$ we some $v^\dprime_1,v^\dprime_2$ and $\intenv^\dprime_1, \intenv^\dprime_2$ such that $\intenv^\dprime_1 \piequiv{\ladv} \intenv^\dprime_2$.
        
        By $\mu_1 \gammaequiv{\ladv} \mu_2$ and \Cref{def:store_equivalence} we have $\mu_1[x \mapsto \val{v^\dprime_1}{z}] \gammaequiv{\ladv} \mu_2[x \mapsto \val{v^\dprime_2}{z}]$.
    \end{description}
        
    \item[Case $c$ is $\texttt{send}(\ch,e)$:]
    By \textsc{T-Send-Aux} we have (1) $\Lambda(\ch) = \ltype{\sigma}{\ltriple{\lval}{\lmode}{q}}$, (2) $\Gamma;\Delta \vdash e : \ltype{\sigma}{\ell_e}$, (3) $\ell_e \sqsubseteq \lval$, and (4) $\pc \sqsubseteq \lmode$.
    
    We case on $\lval \sqsubseteq \ladv$.
    
    \begin{description}
        \item[Case $\lval \sqsubseteq \ladv$:]
        If $\lval \sqsubseteq \ladv$ then by transitivity of $\sqsubseteq$, (2) and \Cref{lemma:noninterferenceexpressions} we have $\eval{e,m_1,\mu_1}{\val{v}{z}}$ and $\eval{e,m_2,\mu_2}{\val{v}{z}}$ for some $\val{v}{z}$.
        As the commands are the same in both runs and as both runs agree on $z$, we have in both runs $\hst' = \hst \Colon \textsf{out}(\ch,e,z)$, hence both runs agree on $\timestamp = \textit{time}(\hst')$.
        
        We now case on $\lmode \sqsubseteq \ladv$
        \begin{description}
            \item[Case $\lmode \sqsubseteq \ladv$:]
            By transitivity of $\sqsubseteq$ we have $\pc \sqsubseteq \ladv$ and hence by \Cref{def:bit_stack_equivalence} we have $\bitstack_1 = \bit \Colon \bitstack'_1$ and $\bitstack_2 = \bit \Colon \bitstack'_2$. We are done as $\alpha_1 = \alpha_2$.
            
            \item[Case $\lmode \not\sqsubseteq \ladv$:]
            We have $\alpha_1 = \ch(\timestamp,\bit_1,\val{v}{z})$ and $\alpha_2 = \ch(\timestamp,\bit_2,\val{v}{z})$, and hence, as $\lmode \not\sqsubseteq \ladv$, have $\alpha_1 \lambdaequiv{\ladv} \alpha_2$ and we are done.
        \end{description}
    
        \item[Case $\lval \not\sqsubseteq \ladv$:]
        If $\lval \not\sqsubseteq \ladv$ then by (2) and \Cref{lemma:noninterferenceexpressions} we have $\eval{e,m_1,\mu_1}{\val{v_1}{z}}$ and $\eval{e,m_2,\mu_2}{\val{v_2}{z}}$.
        As the commands are the same in both runs and as both runs agree on $z$, we have in both runs $\hst' = \hst \Colon \textsf{out}(\ch,e,z)$, hence both runs agree on $\timestamp = \textit{time}(\hst')$.
        
        We now case on $\lmode \sqsubseteq \ladv$
        \begin{description}
            \item[Case $\lmode \sqsubseteq \ladv$:]
            By transitivity of $\sqsubseteq$ we have $\pc \sqsubseteq \ladv$ and hence by \Cref{def:bit_stack_equivalence} we have $\bitstack_1 = \bit \Colon \bitstack'_1$ and $\bitstack_2 = \bit \Colon \bitstack'_2$. We have $\alpha_1 = \overrightarrow{\ch}(\timestamp,\bit,\val{v_1}{z})$ and $\alpha_2 = \overrightarrow{\ch}(\timestamp,\bit,\val{v_2}{z})$ and hence by \Cref{def:event_equivalence} have $\alpha_1 \lambdaequiv{\ladv} \alpha_2$.
            
            \item[Case $\lmode \not\sqsubseteq \ladv$:]
            We have $\alpha_1 = \overrightarrow{\ch}(\timestamp,\bit_1,\val{v}{z})$ and $\alpha_2 = \overrightarrow{\ch}(\timestamp,\bit_2,\val{v}{z})$, and hence, as $\lmode \not\sqsubseteq \ladv$, have $\alpha_1 \lambdaequiv{\ladv} \alpha_2$ and we are done.
        \end{description}
    \end{description}
        
    \item[Case $c$ is $\texttt{if $e$ then $c_1$ else $c_2$}$:]
    By \textsc{T-If-Aux} we have $\Gamma;\Delta \vdash e : \ltype{\inttype}{\bot}$ hence by \Cref{lemma:noninterferenceexpressions} we have $\eval{e,m_1,\mu_1}{\val{v}{z}}$ and $\eval{e,m_2,\mu_2}{\val{v}{z}}$. Hence, both runs step to the same command $c'$ emitting events $\alpha_1 = \epsilon = \alpha_2$.
        
    \item[Case $c$ is $\texttt{while $e$ do $c$}$:]
    By \textsc{T-While-Aux} we have $\Gamma;\Delta \vdash e : \ltype{\inttype}{\bot}$ hence by \Cref{lemma:noninterferenceexpressions} we have $\eval{e,m_1,\mu_1}{\val{v}{z}}$ and $\eval{e,m_2,\mu_2}{\val{v}{z}}$. Hence, both runs step to the same command $c'$ emitting events $\alpha_1 = \epsilon = \alpha_2$.
        
    \item[Case $c$ is $\texttt{oblif $e$ then $c_1$ else $c_2$}$:]
    By \textsc{T-Oblif-Aux} we have (1) $\Gamma;\Delta \vdash e : \ltype{\inttype}{\ell}$, $q = q_1+q_2$ such that (2) $\Ctx{\Gamma}{\Pi}{\Lambda}{\Delta}{\pc \sqcup \ell \Colon \pcstack} \vdash^{q_1}_{\text{aux}} c_1 : \pc \sqcup \ell  \Colon \pcstack$, (3) $\Ctx{\Gamma}{\Pi}{\Lambda}{\Delta}{\pc \sqcup \ell \Colon \pcstack} \vdash^{q_2}_{\text{aux}} c_2 : \pc \sqcup \ell \Colon \pcstack$, and (4) $\pcstack = \pc \Colon \pcstack^\dprime$. 
    
    We case on $\ell \sqcup \pc \sqsubseteq \ladv$.
    
    \begin{description}
        \item[Case $\ell \sqcup \pc \sqsubseteq \ladv$:]
        By transitivity we have (5) $\ell \sqsubseteq \ladv$ and (6) $\pc \sqsubseteq \ladv$. By (5) and \Cref{lemma:noninterferenceexpressions} we have $\eval{e,m_1,\mu_1}{\val{v}{z}}$ and $\eval{e,m_2,\mu_2}{\val{v}{z}}$. By the step in the first run we have $\bitstack_1 = \bit \Colon \_$. Hence, by (6) and $\bitstack_1 \approx^{\pcstack}_{\ladv} \bitstack_2$ we have $\bitstack_2 = \bit \Colon \_$. As both runs agree on values $v$ and $\bit$, they agree on $\bit_1,\bit_2$. Hence, we are done by letting $\bitstack'_1 = \bit_1 \Colon \bit_2 \Colon \bitstack_1$, $\bitstack'_2 = \bit_1 \Colon \bit_2 \Colon \bitstack_2$, and $\pcstack' = \ell \sqcup \pc \Colon \ell \sqcup \pc \Colon \pcstack$ as we have $\bitstack'_1 \approx^{\pcstack'}_{\ladv} \bitstack'_2$.
    
        \item[Case $\ell \sqcup \pc \not\sqsubseteq \ladv$:]
        We have $\eval{e,m_1,\mu_1}{\val{v_1}{z_1}}$ and $\eval{e,m_2,\mu_2}{\val{v_2}{z_2}}$. By the step in the first run we have $\bitstack_1 = \bit \Colon \_$, hence by $\bitstack_1 \approx^{\pcstack}_{\ladv} \bitstack_2$ we have $\bitstack_2 = \bit' \Colon \_$. In particular, $\bitstack_2$ non-empty and hence the second run can also step. As $\ell \sqcup \pc \not\sqsubseteq \ladv$, we are done by letting $\pcstack' = \ell \sqcup \pc \Colon \ell \sqcup \pc \Colon \pcstack$, $\bitstack'_1 = \bit_1 \Colon \bit_2 \Colon \bitstack_1$, and $\bitstack'_2 = \bit'_1 \Colon \bit'_2 \Colon \bitstack_2$ for some $\bit_1,\bit_2,\bit'_1,\bit'_2$ as we have $\bitstack'_1 \approx^{\pcstack'}_{\ladv} \bitstack'_2$.
    \end{description}
        
    \item[Case $c$ is $\texttt{pop}$:]
    By \Cref{def:bit_stack_equivalence} we have $\bit_1 :: \bitstack'_1 \approx^{\pc \Colon \pcstack}_{\ladv} \bit_2 \Colon \bitstack'_2$ s.t. $\bitstack'_1 \approx^{\pcstack}_{\ladv} \bitstack'_2$, hence we are done.
\end{description}
\end{IEEEproof}

We define two systems to be equivalence (\Cref{def:equivalence_systems_aux}) if they are in the same state and are equivalent at each component.

\begin{definition}[Equivalence of auxiliary systems] Define $\Configurationaux_1 \approx_{\ladv} \Configurationaux_2$ with the following rules:
\label{def:equivalence_systems_aux}
\[
\inferrule {
    \mu_1 \gammaequiv{\ladv} \mu_2 \\
    \intenv_1 \piequiv{\ladv} \intenv_2 \\
    \networkstrategy_1 \lambdaequiv{\ladv} \networkstrategy_2 \\
    \tau_1 \lambdaequiv{\ladv} \tau_2
} {
    \consumer{\ctx{\Gamma}{\Pi}{\Lambda} \mid p,\mu_1,\intenv_1,\networkstrategy_1,\hst,\tau_1}
    \approx_{\ladv}
    \consumer{\ctx{\Gamma}{\Pi}{\Lambda} \mid p,\mu_2,\intenv_2,\networkstrategy_2,\hst,\tau_2}
}
\]
\[
\inferrule {
    \bitstack_1 \approx^{\pcstack}_{\ladv} \bitstack_2 \\
    m_1 \deltaequiv{\ladv} m_2 \\
    \mu_1 \gammaequiv{\ladv} \mu_2 \\
    \intenv_1 \piequiv{\ladv} \intenv_2 \\
    \networkstrategy_1 \lambdaequiv{\ladv} \networkstrategy_2 \\
    \tau_1 \lambdaequiv{\ladv} \tau_2
} {
    \producer{\Ctxaux{\Gamma}{\Pi}{\Lambda}{\Delta}{\pcstack}{q} \mid p,\bitstack_1,c,m_1,\mu_1,\intenv_1,\networkstrategy_1,\hst,\tau_1}
    \approx_{\ladv}
    \producer{\Ctxaux{\Gamma}{\Pi}{\Lambda}{\Delta}{\pcstack}{q} \mid p,\bitstack_2,c,m_2,\mu_2,\intenv_2,\networkstrategy_2,\hst,\tau_2}
}
\]
\end{definition}

We can now define single-step noninterference of systems (\Cref{lemma:noninterference}).

\begin{restatable}[Noninterference system]{lemma}{noninterference}
\label{lemma:noninterference}
Consider $\Configurationaux_1$ and $\Configurationaux_2$ such that $\vdash_\text{aux} \Configurationaux_1$ and $\Configurationaux_1 \approx_{\ladv} \Configurationaux_2$. If
\[
    \Configurationaux_1
    \longrightarrowdbl
    \Configurationaux'_1
\]
then
\[
    \Configurationaux_2
    \longrightarrowdbl
    \Configurationaux'_2
\]
and $\Configurationaux'_1 \approx_{\ladv} \Configurationaux'_2$.
\end{restatable}
\begin{IEEEproof}
We case on step $\longrightarrowdbl$ in the first run:
\begin{description}
    \item[Case \textsc{CC-Aux}:]
    By $\consumer{\ctx{\Gamma}{\Pi}{\Lambda} \mid p,m_1,\intenv_1,\networkstrategy_1,\hst,\tau_1} \approx_{\ladv} \consumer{\ctx{\Gamma}{\Pi}{\Lambda} \mid p,\mu_2,\intenv_2,\networkstrategy_2,\hst,\tau_2}$ we have (1a) $\tau_1 \lambdaequiv{\ladv} \tau_2$ and (1b) $\networkstrategy_1 \lambdaequiv{\ladv} \networkstrategy_2$. By the step in the first run, we have that $\networkstrategy_1(\tau_1) = \ch(\timestamp,\bit_1,\val{v_1}{z})$ and $\hst' = \hst$. By (1a) and (1b) we get that $\networkstrategy_2(\tau_2) = \ch(\timestamp,\bit_2,\val{v_2}{z})$.
    From the first run we have $(p)(\ch)\not\Downarrow$, hence the second run makes a step by \textsc{CC-Aux} to $\consumer{\ctx{\Gamma}{\Pi}{\Lambda} \mid p,\mu_2,\intenv_2,\networkstrategy_2,\hst,\tau_2 \cdot \ch(\timestamp,\bit_2,\val{v_2}{z})}$ s.t. $\consumer{\ctx{\Gamma}{\Pi}{\Lambda} \mid p,\mu_1,\intenv_1,\networkstrategy_1,\hst,\tau_1 \cdot \widetilde{\ch}(\timestamp,\bit_1,\val{v_1}{z})}
    \approx_{\ladv}
    \consumer{\ctx{\Gamma}{\Pi}{\Lambda} \mid p,\mu_2,\intenv_2,\networkstrategy_2,\hst,\tau_2 \cdot \widetilde{\ch}(\timestamp,\bit_2,\val{v_2}{z})}$.
    
    \item[Case \textsc{CP-Aux}:]
    By $\consumer{\ctx{\Gamma}{\Pi}{\Lambda} \mid p,\mu_1,\intenv_1,\networkstrategy_1,\hst,\tau_1} \approx_{\ladv} \consumer{\ctx{\Gamma}{\Pi}{\Lambda} \mid p,\mu_2,\intenv_2,\networkstrategy_2,\hst,\tau_2}$ we have (1a) $\tau_1 \lambdaequiv{\ladv} \tau_2$ and (1b) $\networkstrategy_1 \lambdaequiv{\ladv} \networkstrategy_2$.
    
    By the step in the first run, we have $\networkstrategy_1(\tau_1) = \ch(\timestamp,\bit_1,\val{v_1}{z})$ and $(p)(\ch) \Downarrow c,x$ s.t. 
    $\Delta = [x \mapsto \ltype{\sigma}{\ell}]$, $m_1 = [x \mapsto \ch(\val{v_1}{z}]$.
    $\Lambda(\ch) = \ltype{\sigma}{\ltriple{\lval}{\lmode}{q}}$. We have
    \[
        \Configurationaux'_1 = \producer{\Ctxaux{\Gamma}{\Pi}{\Lambda}{[x \mapsto \ltype{\sigma}{\lval}]}{[\lmode]}{q} \mid p,[\bit],c,m_1, \mu_1,\intenv_1,\networkstrategy_1,\hst \Colon \textsf{hl}(\ch,\timestamp,z), \tau_1 \cdot \overleftarrow{\ch}(\timestamp,\bit_1,\val{v_1}{z})}
    \]
    
    We case on $\lval \sqsubseteq \ladv$.
    \begin{description}
        \item[Case $\lval \sqsubseteq \ladv$:]
        By (1a) and (1b) we have that $\networkstrategy_2(\tau_2) = \ch(\timestamp,\bit_2,\val{v_1}{z})$
        and therefore have that the second run steps by $\textsc{CP-Aux}$ to 
        \[
            \Configurationaux'_2 = \producer{\Ctxaux{\Gamma}{\Pi}{\Lambda}{[x \mapsto \ltype{\sigma}{\lval}]}{[\lmode]}{q} \mid p,[\bit_2],c,m_1,\mu_2,\intenv_2,\networkstrategy_2,\hst \Colon \textsf{hl}(\ch,\timestamp,z),\tau_2 \cdot \overleftarrow{\ch}(\timestamp,\bit_2,\val{v_1}{z})}.
        \]
        
        We case on $\lmode \sqsubseteq \ladv$. In both cases we have by (1) and \Cref{def:trace_equivalence,def:networkstrategyequivalence} that
        $
            \Configurationaux'_1
            \approx_{\ladv}
            \Configurationaux'_2
        $
        and we are done.
        
        \item[Case $\lval \not\sqsubseteq \ladv$:]
        By (1a) and (1b) we get that $\networkstrategy_2(\tau_2) = \ch(\timestamp,\bit_2,\val{v_2}{z})$
        and therefore have that the second run steps by $\textsc{CP-Aux}$ to
        \[
            \Configurationaux'_2 = \producer{\Ctxaux{\Gamma}{\Pi}{\Lambda}{[x \mapsto \ltype{\sigma}{\lval}]}{[\lmode]}{q} \mid p,[\bit_2],c,m_2,\mu_2,\intenv_2,\networkstrategy_2,\hst \Colon \textsf{hl}(\ch,\timestamp,z),\tau_2 \cdot \overleftarrow{\ch}(\timestamp,\bit_2,\val{v_1}{z})}.
        \]
        
        We case on $\lmode \sqsubseteq \ladv$. In both cases we have by (1) and \Cref{def:trace_equivalence,def:networkstrategyequivalence} that
        $
            \Configurationaux'_1
            \approx_{\ladv}
            \Configurationaux'_2
        $
        and we are done.
    \end{description}

    \item[Case \textsc{PP-Aux}:]
    By $\Configurationaux_1 \approx_{\ladv} \Configurationaux_2$ and \Cref{lemma:noninterferencecommands}.
    
    \item[Case \textsc{PC-Aux}:]
    Trivial.
\end{description}
\end{IEEEproof}

Finally, we are ready to prove our soundness of the standard type system (\Cref{theorem:soundness}) presented in section \Cref{sec:enforcement}. By the adequacy of system typing and semantics (\Cref{lemma:typingadequacy,lemma:adequacy}), we conduct the proof using the auxiliary model, and can conclude using \Cref{lemma:preservation} (Preservation) and \Cref{lemma:noninterference} (Noninterference system).

\soundness*
\begin{IEEEproof}
Unfolding the definition of the consumer state we have $\Consumer = \consumer{p,\mu,\intenv,\networkstrategy,\hst,\tau_0}$. Let $\Consumeraux = \consumer{\ctx{\Gamma}{\Pi}{\Lambda} \mid p,\mu,\intenv,\networkstrategy,\hst,\tau_0}$. By \Cref{lemma:typingadequacy} we have $\vdash_{\text{aux}} \Consumeraux$ and thus by \Cref{lemma:adequacy} we have
\[
    \Consumeraux
    \longrightarrowdbl {\phantom{I}\mkern-18mu}^*\;
    \Configurationaux
\]
such that $\textit{trace}(\Configurationaux) = \tau \cdot \alpha$.
Let $\Consumer' \in k_{\Gamma,\Pi,\Lambda}(\Consumer,\tau,\ladv)$. By \Cref{def:equivalence_systems_aux} we have
\[
    \Consumeraux \approx_{\ladv} \Consumeraux'
\]
We show that $\vdash_{\text{aux}} \Configurationaux$ and
\[
    \Consumeraux'
    \longrightarrowdbl {\phantom{I}\mkern-18mu}^*\;
    \Configurationaux'
\]
such that $\Configurationaux \approx_{\ladv} \Configurationaux'$ which implies \Cref{def:security_condition}.

If $\textit{trace}(\Consumeraux) = \textit{trace}(\Configurationaux)$ we are done by assumption. Otherwise, we $\tau = \tau_0 \cdot \tau_1$ such that $\textit{trace}(\Configurationaux) = \tau_0 \cdot \tau_1 \cdot \alpha$.

\noindent The proof is by induction in the structure of $\tau_1$.
\begin{description}
    \item[Case $\tau_1$ is $\epsilon$:]
        We are done by \Cref{lemma:preservation} and \Cref{lemma:noninterference}.
    \item[Case $\tau_1$ is $\tau'_1 \cdot \alpha'$:]
        We are done by \Cref{lemma:preservation} and \Cref{lemma:noninterference} and the induction hypothesis.
\end{description}
\end{IEEEproof}
\clearpage
\section{Overhead}
\label{appendix:overhead}
This section bounds the number of dummy messages of \oblivio{} programs (\Cref{theorem:enforcementoverhead}).
First, we define predicate $\text{Phantom}$ over bit-stacks expressing that ever element of the stack is $\xbit{0}$ (\Cref{def:phantombitstack}).
\begin{definition}[Phantom bit-stack] Define predicate $\text{Phantom}$ over bit-stacks by the following rules:
\label{def:phantombitstack}
\begin{mathparpagebreakable}
\inferrule{
}{
    \text{Phantom}([\xbit{0}])
}
\and
\inferrule{
    \text{Phantom}(\bitstack)
}{
    \text{Phantom}(\xbit{0} \Colon \bitstack)
}
\end{mathparpagebreakable}
\end{definition}

Next, we define extension relations on stores (\Cref{def:extensionstore}), memories (\Cref{def:extensionmemory}), and auxiliary configurations (\Cref{def:traceofauxiliaryconfiguration}). These definitions are straightforward by lifting extension of values (\Cref{def:valueextension}), phantom extension of traces (\Cref{def:tracephantomextension}), and phantom extension of network strategies with respect to $\Lambda$ (\Cref{def:networkstrategylambdaphantomextension}).
\begin{definition}[Extension of stores]
\label{def:extensionstore}
Define store $\mu'$ to extend store $\mu$, written $\mu \extends \mu'$, if $\textit{dom}(\mu) = \textit{dom}(\mu)'$ and for all $x \in \textit{dom}(\mu)$ we have
\[
    \mu(x) = \val{v}{z} \implies \exists z' \geq z: \mu'(x) = \val{v}{z'}
\]
\end{definition}

\begin{definition}[Extension of memories]
\label{def:extensionmemory}
Define memory $m'$ to extend memory $m$, written $u \extends u'$, if $\textit{dom}(m) = \textit{dom}(m)'$ and for all $x \in \textit{dom}(m)$ we have
\[
    m(x) = \val{v}{z} \implies \exists z' \geq z: m'(x) = \val{v}{z'}
\]
\end{definition}

\begin{definition}[Extension of auxiliary configurations] Define configuration $\Configurationaux_2$ to extend configuration $\Configurationaux_1$, written $\Configurationaux_1 \lambdaphantomextends \Configurationaux_2$, by the following rules:
\begin{mathparpagebreakable}
\inferrule{
    \mu_1 \extends \mu_2 \\
    \networkstrategy_1 \lambdaphantomextends \networkstrategy_2 \\
    \tau_1 \phantomextends \tau_2
}{
    \consumer{\ctx{\Gamma}{\Pi}{\Lambda} \mid p,\mu_1,\intenv,\networkstrategy_1,\hst_1,\tau_1}
    \lambdaphantomextends
    \consumer{\ctx{\Gamma}{\Pi}{\Lambda} \mid p,\mu_2,\intenv,\networkstrategy_2,\hst_2,\tau_2}
}
\and
\inferrule{
    m_1 \extends m_2 \\
    \mu_1 \extends \mu_2 \\
    \networkstrategy_1 \lambdaphantomextends \networkstrategy_2 \\
    \tau_1 \phantomextends \tau_2
}{
    \producer{\Ctxaux{\Gamma}{\Pi}{\Lambda}{\Delta}{\pcstack}{q} \mid p,\bitstack,c,m_1,\mu_1,\intenv,\networkstrategy_1,\hst_1,\tau_1}
    \lambdaphantomextends
    \producer{\Ctxaux{\Gamma}{\Pi}{\Lambda}{\Delta}{\pcstack}{q} \mid p,\bitstack,c,m_2,\mu_2,\intenv,\networkstrategy_2,\hst_2,\tau_2}
}
\end{mathparpagebreakable}
\end{definition}

We next show \Cref{lemma:phantomstep} which says that if we have a step under phantom mode in the auxiliary semantics of commands, then any message produced is a dummy message on a non-public context channel and consumes potential, memory and input environment are unchanged, and the stores are related by extension.
\begin{restatable}[Phantom step]{lemma}{phantomstep}
\label{lemma:phantomstep}
Consider $\xangle{\Ctxaux{\Gamma}{\Pi}{\Lambda}{\Delta}{\pcstack}{q} \mid \bitstack,c,m,\mu,\intenv,\hst}$ s.t. $\pcstack \vdash \bitstack$ and $\Ctx{\Gamma}{\Pi}{\Lambda}{\Delta}{\pcstack} \vdash^{q}_{\textit{aux}} c : \pcstack''$. If $\bitstack = \xbit{0} \Colon \bitstack''$ and
\[
    \xangle{\Ctxaux{\Gamma}{\Pi}{\Lambda}{\Delta}{\pcstack}{q} \mid \xbit{0} \Colon \bitstack,c,m,\mu,\intenv,\hst}
    \xrightarrowdbl{\alpha}
    \xangle{\Ctxaux{\Gamma}{\Pi}{\Lambda}{\Delta}{\pcstack'}{q'} \mid \bitstack',c',m',\mu',\intenv',\hst'}
\]
then
\begin{enumerate}
    \item if $\alpha = \overrightarrow{\ch}(\timestamp,\bit,\val{v}{z})$ then $\bit = \xbit{0}$ and there exists $r$ and $\lmode \neq \bot$ such that
    \begin{enumerate}
        \item $q'+1+r \leq q$
        \item $\Lambda(\ch) = \ltype{\_}{\ltriple{\lmode}{\_}{r}}$
    \end{enumerate}
    \item $m=m'$
    \item $\mu \preceq \mu'$
    \item $\intenv = \intenv'$
\end{enumerate}
\end{restatable}
\begin{IEEEproof}
The proof is by induction in $c$.
\begin{description}
    \item[Case $c$ is $\texttt{skip}$:]
    Immediate.
    
    \item[Case $c$ is $c_1;c_2$:]
    By the induction hypothesis.

    \item[Case $c$ is $x \assign e$:]
    By $\pcstack \vdash \bitstack$ and $\textsc{T-Assign-Aux}$ we have $\bit = \xbit{1}$ and we are done.

    \item[Case $c$ is $x \oblivassign e$:]
    By \textsc{OblivAssign-Aux} we have $\mu(x) = \val{v_0}{z_0}$ and $\eval{e,m,\mu}{\val{v_1}{z_1}}$. If $\bit = \xbit{0}$ we have $\mu' = \mu[x \mapsto \val{v_0}{\textit{max}(z_0,z_1)}]$ and we are done.
        
    \item[Case $c$ is $x \oblivassign \texttt{input}(\ch,e)$:]
    By \textsc{LocalInput-Aux} and $\text{Phantom}(\bitstack)$ we have $m=m'$, $\intenv=\intenv'$, and $\alpha = \epsilon$. We additionally have $\mu(x)=\val{v_x}{z_x}$ and $z'=\textit{max}(v_x,n)$ such that $\mu' = \mu[x \mapsto \val{v_x}{z'}]$ and we are done.
        
    \item[Case $c$ is $\texttt{send}(\ch,e)$:]
    Immediate.
        
    \item[Case $c$ is $\texttt{if $e$ then $c$ else $c$}$:]
    Immediate.
        
    \item[Case $c$ is $\texttt{while $e$ do $c$}$:]
    Immediate.
        
    \item[Case $c$ is $\texttt{oblif $e$ then $c$ else $c$}$:]
    Immediate.
        
    \item[Case $c$ is $\texttt{pop}$:]
    Immediate.
\end{description}
\end{IEEEproof}

\Cref{lemma:phantompreservation} says that a bit-stack consisting only of phantom is preserved for any step of a well-typed auxiliary command. We shall shortly use this lemma for showing termination of commands executed under phantom mode, which allows us to reason that \oblivio{} programs do not get stuck while handling incoming dummy messages.
\begin{restatable}[Phantom preservation]{lemma}{phantompreservation}
\label{lemma:phantompreservation}
Consider $\xangle{\Ctxaux{\Gamma}{\Pi}{\Lambda}{\Delta}{\pcstack}{q} \mid \bitstack,c,m,\mu,\intenv,\hst}$ s.t. $\pcstack \vdash \bitstack$ and $\Ctx{\Gamma}{\Pi}{\Lambda}{\Delta}{\pcstack} \vdash^{q}_{\textit{aux}} c : \pcstack''$. If $\text{Phantom}(\bitstack)$ and
\[
    \xangle{\Ctxaux{\Gamma}{\Pi}{\Lambda}{\Delta}{\pcstack}{q} \mid \bitstack,c,m,\mu,\intenv,\hst}
    \xrightarrowdbl{\alpha}
    \xangle{\Ctxaux{\Gamma}{\Pi}{\Lambda}{\Delta}{\pcstack'}{q'} \mid \bitstack',c',m',\mu',\intenv',\hst',\tau'}
\]
then $\text{Phantom}(\bitstack')$.
\end{restatable}
\begin{IEEEproof}
The proof is by induction in $c$.
\begin{description}
    \item[Case $c$ is $\texttt{skip}$:]
    Immediate.
    
    \item[Case $c$ is $c_1;c_2$:]
    By induction hypothesis.

    \item[Case $c$ is $x \assign e$:]
    By $\textsc{T-Assign-Aux}$ and $\pcstack \vdash \bitstack$ we have $\bitstack = [\xbit{1}]$ hence contradiction.

    \item[Case $c$ is $x \oblivassign e$:]
    Immediate.
        
    \item[Case $c$ is $x \oblivassign \texttt{input}(\ch,e)$:]
    Immediate.
        
    \item[Case $c$ is $\texttt{send}(\ch,e)$:]
    Immediate.
        
    \item[Case $c$ is $\texttt{if $e$ then $c$ else $c$}$:]
    Immediate.
        
    \item[Case $c$ is $\texttt{while $e$ do $c$}$:]
    By $\textsc{T-While-Aux}$ and $\pcstack \vdash \bitstack$ we have $\bitstack = [\xbit{1}]$ hence contradiction.
        
    \item[Case $c$ is $\texttt{oblif $e$ then $c$ else $c$}$:]
    By the step and $\text{Phantom}(\bitstack)$ we have $\bitstack' = \xbit{0} \Colon \xbit{0} \Colon \bitstack$ and therefore have $\text{Phantom}(\bitstack')$.
        
    \item[Case $c$ is $\texttt{pop}$:]
    By the step and $\text{Phantom}(\bitstack)$ we have $\bitstack = \xbit{0} \Colon \bitstack'$ such that $\text{Phantom}(\bitstack')$ and we are done.
\end{description}
\end{IEEEproof}

\Cref{lemma:messagemode} relates the execution mode of an auxiliary command to the mode of generated events.
\begin{restatable}[Message mode]{lemma}{messagemode}
\label{lemma:messagemode}
Consider $\xangle{\Ctxaux{\Gamma}{\Pi}{\Lambda}{\Delta}{\pcstack}{q} \mid \bitstack,c,m,\mu,\intenv,\hst}$ s.t. $\pcstack \vdash \bitstack$ and $\Ctx{\Gamma}{\Pi}{\Lambda}{\Delta}{\pcstack} \vdash^{q}_{\textit{aux}} c : \pcstack''$. If $\bitstack = \bit \Colon \bitstack''$ and
\[
    \xangle{\Ctxaux{\Gamma}{\Pi}{\Lambda}{\Delta}{\pcstack}{q} \mid \xbit{0} \Colon \bitstack,c,m,\mu,\intenv,\hst}
    \xrightarrowdbl{\alpha}
    \xangle{\Ctxaux{\Gamma}{\Pi}{\Lambda}{\Delta}{\pcstack'}{q'} \mid \bitstack',c',m',\mu',\intenv',\hst'}
\]
then if $\alpha = \overrightarrow{\ch}(\timestamp,\bit',\val{v}{z})$ then $\bit' = \bit$.
\end{restatable}
\begin{IEEEproof}
The proof is by induction in $c$.
\begin{description}
    \item[Case $c$ is $\texttt{skip}$:]
    Trivial.
    
    \item[Case $c$ is $c_1;c_2$:]
    By the induction hypothesis.

    \item[Case $c$ is $x \assign e$:]
    Trivial.

    \item[Case $c$ is $x \oblivassign e$:]
    Trivial.
        
    \item[Case $c$ is $x \oblivassign \texttt{input}(\ch,e)$:]
    Trivial.
        
    \item[Case $c$ is $\texttt{send}(\ch,e)$:]
    By \textsc{Send-Aux}.
        
    \item[Case $c$ is $\texttt{if $e$ then $c$ else $c$}$:]
    Trivial.
        
    \item[Case $c$ is $\texttt{while $e$ do $c$}$:]
    Trivial.
        
    \item[Case $c$ is $\texttt{oblif $e$ then $c$ else $c$}$:]
    Trivial.
        
    \item[Case $c$ is $\texttt{pop}$:]
    Trivial.
\end{description}
\end{IEEEproof}

\Cref{lemma:extensionpreservation} shows that the extension relations are preserved by the auxiliary semantics of commands.
\begin{restatable}[Extension preservation]{lemma}{extensionpreservation}
\label{lemma:extensionpreservation}
Consider $\xangle{\Ctxaux{\Gamma}{\Pi}{\Lambda}{\Delta}{\pcstack}{q} \mid \bitstack,c,m_1,\mu_1,\intenv,\hst}$ such that $\Ctx{\Gamma}{\Pi}{\Lambda}{\Delta}{\pcstack} \vdash^q_{\textit{aux}} c : \pcstack''$. Consider $m_2,\mu_2$ such that $m_1 \extends m_2$, $\mu_1 \extends \mu_2$, if
\[
    \xangle{\Ctxaux{\Gamma}{\Pi}{\Lambda}{\Delta}{\pcstack}{q} \mid \bitstack,c,m_1,\mu_1,\intenv,\hst_1}
    \xrightarrowdbl{\alpha_1}
    \xangle{\Ctxaux{\Gamma}{\Pi}{\Lambda}{\Delta}{\pcstack'}{q'} \mid \bitstack',c',m'_1,\mu'_1,\intenv',\hst'_1}
\]
then
\[
    \xangle{\Ctxaux{\Gamma}{\Pi}{\Lambda}{\Delta}{\pcstack}{q} \mid \bitstack,c,m_2,\mu_2,\intenv,\hst_2}
    \xrightarrowdbl{\alpha_2}
    \xangle{\Ctxaux{\Gamma}{\Pi}{\Lambda}{\Delta}{\pcstack'}{q'} \mid \bitstack',c',m'_2,\mu'_2,\intenv',\hst'_2}
\]
s.t. $m'_1 \extends m'_2$, and $\mu'_1 \extends \mu'_2$ and if $\alpha_1 = \overrightarrow{\ch}(\timestamp_1,\bit,\val{v_1}{z_1})$ then $\alpha_2 = \overrightarrow{\ch}(\timestamp_2,\bit,\val{v_2}{z_2})$ s.t. $\val{v_1}{z_1} \extends \val{v_2}{z_2}$.
\end{restatable}
\begin{IEEEproof}
The proof is by induction in $c$.
\begin{description}
    \item[Case $c$ is $\texttt{skip}$:]
    Immediate.
    
    \item[Case $c$ is $c_1;c_2$:]
    By the induction hypothesis.

    \item[Case $c$ is $x \assign e$:]
    We have $\eval{e,m_1,\mu_1}{\val{v_1}{z_1}}$ and $\eval{e,m_2,\mu_2}{\val{v_2}{z_2}}$.
    By $m_1 \extends m_2$ and $\mu_1 \extends \mu_2$ we have $\val{v_1}{z_1} \extends \val{v_2}{z_2}$ and therefore $\mu_1[x \mapsto \val{v_1}{z_1}] \extends \mu_2[x \mapsto \val{v_2}{z_2}]$.

    \item[Case $c$ is $x \oblivassign e$:]
    By \textsc{OblivAssign-Aux}, $m_1 \extends m_2$, $\mu_1 \extends \mu_2$, and monotonicity of $\textit{max}$, observing that both runs have same bit-stack $\bitstack$.
        
    \item[Case $c$ is $x \oblivassign \texttt{input}(\ch,e)$:]
    By \textsc{LocalInput-Aux}, $\mu_1 \extends \mu_2$, and monotonicity of $\textit{max}$, observing that both runs have same bit-stack $\bitstack$ and local input environment $\intenv$.
        
    \item[Case $c$ is $\texttt{send}(\ch,e)$:]
    By $m_1 \extends m_2$ and $\mu_1 \extends \mu_2$ we have $\eval{e,m_1,\mu_1}{\val{v_1}{z_1}}$ and $\eval{e,m_2,\mu_2}{\val{v_2}{z_2}}$ such that $\val{v_1}{z_1} \extends \val{v_2}{z_2}$. We are done by \Cref{lemma:messagemode}.
        
    \item[Case $c$ is $\texttt{if $e$ then $c$ else $c$}$:]
    Immediate.
        
    \item[Case $c$ is $\texttt{while $e$ do $c$}$:]
    Immediate.
        
    \item[Case $c$ is $\texttt{oblif $e$ then $c$ else $c$}$:]
    Immediate.
        
    \item[Case $c$ is $\texttt{pop}$:]
    Immediate.
\end{description}
\end{IEEEproof}

\Cref{lemma:multistepcomposition,lemma:highcommandsterminate} give us termination of any auxiliary command that is typeable with non-public bottom element of $\pc$-stack $\pcstack$.
\begin{restatable}[Multistep composition]{lemma}{multistepcomposition}
\label{lemma:multistepcomposition}
Consider $\xangle{\Ctxaux{\Gamma}{\Pi}{\Lambda}{\Delta}{\pcstack}{q} \mid \bitstack,c_1;c_2,m,\mu,\intenv,\hst}$ such that $\pcstack \vdash \bitstack$ and $\Ctx{\Gamma}{\Pi}{\Lambda}{\Delta}{\pcstack} \vdash^{q}_{\textit{aux}} c : \pcstack''$. If
\[
    \xangle{\Ctxaux{\Gamma}{\Pi}{\Lambda}{\Delta}{\pcstack}{q} \mid \bitstack,c_1,m,\mu,\intenv,\hst}
    \xrightarrow{\tau}{\phantom{I}\mkern-18mu}^{n}
    \xangle{\Ctxaux{\Gamma}{\Pi}{\Lambda}{\Delta}{\pcstack'}{q'} \mid \bitstack',\texttt{stop},m',\mu',\intenv',\hst'}
\]
then
\[
    \xangle{\Ctxaux{\Gamma}{\Pi}{\Lambda}{\Delta}{\pcstack}{q} \mid \bitstack,c_1;c_2,m,\mu,\intenv,\hst}
    \xrightarrow{\tau}{\phantom{I}\mkern-18mu}^{n}
    \xangle{\Ctxaux{\Gamma}{\Pi}{\Lambda}{\Delta}{\pcstack'}{q'} \mid \bitstack',c_2,m',\mu',\intenv',\hst'}
\]
such that $\pcstack' \vdash \bitstack'$ and $\Ctx{\Gamma}{\Pi}{\Lambda}{\Delta}{\pcstack'} \vdash^{q'}_{\textit{aux}} c_2 : \pcstack''$.
\end{restatable}
\begin{IEEEproof}
The proof is by induction in $n$.
\begin{description}
    \item[Base case $n=0$:] Impossible as \textsc{T-Seq-Aux} implies $c_1 \neq \texttt{stop}$.
    \item[Base case $n=1$:] By \textsc{Seq2-Aux} and \Cref{lemma:preservation}.
    \item[Base case $n=k+1$:]
    By \Cref{lemma:preservation} and the induction hypothesis.
\end{description}    
\end{IEEEproof}

\begin{restatable}[High commands terminate]{lemma}{highcommandsterminate}
\label{lemma:highcommandsterminate}
Consider $\xangle{\Ctxaux{\Gamma}{\Pi}{\Lambda}{\Delta}{\pcstack}{q} \mid \bitstack,c,m,\mu,\intenv,\hst}$ such that $\pc \vdash \pcstack$ and $\Ctx{\Gamma}{\Pi}{\Lambda}{\Delta}{\pcstack} \vdash^{q}_{\textit{aux}} c : \pcstack'$. We have that if $\pc \neq \bot$ then there exists $n \geq 1$ such that
\[
    \xangle{\Ctxaux{\Gamma}{\Pi}{\Lambda}{\Delta}{\pcstack}{q} \mid \bitstack,c,m,\mu,\intenv,\hst}
    \xrightarrow{\tau}{\phantom{I}\mkern-18mu}^{n}
    \xangle{\Ctxaux{\Gamma}{\Pi}{\Lambda}{\Delta}{\pcstack'}{q'} \mid \bitstack',\texttt{stop},m',\mu',\intenv',\hst'}
\]
\end{restatable}
\begin{IEEEproof}
The proof is by induction in $c$.
\begin{description}
    \item[Case $c$ is $\texttt{skip}$:]
    Immediate with $n=1$.
    
    \item[Case $c$ is $c_1;c_2$:]
    By the induction hypothesis on $c_1$ we have
    \[
        \xangle{\Ctxaux{\Gamma}{\Pi}{\Lambda}{\Delta}{\pcstack}{q} \mid \bitstack,c_1,m,\mu,\intenv,\hst}
        \xrightarrow{\tau}{\phantom{I}\mkern-18mu}^{n}
        \xangle{\Ctxaux{\Gamma}{\Pi}{\Lambda}{\Delta}{\pcstack''}{q''} \mid \bitstack'',\texttt{stop},m'',\mu'',\intenv'',\hst''}
    \]
    and therefore by \Cref{lemma:multistepcomposition} that
    \[
        \xangle{\Ctxaux{\Gamma}{\Pi}{\Lambda}{\Delta}{\pcstack}{q} \mid \bitstack,c_1;c_2,m,\mu,\intenv,\hst}
        \xrightarrow{\tau}{\phantom{I}\mkern-18mu}^{n}
        \xangle{\Ctxaux{\Gamma}{\Pi}{\Lambda}{\Delta}{\pcstack''}{q''} \mid \bitstack'',c_2,m'',\mu'',\intenv'',\hst''}
    \]
    and $\pcstack'' \vdash \bitstack''$ and $\Gamma,\Delta,\Pi,\Lambda,\pcstack'' \vdash^{q''}_{\textit{aux}} c_2 : \pc \Colon \pcstack'$.
    We are done by the induction hypothesis on $c_2$.

    \item[Case $c$ is $x \assign e$:]
    Immediate with $n=1$.

    \item[Case $c$ is $x \oblivassign e$:]
    Immediate with $n=1$.
        
    \item[Case $c$ is $x \oblivassign \texttt{input}(\ch,e)$:]
    Immediate with $n=1$.
        
    \item[Case $c$ is $\texttt{send}(\ch,e)$:]
    Immediate with $n=1$.
        
    \item[Case $c$ is $\texttt{if $e$ then $c$ else $c$}$:]
    Immediate with $n=1$.
        
    \item[Case $c$ is $\texttt{while $e$ do $c$}$:]
    Impossible by assumptions $\Ctx{\Gamma}{\Pi}{\Lambda}{\Delta}{\pcstack} \vdash^q c$, $\pc \vdash \pcstack$, and $\pc \neq \bot$.
        
    \item[Case $c$ is $\texttt{oblif $e$ then $c$ else $c$}$:]
    By the induction hypothesis on $c_1$ and $c_2$, \textsc{Pop}, and \Cref{lemma:multistepcomposition}.
        
    \item[Case $c$ is $\texttt{pop}$:]
    Immediate with $n=1$.
\end{description}
\end{IEEEproof}

As well-formedness of network strategies (\Cref{def:wellformed_networkstrategy}) gives us that dummy messages are only produced on non-public context channels, we can use the above lemma to conclude that \oblivio{} programs do not get stuck handling dummy messages (\Cref{lemma:phantomtermination}).
\begin{restatable}[Phantom termination]{lemma}{phantomtermination}
\label{lemma:phantomtermination}
Consider $\Produceraux = \producer{\Ctxaux{\Gamma}{\Pi}{\Lambda}{\Delta}{\pcstack}{q} \mid p,\bitstack,c,m,\mu,\intenv,\networkstrategy,\hst,\tau}$ such that $\vdash_{\textit{aux}} \Produceraux$. We have that if $\text{Phantom}(\bitstack)$ then
\[
    \producer{\Ctxaux{\Gamma}{\Pi}{\Lambda}{\Delta}{\pcstack}{q} \mid p,\bitstack,c,m,\mu,\intenv,\networkstrategy,\hst,\tau}
    \longrightarrowdbl {\phantom{I}\mkern-18mu}^{*}
    \consumer{\ctx{\Gamma}{\Pi}{\Lambda} \mid p,\mu',\intenv,\networkstrategy,\hst',\tau \cdot \tau'}
\]
such that
\begin{enumerate}
    \item $\vdash_{\textit{aux}} \consumer{\ctx{\Gamma}{\Pi}{\Lambda} \mid p,\mu',\intenv,\networkstrategy,\hst',\tau \cdot \tau'}$
    \item $\epsilon \phantomextends \tau'$
    \item $|\tau'| \leq q$
    \item $\mu \extends \mu'$
\end{enumerate}
\end{restatable}
\begin{IEEEproof}
We show that there exists $n$ such that
\begin{align*}
    \producer{\Ctxaux{\Gamma}{\Pi}{\Lambda}{\Delta}{\pcstack}{q} \mid p,\bitstack,c,m,\mu,\intenv,\networkstrategy,\hst,\tau}
    &\longrightarrowdbl {\phantom{I}\mkern-18mu}^{n}
    \producer{\Ctxaux{\Gamma}{\Pi}{\Lambda}{\Delta}{\pcstack'}{q'} \mid p,\bitstack',\texttt{stop},m',\mu',\intenv',\networkstrategy,\hst',\tau \cdot \tau'} \\
    &\longrightarrowdbl
    \consumer{\ctx{\Gamma}{\Pi}{\Lambda} \mid p,\mu',\intenv,\networkstrategy,\hst' \Colon \textsf{ret},\tau \cdot \tau'}
\end{align*}
such that
\begin{enumerate}
    \item $\vdash_{\textit{aux}} \consumer{\ctx{\Gamma}{\Pi}{\Lambda} \mid p,\mu',\intenv,\networkstrategy,\hst' \Colon \textsf{ret},\tau \cdot \tau'}$
    \item $\epsilon \phantomextends \tau'$
    \item $|\tau'| + q' \leq q$
    \item $\mu \extends \mu'$
\end{enumerate}

By $\vdash_{\textit{aux}} \Produceraux$ and $\text{Phantom}(\bitstack)$ we have $\Ctx{\Gamma}{\Pi}{\Lambda}{\Delta}{\pcstack} \vdash^{q}_\text{aux} c : [\pc']$ such that $\pc' \neq \bot$. Thus, by \Cref{lemma:highcommandsterminate} we have the existence of $n$ such that
\[
    \xangle{\Ctxaux{\Gamma}{\Lambda}{\Delta}{\pcstack}{q} \mid \bitstack,c,m,\mu,\intenv,\hst}
    \xrightarrow{\tau}{\phantom{I}\mkern-18mu}^{n}
    \xangle{\Ctxaux{\Gamma}{\Lambda}{\Delta}{\pcstack''}{q''} \mid \bitstack'',\texttt{stop},m'',\mu'',\intenv'',\hst''}
\]
We proceed by induction in $n$
\begin{description}
    \item[Base case $n=0$:]
    By \textsc{PC} and \Cref{lemma:preservation}.
    
    \item[Base case $n=k+1$:]
    Done \Cref{lemma:phantomstep,lemma:phantompreservation,lemma:preservation} and the induction hypothesis.
\end{description}
\end{IEEEproof}

We define function $\textit{potential}(\Configurationaux)$ for projecting the potential of an auxiliary configuration in the straightforward way.
\begin{definition}[Potential of auxiliary configuration] Define the projection of the potential of auxiliary configuration $\Configurationaux$, written $\textit{potential}(\Configurationaux)$, as follows:
\label{def:potentialofauxiliarconfiguration}
\begin{align*}
    \textit{potential}(\Configurationaux) =
    \begin{cases}
        q
            &\textit{if } \Configurationaux = \producer{\Ctxaux{\Gamma}{\Pi}{\Lambda}{\Delta}{\pcstack}{q} \mid p,\bitstack,c,m,\mu,\intenv,\networkstrategy,\hst,\tau} \\
        0
            &\textit{otherwise}
    \end{cases}
\end{align*}
\end{definition}

\Cref{lemma:producersteppotential} relates the steps of two auxiliary producer configurations and
\begin{restatable}[Producer step potential]{lemma}{producersteppotential}
\label{lemma:producersteppotential}
Consider $\Produceraux_1$ and $\Produceraux_2$ such that
\begin{enumerate}
    \item $\vdash_{\textit{aux}} \Produceraux_1$
    \item $\vdash_{\textit{aux}} \Produceraux_2$
    \item $\Produceraux_1 \lambdaphantomextends \Produceraux_2$
    \item $\textit{trace}(\Produceraux_1)=\tau_1$
    \item $\textit{trace}(\Produceraux_2)=\tau_2$
    \item $\Lambda \vdash \tau_1 : q_1$
    \item $\Lambda \vdash \tau_2 : q_2$
    \item $\textit{potential}(\Produceraux_1) \leq q_1$
    \item $\textit{potential}(\Produceraux_2) \leq q_2$
\end{enumerate}
If
\(
    \Produceraux_1
    \xrightarrowdbl[\textit{unsafe}]{}
    \Produceraux'_1
\)
with $\textit{trace}(\Produceraux'_1)=\tau'_1$ then $\tau'_1 = \tau_1 \cdot \alpha_1$ and
\(
    \Produceraux_2
    \longrightarrowdbl
    \Produceraux'_2
\)
with $\textit{trace}(\Produceraux'_2)=\tau'_2$ such that $\tau'_2 = \tau_2 \cdot \alpha_2$ and
\begin{enumerate}
    \item $\vdash_{\textit{aux}} \Produceraux'_1$
    \item $\vdash_{\textit{aux}} \Produceraux'_2$
    \item $\Produceraux'_1 \lambdaphantomextends \Produceraux'_2$
    \item $\Lambda \vdash \tau_1 \cdot \alpha_1 : q'_1$
    \item $\Lambda \vdash \tau_2 \cdot \alpha_2 : q'_2$
    \item $\textit{potential}(\Produceraux'_1) \leq q'_1$
    \item $\textit{potential}(\Produceraux'_2) \leq q'_2$
    \item $\alpha_1 = \epsilon \land \alpha_2 \neq \epsilon \implies q'_2 < q_2$
\end{enumerate}
\end{restatable}
\begin{IEEEproof}
    By \textsc{PP-Unsafe} we have $\tau'_1=\tau_1 \cdot \alpha_1$ and by \textsc{PP} we have $\tau_2=\tau_2 \cdot \alpha_2$.
    
    By \Cref{lemma:extensionpreservation,lemma:preservation} it remains to show that if
    $\alpha_1 = \epsilon$ and $\alpha_2 \neq \epsilon$ then $\Lambda \vdash \tau_2 : q_2$ and $\Lambda \vdash \tau_2 \cdot \alpha_2 : q'_2$ such that $q_2 > q'_2$.
    This follows by \Cref{lemma:messagemode,lemma:phantomstep}.
\end{IEEEproof}

As the final step before showing the overhead theorem, we show \Cref{lemma:singlestepoverhead}. The lemma maintains the invariant needed for the theorem for each step in the unsafe, suppressing system semantics.
\begin{restatable}[Single step overhead]{lemma}{singlestepoverhead}
\label{lemma:singlestepoverhead}
Consider $\Configurationaux_1,\Configurationaux_2$ such that
\begin{enumerate}
    \item $\vdash_{\textit{aux}} \Configurationaux_1$
    \item $\vdash_{\textit{aux}} \Configurationaux_2$
    \item $\Configurationaux_1 \lambdaphantomextends \Configurationaux_2$
    \item $\textit{trace}(\Configurationaux_1) = \tau_1$
    \item $\textit{trace}(\Configurationaux_2) = \tau_2$
    \item $\Lambda \vdash \tau_1 : q_1$
    \item $\Lambda \vdash \tau_2 : q_2$
    \item $\textit{potential}(\Configurationaux_1) \leq q_1$
    \item $\textit{potential}(\Configurationaux_2) \leq q_2$
    \item $|\tau_2| + q_2 \leq |\tau_1| * (1 + \maxpotential)$
\end{enumerate}
We have that if
\(
    \Configurationaux_1
    \xrightarrowdbl[\textit{unsafe}]{}
    \Configurationaux'_1
\)
with $\textit{trace}(\Configurationaux'_1) = \tau'_1$
\begin{enumerate}
    \item $\tau'_1 = \tau_1 \cdot \alpha_1$
    \item $\Lambda \vdash \tau_1 \cdot \alpha_1 : q'_1$
    \item $\textit{potential}(\Configurationaux'_1) \leq q'_1$
    \item $\alpha_1 \neq \epsilon \implies q'_1 \leq q_1 + \maxpotential$
    \item and
    \begin{enumerate}
        \item either
        \(
            \Configurationaux_2
            \longrightarrowdbl
            \Configurationaux'_2
        \)
        with $\textit{trace}(\Configurationaux'_2) = \tau_2 \cdot \alpha_2$ and
        \begin{enumerate}
            \item $\Configurationaux'_1 \lambdaphantomextends \Configurationaux'_2$
            \item $\Lambda \vdash \tau_2 \cdot \alpha_2 : q'_2$
            \item $\textit{potential}(\Configurationaux'_2) \leq q'_2$
            \item $|\tau_2 \cdot \alpha_2| + q'_2 \leq |\tau_1 \cdot \alpha_1| * (1 + \maxpotential)$
        \end{enumerate}
        \item or $\alpha_1 \neq \epsilon$ and
        \(
            \Configurationaux_2
            \longrightarrowdbl {\phantom{I}\mkern-18mu}^{*}
            \Configurationaux'_2
        \)
        with $\textit{trace}(\Configurationaux'_2) = \tau_2 \cdot \tau'_2 \cdot \alpha_2$ and
        \begin{enumerate}
            \item $\Configurationaux'_1 \lambdaphantomextends \Configurationaux'_2$
            \item $|\tau'_2| \leq q_2$
            \item $\alpha_2 \neq \epsilon$
            \item $\Lambda \vdash \tau_2 \cdot \tau'_2 \cdot \alpha_2 : q'_2$
            \item $\textit{potential}(\Configurationaux'_2) \leq q'_2$
            \item $|\tau_2 \cdot \tau'_2 \cdot \alpha_2| + q'_2 \leq |\tau_1 \cdot \alpha_1| * (1 + \maxpotential)$
        \end{enumerate}
    \end{enumerate}
\end{enumerate}
\end{restatable}
\begin{IEEEproof}
We begin by casing on the step in the unsafe run
\begin{description}
    \item[Case CC-Unsafe:]
    By the step in the unsafe run we have $\networkstrategy_1(\tau_1)=\ch(\timestamp,\xbit{1},\val{v_1}{z_1})$ such that $(p)(\ch) \not\Downarrow$ and therefore $\Configurationaux'_1 = \consumer{\ctx{\Gamma}{\Pi}{\Lambda} \mid p,\mu_1,\intenv,\networkstrategy_1,\hst_1,\tau_1 \cdot \widetilde{\ch}(\timestamp,\xbit{1},\val{v_1}{z_1})}$. Let $\networkstrategy_2(\tau_2)=\ch'(\timestamp',\bit,\val{v_2}{z_2})$. We have $\Lambda \vdash \tau_2 : q_2$ and proceed by strong induction on $q_2$.
    \begin{description}
        \item [Base case $q_2=0$:]
        By well-formedness of $\networkstrategy_2$ we have $\bit=\xbit{1}$ and hence by $\networkstrategy_1 \lambdaphantomextends \networkstrategy_2$ we have $\ch'=\ch$, $v_2 = v_1$, and $z_2 \geq z_1$.
        We are done by \Cref{def:maximumpotential} showing item 4.a.
        
        \item [Inductive step $0 \leq q' < q_2$:]
        We case $\bit$.
        \begin{description}
            \item[Case $\bit=\xbit{1}$:]
            Done by $\networkstrategy_1 \lambdaextends \networkstrategy_2$ and \Cref{def:maximumpotential} showing item 4.a.
            
            \item[Case $\bit=\xbit{0}$:]
            By well-formedness of $\networkstrategy_2$ we have $\Lambda(\ch') = \ltype{\sigma}{\ltriple{\lmode}{\lval}{r}}$ s.t. $\lmode \neq \bot$ and $q_2 \geq 1 + r$.
            
            We case on $(p)(\ch') \Downarrow$.
            \begin{description}
                \item [Case $(p)(\ch') \not\Downarrow$:]
                By \Cref{def:tracepotential} we have $\Lambda \vdash \tau_2 \cdot \tau'_2 : q'_2$ such that $q_2 > q'_2$ and we conclude by the induction hypothesis showing item 4.b.
                
                \item [Case $(p)(\ch') \Downarrow c,x$:]
                We have a step $\Consumeraux_2 \longrightarrowdbl \producer{\Ctxaux{\Gamma}{\Pi}{\Lambda}{\Delta}{\pcstack}{r} \mid p,[\xbit{0}],c,m,\mu,\intenv,\networkstrategy,\hst,\tau}$. We have $\text{Phantom}(\xbit{0})$ and therefore by \Cref{lemma:preservation,lemma:phantomtermination} we have
                \begin{align*}
                    &\producer{\Ctxaux{\Gamma}{\Pi}{\Lambda}{\Delta}{\pcstack}{r} \mid p,[\xbit{0}],c,m,\mu,\intenv,\networkstrategy,\hst,\tau \cdot \overrightarrow{\ch'}(\timestamp',\xbit{0},\val{v_2}{z_2})}
                    \\
                    &\longrightarrowdbl {\phantom{I}\mkern-18mu}^{*}
                    \consumer{\ctx{\Gamma}{\Pi}{\Lambda} \mid p,\mu_2,\intenv,\networkstrategy_2,\hst'_2,\tau_2 \cdot \overrightarrow{\ch'}(\timestamp',\xbit{0},\val{v_2}{z_2}) \cdot \tau'_2}
                \end{align*}
                such that
                \begin{enumerate}
                    \item $\vdash_{\textit{aux}} \consumer{\ctx{\Gamma}{\Pi}{\Lambda} \mid p,\mu',\intenv,\networkstrategy,\hst',\tau_2 \cdot \overrightarrow{\ch'}(\timestamp',\xbit{0},\val{v_2}{z_2}) \cdot \tau'_2}$
                    \item $|\tau'_2| \leq r$
                    \item $\epsilon \phantomextends \tau'_2$
                    \item $\mu \extends \mu'$
                \end{enumerate}
                From $\tau_1 \phantomextends \tau_2$ and $\epsilon \phantomextends \tau'_2$ we have $\tau_1 \phantomextends \tau_2 \cdot \overrightarrow{\ch'}(\timestamp',\xbit{0},\val{v_2}{z_2}) \cdot \tau'_2$ and from $q_2 \geq 1 + r$ and $|\tau'_2| \leq r$ we conclude there exists $q''_2 < q_2$ such that $\Lambda \vdash \tau_2 \cdot \overrightarrow{\ch'}(\timestamp',\xbit{0},\val{v_2}{z_2}) \cdot \tau'_2 : q''_2$.
                By transitivity of $\extends$ on stores we have $\Configurationaux_1 \lambdaphantomextends \Consumeraux_2$. We are done by the induction hypothesis showing item 4.b.
            \end{description}
        \end{description}
    \end{description}
    
    \item[Case CP-Unsafe:]
    By the step in the unsafe run we have $\networkstrategy_1(\tau_1)=\ch(\timestamp,\xbit{1},\val{v_1}{z_1})$ such that $(p)(\ch) \Downarrow c,x$.
    By well-formedness of $\networkstrategy_1$ we have $\Lambda(\ch)=\ltype{\sigma}{\ltriple{\lmode}{\lval}{r}}$ such that
    $\Configurationaux'_1 = \producer{\Ctxaux{\Gamma}{\Pi}{\Lambda}{[x \mapsto \lval]}{[\lmode]}{r} \mid p,\mu_1,\intenv,\networkstrategy_1,\hst_1,\tau_1 \cdot \overleftarrow{\ch}(\timestamp,\xbit{1},\val{v_1}{z_1})}$.
    Let $\networkstrategy_2(\tau_2)=\ch'(\timestamp',\bit,\val{v_2}{z_2})$. We have $\Lambda \vdash \tau_2 : q_2$ and proceed by strong induction on $q_2$.
    \begin{description}
        \item [Base case $q_2=0$:]
        By well-formedness of $\networkstrategy_2$ we have $\bit=\xbit{1}$ and hence by $\networkstrategy_1 \lambdaphantomextends \networkstrategy_2$ we have $\ch'=\ch$, $v_2 = v_1$, and $z_2 \geq z_1$.
        We are done by \Cref{def:maximumpotential} showing item 4.a.
        
        \item [Inductive step $0 \leq q' < q_2$:]
        We case $\bit$.
        \begin{description}
            \item[Case $\bit=\xbit{1}$:]
            Done by $\networkstrategy_1 \lambdaextends \networkstrategy_2$ and \Cref{def:maximumpotential} showing item 4.a.
            
            \item[Case $\bit=\xbit{0}$:]
            By well-formedness of $\networkstrategy_2$ we have $\Lambda(\ch') = \ltype{\sigma'}{\ltriple{\lmode'}{\lval'}{r'}}$ s.t. $\lmode \neq \bot$ and $q_2 \geq 1 + r'$.
            
            We case on $(p)(\ch') \Downarrow$.
            \begin{description}
                \item [Case $(p)(\ch') \not\Downarrow$:]
                By \Cref{def:tracepotential} we have $\Lambda \vdash \tau_2 \cdot \alpha_2 : q'_2$ such that $q_2 > q'_2$ and we conclude by the induction hypothesis showing item 4.b.
                
                \item [Case $(p)(\ch') \Downarrow c,x$:]
                We have a step $\Consumeraux_2 \longrightarrowdbl \producer{\Ctxaux{\Gamma}{\Pi}{\Lambda}{\Delta}{\pcstack}{r'} \mid p,[\xbit{0}],c,m,\mu,\intenv,\networkstrategy,\hst,\tau}$. We have $\text{Phantom}(\xbit{0})$ and therefore by \Cref{lemma:preservation,lemma:phantomtermination} we have
                \begin{align*}
                    &\producer{\Ctxaux{\Gamma}{\Pi}{\Lambda}{\Delta}{\pcstack}{r'} \mid p,[\xbit{0}],c,m,\mu,\intenv,\networkstrategy,\hst,\tau \cdot \overleftarrow{\ch'}(\timestamp',\xbit{0},\val{v_2}{z_2})}
                    \\
                    &\longrightarrowdbl {\phantom{I}\mkern-18mu}^{*}
                    \consumer{\ctx{\Gamma}{\Pi}{\Lambda} \mid p,\mu',\intenv,\networkstrategy,\hst',\tau_2 \cdot \overleftarrow{\ch'}(\timestamp',\xbit{0},\val{v_2}{z_2}) \cdot \tau'_2}
                \end{align*}
                such that
                \begin{enumerate}
                    \item $\vdash_{\textit{aux}} \consumer{\ctx{\Gamma}{\Pi}{\Lambda} \mid p,\mu',\intenv,\networkstrategy,\hst',\tau_2 \cdot \overleftarrow{\ch'}(\timestamp',\xbit{0},\val{v_2}{z_2}) \cdot \tau'_2}$
                    \item $|\tau'_2| \leq r'$
                    \item $\epsilon \phantomextends \tau'_2$
                    \item $\mu \extends \mu'$
                \end{enumerate}
                From $\tau_1 \phantomextends \tau_2$ and $\epsilon \phantomextends \tau'_2$ we have $\tau_1 \phantomextends \tau_2 \cdot \overleftarrow{\ch'}(\timestamp',\xbit{0},\val{v_2}{z_2}) \cdot \tau'_2$ and from $q_2 \geq 1 + q$ and $|\tau'_2| \leq r'$ we conclude there exists $q''_2 < q_2$ such that
                $\Lambda \vdash \tau_2 \cdot \ch'(\timestamp',\xbit{0},\val{v_2}{z_2}) \cdot \tau'_2 : q''_2$.
                By transitivity of $\extends$ on stores we have $\Configurationaux_1 \lambdaphantomextends \Consumeraux_2$. We are done by the induction hypothesis showing item 4.b.
            \end{description}
        \end{description}
    \end{description}
    
    \item[Case PP-Unsafe:]
    By \Cref{lemma:producersteppotential} and \Cref{def:maximumpotential} showing item 4.a.
    
    \item[Case PC-Unsafe:]
    Immediate.
\end{description}
\end{IEEEproof}

We are now ready to show the overhead theorem (\Cref{theorem:enforcementoverhead}).
\enforcementoverhead*
\begin{IEEEproof}
By \Cref{lemma:adequacy} we conduct the proof in the auxiliary semantics.

We additionally show that if
\(
    \consumer{\ctx{\Gamma}{\Pi}{\Lambda} \mid p,\mu_1,\intenv,\networkstrategy_1,\hst,\epsilon}
    \xrightarrowdbl[\textit{unsafe}]{}{\phantom{I}\mkern-18mu}^*
    \Configurationaux_1
\)
with $\textit{trace}(\Configurationaux_1) = \tau_1$ then
\(
    \consumer{\ctx{\Gamma}{\Pi}{\Lambda} \mid p,\mu_2,\intenv,\networkstrategy_2,\hst,\epsilon}
    \longrightarrowdbl{\phantom{I}\mkern-18mu}^*
    \Configurationaux_2
\)
with $\textit{trace}(\Configurationaux_2) = \tau_2$ such that
\begin{enumerate}
    \item $\vdash_{\textit{aux}} \Configurationaux_1$
    \item $\vdash_{\textit{aux}} \Configurationaux_2$
    \item $\Configurationaux_1 \lambdaphantomextends \Configurationaux_2$
    \item $\Lambda \vdash \tau_1 : q_1$
    \item $\Lambda \vdash \tau_2 : q_2$
    \item $|\tau_2| + q_2 \leq |\tau_1| * (1 + \maxpotential)$
\end{enumerate}

The proof is by induction in $\tau_1$.
\begin{description}
    \item[Base case $\tau_1=\epsilon$:]
    Done by \Cref{lemma:singlestepoverhead}.
    
    \item[Inductive step $\tau_1=\tau'_1 \cdot \alpha_1$:]
    Done by the induction hypothesis and \Cref{lemma:singlestepoverhead}.
\end{description}
\end{IEEEproof}
\clearpage
\section{Examples}
\label{appendix:examples}
This section provides additional example programs written in \oblivio{}.

\subsection{Dummy message bounds}
This simple example demonstrates the annotated potentials by considering two nodes, \texttt{ALICE}, \texttt{BOB}, that send messages to each other. The handler for \texttt{BOB/B2} sends no messages, hence it is annotated with potential $0$. The handler for \texttt{ALICE/A2} sends two messages to \texttt{BOB/B2} and is annotated with potential $(1+0)+(1+0)=2$, accounting for the two sends and the 0 potential of \texttt{BOB/B2}. The handler for \texttt{BOB/B1} sends two messages to \texttt{ALICE/A2} and is therefore annotated with potential $(1+2)+(1+2)=6$, accounting for the two sends and the potential of \texttt{ALICE/A2}. Similarly, the handler for \texttt{ALICE/A1} sends to messages to \texttt{BOB/B1} and is annotated with potential $(1+6)+(1+6)=14$.

\begin{multicols}{2}
\begin{lstlisting}[caption=Alice,label=lst:alice_exponential,mathescape]
ALICE

A1$_H$ $\$14$ (secret : int$_H$) {
    oblif secret
    then send(BOB/B1, secret+1);
    else send(BOB/B1, secret-1);
}


A2$_H$ $\$2$ (secret : int$_H$) {
    oblif secret
    then send(BOB/B2, secret+1);
    else send(BOB/B2, secret-1);
}
\end{lstlisting}

\columnbreak

\begin{lstlisting}[caption=bob,label=lst:bob_exponential,mathescape]
BOB

local channel STDOUT: int$_L$

B1$_H$ $\$6$ (secret : int$_H$) {
    oblif secret
    then send(ALICE/A2, secret+1);
    else send(ALICE/A2, secret-1);
}

B2$_H$ (secret : int$_H$) {
    output(STDOUT, secret);
}
\end{lstlisting}
\end{multicols}

\subsection{Chat}
The remaining example in this section use a version of \oblivio{} extended a command for local output and pair and array types.

We present a simple chat between users Alice and Bob. Alice (\Cref{lst:alice_chat}) receives messages on channel \texttt{ALICE/CHAT} and obliviously branches on whether the message is the empty string. If not, she outputs the message to \texttt{STDOUT}. She then inputs a message from \texttt{STDIN} and sends it to Bob, before setting variable \texttt{msg\_out} to the empty string. The code for Bob (\Cref{lst:bob_chat}) is analogous.

\begin{multicols}{2}
\begin{lstlisting}[caption=Alice,label=lst:alice_chat,mathescape]
ALICE

local channel STDIN : string$_H$;
local channel STDOUT : string$_H$;

var msg_out: string$_H$ = "";

CHAT$_L$ (msg_in: string$_H$) {
    oblif msg_in != ""
    then output(STDOUT,"Bob says: " ^ msg_in);
    else skip;

    msg_out ?= input(STDIN,32);
    send(BOB/CHAT,msg_out);
    msg_out = "";
}
\end{lstlisting}
\columnbreak
\begin{lstlisting}[caption=Bob,label=lst:bob_chat,mathescape]
BOB

local channel STDIN : string$_H$;
local channel STDOUT : string$_H$;

var msg_out: string$_H$ = "";

CHAT$_L$ (msg_in: string$_H$) {
    oblif msg_in != ""
    then output(STDOUT,"Alice says: " ^ msg_in);
    else skip;
    
    msg_out ?= input(STDIN,32);
    send(ALICE/CHAT,msg_out);
    msg_out = "";
}
\end{lstlisting}
\end{multicols}

\subsection{Auction}
We present a modified version of the auction example from \Cref{sec:example} that makes use of pair types.
The server code (\Cref{lst:auction_timer_extended}) now only has a single handler \texttt{BID} for accepting bids, and clients are required to provide their name as part of the bid. The code for handler \texttt{TICK} has been altered to unconditionally send auction status updates to Alice and Bob with the current winner and winning bid.
\begin{multicols}{2}
\begin{lstlisting}[caption=Auction server,label=lst:auction_server_extended,mathescape]
AUCTIONHOUSE // Node ID

var winning_bid : (string$_H$*int$_H$)$_H$ = ("",0);
var round_counter : int$_L$ = 500;

BID$_H$ (bid: (string$_L$*int$_H$)$_L$) {
    oblif snd winning_bid < snd bid
    then winning_bid ?= bid;
    else skip;
}

TICK$_L$ (dmy: int$_L$) {
    if round_counter > 0
    then {
        round_counter = round_counter - 1;
        send(ALICE/AUCTION_STATUS, winning_bid);
        send(BOB/AUCTION_STATUS, winning_bid);
        send(AUCTIONTIMER/BEGIN, 1);
    } else {
        send(ALICE/AUCTION_OVER,winning_bid);
        send(BOB/AUCTION_OVER,winning_bid);
    }
}
\end{lstlisting}

\columnbreak

\begin{lstlisting}[caption=Auction timer,label=lst:auction_timer_extended,mathescape]
AUCTIONTIMER

var c : int$_L$ = 0;

BEGIN$_L$ (i : int$_L$) {
    c = i * 2000;
    while (c > 0) do {
        c = c - 1;
    }
    send(AUCTIONHOUSE/TICK, 0);
}
\end{lstlisting}
\end{multicols}

The code for the auction timer (\Cref{lst:auction_timer_extended}) has not been modified and is the same as in \Cref{sec:example}.
Alice (\Cref{lst:auction_alice_extended}) receives auction status messages on channel \texttt{AUCTION\_STATUS} with public context label $L$. If Alice is not currently leading, and if the current winning bid is strictly less than the maximum bid Alice is willing to make, she submits a new, higher bid. Finally, channel \texttt{AUCTION\_OVER} receives a message with the final results of the auction. Bob's code (\Cref{lst:auction_bob_extended}) is equivalent to the code of Alice, though with a different maximum bid.
\begin{multicols}{2}
\begin{lstlisting}[caption=Alice,label=lst:auction_alice_extended,mathescape,escapechar=|]
ALICE

local channel WINNER: (string$_H$*int$_H$)$_H$;

var max_bid : int$_H$ = 432;

AUCTION_STATUS$_L$ $\$1$ (bid : (string$_H$*int$_H$)$_H$) {
    oblif snd bid < max_bid && fst bid != "Alice"
    then send(AUCTIONHOUSE/BID,
        ("Alice", snd bid + 1));
    else skip;
}

AUCTION_OVER$_L$ (winner : (string$_H$*int$_H$)$_H$) {
    output(WINNER, winner);
}
\end{lstlisting}

\columnbreak

\begin{lstlisting}[caption=Bob,label=lst:auction_bob_extended,mathescape,escapechar=|]
BOB

local channel WINNER: (string$_H$*int$_H$)$_H$;

var max_bid : int@$_H$ = 350;

AUCTION_STATUS$_L$ $\$1$ (bid : (string$_H$*int$_H$)$_L$) {
    oblif snd bid < max_bid && fst bid != "Bob"
    then send(AUCTIONHOUSE/BID,
        ("Bob", snd bid + 1));
    else skip;
}

AUCTION_OVER$_L$ (winner : (string$_H$*int$_H$)$_H$) {
    output(WINNER, winner);
}
\end{lstlisting}
\end{multicols}

\subsection{Shopping}
We present a small shopping example consisting of Alice, a luxury shop, and a bank. We let Alice's decision to buy from the shop be public, but if the payment is declined due to lack of funds in her bank account, it would be a huge embarrassment that she is keen to avoid. In this example, all messages except the final confirmation and error message are genuine and sent unconditionally. The semantics of \oblivio{} still ensures that the time at which messages are sent is not influenced by secrets.

The interaction starts by handler \texttt{ALICE/START}, which sends an order request to \texttt{DE\_LUXES\_UNIQUES/ORDER} containing the secret id of the item Alice wants to buy and Alice's secret bank id. The handler for \texttt{DE\_LUXES\_UNIQUES/ORDER} looks up the item and stores it an order buffer before sending a transfer request to the bank on \texttt{BANK/TRANSFER} containing Alice's bank id, the amount to transfer, and the shop's bank id. The handler for \texttt{BANK/TRANSFER} looks whether there is enough money in the source account to make the transfer. If so, it performs the transfer, otherwise, it changes the code to 200. Finally, it sends the code to \texttt{DE\_LUXES\_UNIQUES/RESPONSE}, which then informs Alice whether payment was successful or an error occured.

\begin{center}
\begin{tabular}{c}
\begin{lstlisting}[caption=Alice,label=lst:shopping_alice,mathescape,escapechar=|,linewidth=.8\linewidth]
ALICE

local channel STDOUT : (string$_H$*int$_H$)$_H$;

var item_id : int$_H$;
var bank_id : int$_H$;

START$_L$ (x : int$_L$) {
    send(DE_LUXES_UNIQUES/ORDER,(item_id,bank_id));
}

OK$_H$ (item : (string$_H$*int$_H$)$_H$) {
    output(STDOUT, ("Order accepted! Item: " ^ fst item ^ "Price: ", snd item));
}

ERROR$_H$ (code : int$_H$) {
    output(STDOUT, ("Bank returned error code: ", code));
}
\end{lstlisting}
\end{tabular}
\end{center}

\begin{multicols}{2}
\begin{lstlisting}[caption=Luxury shop,label=lst:shopping_shop,mathescape,escapechar=|]
DE_LUXES_UNIQUES

var bank_id : int$_L$;

var items : (string$_L$*int$_L$)$_L$[]$_L$ =
    [   ("Necklace", 2200)
    ;   ("Earrings", 1350)
    ;   ("Perfume", 475)
    ;   ("Handbag", 699)
    ;   ("Fascinator", 978)
    ;   ("Ring", 1450)
    ];

var order_buf : (string$_H$*int$_H$)$_H$ = ("",0);

ORDER$_L$ (order : (int$_H$*int$_H$)$_L$) {
    order_buf = items[fst order];
    send(BANK/TRANSFER,
        (snd order,(snd order_buf,bank_id)));
}

RESPONSE$_L$ $\$2$ (code : int$_H$) {
    oblif code == 0
    then send(ALICE/OK,order_buf);
    else send(ALICE/ERROR,code);
}
\end{lstlisting}

\columnbreak

\begin{lstlisting}[caption=Bank,label=lst:shopping_bank,mathescape]
BANK

var balance : int$_H$[]$_H$;

var from : int$_H$;
var amount : int$_H$;
var to : int$_H$;
var code : int$_H$;

TRANSFER$_L$(msg : (int$_H$*(int$_H$*int$_H$)$_L$)$_L$) {
    from = fst msg;
    amount = fst snd msg;
    to = snd snd msg;
    code = 0;
    
    oblif amount <= balance[from]
    then {
        balance[from] ?= balance[from] - amount;
        balance[to] ?= balance[to] + amount;
    }
    else code ?= 200;

    send(DE_LUXES_UNIQUES/RESPONSE,code);
}
\end{lstlisting}
\end{multicols}

\subsection{Ring broadcast}
This example demonstrates a small chat where users Alice, Bob, and Carol send messages to each other in a ring structure, such that Alice sends to Bob, Bob sends to Carol, and Carol sends to Alice. Alice can prepare messages on local input channel \texttt{STDIN} which are read and broadcast if the received message is empty or if Alice was the sender of the message. In this way, nodes maintain constant traffic, but only send non-empty string messages if a message is available on \texttt{STDIN}.

\noindent
\begin{minipage}{\linewidth}
\begin{multicols}{2}
\begin{lstlisting}[caption=Alice,label=lst:alice_ring_broadcast,mathescape]
ALICE

local channel STDIN : string$_H$;
local channel STDOUT : string$_H$;

var my_msg : string$_H$ = "";
var msg_out : (string$_H$*string$_H$)$_H$ = ("","");

RCV$_L$ (msg_in: (string$_H$*string$_H$)$_H$) {
    oblif fst msg_in == "" || fst msg_in == "Alice"
    then {
        my_msg ?= input(STDIN, 64);
        oblif my_msg != ""
        then msg_out ?= ("Alice", my_msg);
        else msg_out ?= ("","");
        my_msg ?= "";
    } else {
        output(STDOUT,
            fst msg_in ^ ": " ^ snd msg_in);
        msg_out ?= msg_in;
    }
    
    send(BOB/RCV, msg_out);
}
\end{lstlisting}

\columnbreak

\begin{lstlisting}[caption=Bob,label=lst:bob_ring_broadcast,mathescape]
BOB

local channel STDIN : string$_H$;
local channel STDOUT : string$_H$;

var my_msg : string$_H$ = "";
var msg_out : (string$_H$*string$_H$)$_H$ = ("","");

RCV$_L$ (msg_in: (string$_H$*string$_H$)$_H$) {
    oblif fst msg_in == "" || fst msg_in == "Bob"
    then {
        my_msg ?= input(STDIN, 64);
        oblif my_msg != ""
        then msg_out ?= ("Bob", my_msg);
        else msg_out ?= ("","");
        my_msg ?= "";
    } else {
        output(STDOUT,
            fst msg_in ^ ": " ^ snd msg_in);
        msg_out ?= msg_in;
    }
    
    send(CAROL/RCV, msg_out);
}
\end{lstlisting}
\end{multicols}
\end{minipage}
\begin{center}
\begin{tabular}{c}
\begin{lstlisting}[caption=Carol,label=lst:carol_ring_broadcast,mathescape,linewidth=0.5\linewidth]
CAROL

local channel STDIN : string$_H$;
local channel STDOUT : string$_H$;

var my_msg : string$_H$ = "";
var msg_out : (string$_H$*string$_H$)$_H$ = ("","");

RCV$_L$ (msg_in: (string$_H$*string$_H$)$_H$) {
    oblif fst msg_in == "" || fst msg_in == "Carol"
    then {
        my_msg ?= input(STDIN, 64);
        oblif my_msg != ""
        then msg_out ?= ("Carol", my_msg);
        else msg_out ?= ("","");
        my_msg ?= "";
    } else {
        output(STDOUT, fst msg_in ^ ": " ^ snd msg_in);
        msg_out ?= msg_in;
    }
    
    send(ALICE/RCV, msg_out);
}
\end{lstlisting}
\end{tabular}
\end{center}

\subsection{Dating}
In this example, Alice has set up a small service that implements a dating agent. It can receive dating profiles on \texttt{ALICE/ASK} and checks whether the profile's age is within a pre-specified range and whether any of the profile's interests match Alice's pre-specified condition. As we iterate over the interests in a while-loop, The profile is required to state a public upper bound on the number of interests.

\noindent
\begin{minipage}{\linewidth}
\begin{lstlisting}[caption=Alice,label=lst:alice_dating,mathescape]
ALICE

var min_age : int$_H$;
var max_age : int$_H$;
var condition : string$_H$;

var name_buffer : string$_H$;
var age_buffer : int$_H$;
var n_interests_buffer : int$_L$;
var interests_buffer : string$_H$[]$_H$;

var result : string$_H$ = "";
var score : int$_H$ = 0;
var counter : int$_L$ = 0;

var match_message : string$_H$;

ASK$_L$ $\$1$ (profile: (string$_H$*(int$_H$*(int$_L$*string$_H$[]$_H$)$_L$)$_L$)$_L$) {
    name_buffer = fst profile;
    age_buffer = fst snd profile;
    n_interests_buffer = fst snd snd profile;
    interests_buffer = snd snd snd profile;
    
    oblif min_age <= age_buffer $\andexp$ age_buffer <= max_age
    then score ?= score + 1;
    else skip;
  
    while counter < n_interests_buffer
    do {
        oblif interests_buffer[counter] == condition
        then score ?= score + 1;
        else skip;
        counter = counter + 1;
    }
    counter = 0;

    oblif score >= 2
    then {
        output(MATCH, (name_buffer,(age_buffer,interests_buffer)));
        send(BOB/REPLY, match_message);
    }
    else skip;
    
    name_buffer = "";
    age_buffer = 0;
    n_interests_buffer = 0;
    interests_buffer = [""];
    score = 0;
}
\end{lstlisting}
\end{minipage}

\begin{lstlisting}[caption=Bob,label=lst:bob_dating,mathescape]
BOB

local channel MATCH : string$_H$;

var name : string$_H$;
var age : int$_H$;
var interests : string$_H$[]$_H$;

START$_L$ (ignored: int$_L$) {
    send(ALICE/ASK,(name,(age,(3,interests))));
}

REPLY$_H$ (msg: string$_H$) {
    output(MATCH, msg);
}
\end{lstlisting}

\fi

\end{document}